\documentclass[12pt]{article}
\pdfoutput=1

\usepackage{textcomp}
\usepackage{color}

\usepackage{putex}
\usepackage{autobreak}
\usepackage{graphicx}
\graphicspath{{plots/}}
\usepackage{tabularx}
\usepackage{caption}
\usepackage{amsmath}
\usepackage{array}
\usepackage{subcaption}
\usepackage{epstopdf}
\usepackage{enumerate}
\usepackage{cite}
\usepackage{youngtab}
\usepackage{tensor}
\usepackage{slashed}
\usepackage[aligntableaux=center]{ytableau}
\usepackage[utf8]{inputenc}
\usepackage{rotating}
\usepackage{multirow}
\usepackage[
      colorlinks=true,
      linkcolor=blue,
      urlcolor=blue,
      filecolor=black,
      citecolor=red,
      ]{hyperref}

\newcommand{\grSU}{SU}

\newcommand{\grSL}{SL}

\newcommand{\abs}[1]{\left\lvert #1 \right\rvert}

\newcommand {\be} {\begin {equation}}
\newcommand {\ee} {\end {equation}}

\newcommand {\bes} {\begin {equation*}}
\newcommand {\ees} {\end {equation*}}

\newcommand{\es}[2] {\begin{equation} \label{#1} \begin{split} #2 \end{split} \end{equation}}

\newcommand{\Z}{\mathbb{Z}}

\newcommand{\R}{\mathbb{R}}

\newcommand{\cF}{{\mathcal F}}
\newcommand{\cG}{{\mathcal G}}

\newcommand{\cN}{{\mathcal N}}

\newcommand{\cT}{{\mathcal T}}

\newcommand{\beq}{\begin{equation}}
\newcommand{\eeq}{\end{equation}}

\def\ie{\begin{equation}\begin{aligned}}
\def\fe{\end{aligned}\end{equation}}

\newcommand{\A}{{\alpha}}
\newcommand{\B}{{\beta}}

\newcommand{\mZ}{{\mathbb Z}}

\numberwithin{equation}{section}

\def\<{\langle}
\def\>{\rangle}

\newcommand{\gym}{g_\text{YM}}

\begin{document}

\preprint{PUPT-2627}

\institution{Exile}{Department of Particle Physics and Astrophysics, Weizmann Institute of Science, \cr Rehovot, Israel}
\institution{PU}{Joseph Henry Laboratories, Princeton University, Princeton, NJ 08544, USA}

\title{
Bootstrapping $\mathcal{N}=4$ super-Yang-Mills on the conformal manifold
}

\authors{Shai M.~Chester,\worksat{\Exile} Ross Dempsey,\worksat{\PU} and Silviu S. Pufu\worksat{\PU}  }

\abstract{
We combine supersymmetric localization results with numerical bootstrap techniques to compute upper bounds on the low-lying CFT data of ${\cal N} = 4$ super-Yang-Mills theory as a function of the complexified gauge coupling $\tau$.  In particular, from the stress tensor multiplet four-point function, we extract the scaling dimension of the lowest-lying unprotected scalar operator and its OPE coefficient.  While our method can be applied in principle to any gauge group $G$, we focus on $G = SU(2)$ and $SU(3)$ for simplicity.  At weak coupling, the upper bounds we find are very close to the corresponding four-loop results.  We also give preliminary evidence that these upper bounds become small islands under reasonable assumptions.
}
\date{November 2021}

\maketitle

\tableofcontents

\section{Introduction and Summary}
\label{intro}

Four-dimensional $\mathcal{N}=4$ super-Yang-Mills (SYM) theory is one of the most well-studied models of gauge dynamics in theoretical high energy physics.  Part of the reason for the continued interest is that this theory has a large amount of symmetry.  Indeed,   not only does ${\cal N} = 4$ SYM have the maximal amount of supersymmetry possible in four dimensions for a non-gravitational theory, but it also possesses conformal symmetry as its beta function vanishes identically.   In fact, for any simple gauge group $G$, this theory possesses a conformal manifold parameterized by the complexified gauge coupling $\tau\equiv\frac{\theta}{2\pi}+\frac{4\pi i}{g_\text{YM}^2}$, with $g_\text{YM}$ being the Yang-Mills coupling and $\theta$ the theta-angle.  The case $G = SU(N)$  also features prominently in gauge/gravity duality \cite{Maldacena:1997re, Witten:1998qj,Gubser:1998bc}, where in the combined large $N$ and large 't Hooft coupling $\lambda = g_\text{YM}^2 N$ limit, ${\cal N} = 4$ SYM has a dual description in terms of weakly coupled string theory on $AdS_5 \times S^5$.

$\mathcal{N}=4$ SYM has been studied using a variety of methods in various limits. Indeed, for any gauge group $G$, perturbatively at small $g_\text{YM}$ one can calculate various observables using Feynman diagrams---see, for instance, \cite{Velizhanin:2009gv,Eden:2012rr,Fleury:2019ydf,Eden:2016aqo,Goncalves:2016vir} for cutting-edge results at four loops.   As mentioned above, the $G = SU(N)$ theory can be studied in the combined large $N$ and large $\lambda$ limit using the holographic duality.  Certain observables at large $N$ and finite $\lambda$ have also been computed using the technique of integrability 
(see, for instance,  \cite{Beisert:2010jr,Gromov:2017blm} for reviews and references).  Lastly, exact results at finite $N$ and finite $\lambda$ so far have been restricted to properties of supersymmetry-protected operators that can be deduced either from anomaly considerations, the 2d chiral algebra subsector \cite{Beem:2013sza}, or using the technique of supersymmetric localization (see for instance \cite{Pestun:2007rz,Hama:2012bg,Giombi:2009ds,Giombi:2009ek}, as well as \cite{Pestun:2016zxk} for a collection and reviews and references).  In addition to these exact results, at finite $N$ and finite $\lambda$, one can also obtain bounds on the low-lying local CFT data using the technique of conformal bootstrap \cite{Beem:2013qxa,Beem:2016wfs,Alday:2013opa,Bissi:2020jve,Alday:2021peq}.  In these studies, however, it was not known how to make the bootstrap equations sensitive to the coupling $\tau$, so these bounds necessarily apply to the entire conformal manifold.

In this paper, we will combine supersymmetric localization with the numerical conformal bootstrap to put bounds on local CFT data for any $\tau$.\footnote{See \cite{Baggio:2017mas} for a previous numerical bootstrap study of a conformal manifold for a 3d $\mathcal{N}=2$ theory. In this 3d case, the conformal manifold parameter appeared in the superpotential and so could be related to local CFT data via chiral ring relations.}  We study explicitly the $G = SU(2)$ and $SU(3)$ theories, but our analysis generalizes straightforwardly to any $G$.   In \cite{Binder:2019jwn,Chester:2020dja}, exact relations were derived between certain integrated stress tensor multiplet four-point functions and derivatives $\partial_\tau\partial_{\bar\tau}\partial_m^2F\vert_{m=0}$ and $\partial_m^4F\vert_{m=0}$ of the sphere free energy $F(m,\tau, \bar \tau)$ of the $\mathcal{N}=2^*$ theory,\footnote{The ${\cal N} = 2^*$ theory is a mass deformation of ${\cal N} = 4$ super-Yang-Mills that preserves ${\cal N} = 2$ supersymmetry.} which was computed using localization in terms of a $\text{rank}(G)$-dimensional  integral that depends nontrivially on the full complex $\tau$ \cite{Pestun:2007rz} (see also \cite{Russo:2012ay,Russo:2013qaa,Russo:2013kea}). These two integrated constraints were used in \cite{Binder:2019jwn,Chester:2019pvm,Chester:2020dja,Chester:2019jas,Chester:2020vyz} to constrain the large $N$ expansion of the correlator.  In particular, by combining these constraints with analyticity in Mellin space and crossing symmetry, Refs.~\cite{Binder:2019jwn,Chester:2019pvm,Chester:2020dja,Chester:2019jas,Chester:2020vyz} determined the first few $1/N^2$ corrections at finite $\tau$, which correspond to protected higher-derivative corrections to the supergravity action on $AdS_5$, and matched the type IIB S-matrix in the flat space limit. 

Here, we explain how to combine these two integrated constraints with the infinitely many constraints from crossing symmetry and unitarity. When expanded in conformal blocks, the integrated constraints and the crossing symmetry constraints receive contributions from blocks of large $\Delta$ that differ exponentially in $\Delta$. This leads to difficulties in combining these constraints. In the typical numerical formulation of the bootstrap problem, one would approximate the crossing symmetry constraints by a polynomial in $\Delta$ times a positive factor.  If one were to do the same for the integrated constraints, the difference in the large $\Delta$ behavior would require approximating an exponential by a polynomial. This approach was in fact used in an analogous 2D study \cite{Lin:2015wcg}, but in our case we found that such an approximation made the bootstrap insensitive to the integrated constraints.  Instead, we employed the approach to the conformal bootstrap based on linear programming\cite{Rattazzi:2008pe}, in which the values of $\Delta$ are discretized up to a cutoff. In this way, both the crossing and integrated constraints become a set of linear constraints for which feasibility can be evaluated using standard linear programming methods.

In practice, because the localization input used in the integrated constraints for the $SU(N)$ gauge theory is an $(N-1)$-dimensional integral, we only applied this approach to $\mathcal{N}=4$ SYM with gauge group $SU(2)$ and $SU(3)$. We computed upper bounds on the scaling dimension and OPE coefficient of the lowest dimension scalar unprotected operator (at weak coupling, this is the Konishi operator).   At weak coupling, we found that our upper bounds are nearly saturated by the four-loop results.  When scanning over all couplings, we found that the maximal value of the scaling dimension bound occurs at the self-dual point $\tau=e^{i\pi/3}$ with enhanced $\mathbb{Z}_3$ symmetry, and that this maximal value was strictly lower than the bound obtained without using the integrated constraints \cite{Beem:2013qxa,Beem:2016wfs,Alday:2013opa,Bissi:2020jve,Alday:2021peq}. Furthermore, after imposing a  gap above the lowest dimension operator, we found lower bounds that converge to the upper bounds as the gap is increased.

The rest of this paper is organized as follows. In Section~\ref{4point} we review the conformal block decomposition of the stress tensor four-point correlator, the definitions of the integrated constraints, and weak coupling results for local CFT data.  In Section~\ref{numBoot}, we show how to combine the integrated constraints with the usual crossing symmetry constraints to numerically bootstrap $\mathcal{N}=4$ SYM as a function of $\tau$ for any gauge group.  We carry out this procedure for the $SU(2)$ and $SU(3)$ theories. Finally, in Section~\ref{disc} we end with a discussion of our results and of future directions. Various technical details are discussed in the Appendices.

\section{Stress tensor four-point function}
\label{4point}

The main object of study in this paper is the stress tensor multiplet four-point function. We begin by discussing general constraints coming from $\mathcal{N}=4$ superconformal invariance, including the expansion in superblocks. We then review the two exact constraints on the correlator that relate certain integrals of the correlator to derivatives of the mass deformed sphere free energy, which can be efficiently computed using supersymmetric localization. Finally, we review analytic weak coupling predictions for CFT data that appears in the correlator, which we will compare to our numerical finite $N$ and $\tau$ results in the next section.

\subsection{Setup}
\label{setup}

Let us denote the bottom component of the stress tensor multiplet by $S$.  This operator is a dimension 2 scalar in the ${\bf 20}'$ irrep of the $\mathfrak{su}(4)_R \cong \mathfrak{so}(6)_R$ R-symmetry algebra, and can thus be represented as a rank-two traceless symmetric tensor $S_{IJ}(\vec{x})$, with indices $I, J = 1, \ldots, 6$.  For simplicity we will contract these indices with null polarization vectors $Y^I$, where $Y \cdot Y = 0$.  We are interested in studying the four-point function $\langle SSSS\rangle$, which is fixed by conformal and $\mathfrak{su}(4)$ symmetry to take the form
 \es{2222}{
 & \langle S(\vec x_1,Y_1) S(\vec x_2,Y_2) S(\vec x_3,Y_3) S(\vec x_4,Y_4) \rangle = \frac{Y^2_{12}Y^2_{34}}{{x}_{12}^4 {x}_{34}^{4}} \mathcal{S}(U,V;\sigma,\tau)\,,
    }
where we define $\vec{x}_{ij} \equiv \vec{x}_i - \vec{x}_j$ and
 \es{uvsigmatauDefs}{
  U \equiv \frac{{x}_{12}^2 {x}_{34}^2}{{x}_{13}^2 {x}_{24}^2} \,, \qquad
   V \equiv \frac{{x}_{14}^2 {x}_{23}^2}{{x}_{13}^2 {x}_{24}^2}  \,, \qquad
   \sigma\equiv\frac{(Y_1\cdot Y_3)(Y_2\cdot Y_4)}{(Y_1\cdot Y_2)(Y_3\cdot Y_4)}\,,\qquad \tau\equiv\frac{(Y_1\cdot Y_4)(Y_2\cdot Y_3)}{(Y_1\cdot Y_2)(Y_3\cdot Y_4)} \,.
 }
 The constraints of superconformal symmetry are given by the Ward identity in \cite{Dolan:2001tt}, whose solution can be formally written in two different ways. The first expression is
  \es{T}{
 \mathcal{S}(U,V;\sigma,\tau)&=\mathcal{S}_\text{free}(U,V;\sigma,\tau)+\Theta(U,V;\sigma,\tau)\mathcal{T}(U,V)\,,\\
 \Theta(U,V;\sigma,\tau)&\equiv\tau+(1-\sigma-\tau)V+\tau(\tau-1-\sigma)U+\sigma(\sigma-1-\tau)UV+\sigma V^2+\sigma\tau U^2\,,
 }
where $\mathcal{S}_\text{free}(U,V;\sigma,\tau)$ is a free theory correlator in a theory of $4c = N^2-1$ scalar fields.  In particular, in such a free theory, $S_{IJ} = (8c)^{-1/2}\left(X_I^a X_J^a - \frac{1}{6} \delta_{IJ} X_K^a X_K^a\right)$, where $a = 1, \ldots, 4c$, and using Wick contractions with the propagator $\langle X_I^a(\vec{x}) X_J^b(0) \rangle = \frac{\delta^{ab} \delta_{IJ} }{\abs{\vec{x}}^2}$, one obtains
\es{free}{
\mathcal{S}_\text{free}(U,V;\sigma,\tau)=1+U^2\sigma^2+\frac{U^2}{V^2}\tau^2+\frac{1}{c}\left(U\sigma+\frac UV\tau+\frac{U^2}{V}\sigma\tau\right)\,.
}
Thus, all information about the interacting ${\cal N} =4$ SYM theory is encoded in the function $\mathcal{T}(U,V)$. 

Another way of solving the superconformal Ward identity is \cite{Dolan:2001tt,Beem:2016wfs}
\es{redSText2}{
&\mathcal{S}(U,V;\sigma,\tau) = \Theta(U,V;\sigma,\tau)\cG(U,V) \\
&{}+ \biggl [ \Phi_1(z,\bar z;\sigma,\tau)f_1(z)+\Phi_2(z,\bar z;\sigma,\tau)f_2(z)
+\Phi_3(z, \bar z; \sigma,\tau) f_3(z)+( z \leftrightarrow \bar z) \biggr] \,,\\
}
where $\Theta$ is the same as in \eqref{T} and we define
\es{Phis}{
U&=z\bar z\,,\qquad V=(1-z)(1-\bar z)\,,\qquad \Phi_1\equiv   \frac{z\bar z}{z-\bar z}+ \frac{z^2\bar z\tau}{(z-\bar z)( 1-z)} + \frac{z\bar z^2\sigma}{\bar z-z}\,,\qquad \Phi_2\equiv  \sigma U \,,\\
\Phi_3&\equiv   \frac{z(\bar z-1)}{z-\bar z}+\frac{z^2\bar z \sigma^2 }{\bar z-z} +\frac{z^2\bar z \tau^2 }{(z-1)(z-\bar z)}+\frac{z^2\bar z(z-2)\sigma\tau }{(z-1)(z-\bar z)}  +\frac{z(z+\bar z-2z\bar z)\tau}{(z-1)(z-\bar z)} +\frac{z(z+\bar z-\bar z^2)\sigma}{z-\bar z}\,.\\
}
If we then equate \eqref{redSText2} to \eqref{T}, we find that the free theory correlator fixes the holomorphic functions $f_j$ to be
 \es{freef}{
  f_1(z) = 2+\frac1c-\frac1z+\frac{1}{z-1}\,, \quad
   f_2(z) = - \frac{3}{2} - \frac{1}{2c} + z + \frac 1z \,, \quad f_3(z)=1+\frac1c-z+\frac{1}{1-z}  \,,
 }
 while the remaining free part contributes to $\cG(U,V)$, which is then related to $\cT(U,V)$ as
 \es{GtoT}{
 \cT(U,V)=\cG(U,V)-(1+V^{-2}+c^{-1}V^{-1})\,.
 }
 We can expand $\cG(U,V)$\footnote{We could also in principle expand $\mathcal{T}(U,V)$ in blocks, but then the OPE coefficients squared would not necessarily be positive, because the free theory correlator in the definition of $\mathcal{T}(U,V)$ in \eqref{T} contributes to both long and short multiplets.} in terms of long and short multiplets as
 \es{Gexp}{
 \cG(U,V)=U^{-2}\sum_{\ell=0,2,\dots}\sum_{\Delta\geq\ell+2}\lambda^2_{\Delta,\ell}G_{\Delta+4,\ell}(U,V)+\mathcal{F}^{(0)}_\text{short}(U,V)+c^{-1}\mathcal{F}^{(1)}_\text{short}(U,V)\,,
 }
 where $G_{\Delta,\ell}(U,V)$ with scaling dimension $\Delta$ and spin $\ell$ are 4d conformal blocks
 \es{4dblock}{
 G_{\Delta,\ell}(U,V) &=\frac{z\bar z}{z-\bar z}(k_{\Delta+\ell}(z)k_{\Delta-\ell-2}(\bar z)-k_{\Delta+\ell}(\bar z)k_{\Delta-\ell-2}( z))\,,\\
k_h(z)&\equiv z^{\frac h2}{}_2F_1(h/2,h/2,h,z)\,.
 }
The long multiplets that appear in the $S\times S$ OPE must have even $\ell$ and satisfy the unitarity bound $\Delta\geq\ell+2$.  The short multiplet OPE coefficients do not depend on the coupling and so can be computed from the free theory to give the exact expressions $\mathcal{F}_\text{short}^{(0)}$ and $\mathcal{F}_\text{short}^{(1)}$ \cite{Beem:2016wfs} for $c\geq\frac34$:\footnote{For $c<\frac34$, unitarity requires that certain short multiplets have modified OPE coefficients, and that in particular higher spin currents appear, so all such theories must be free theories \cite{Beem:2016wfs}. We will not consider $c<\frac34$ in this paper. For $c\geq\frac34$, these formulae apply to both free and interacting theories, where for the free theory there are long multiplets that saturate the unitarity bound with OPE coefficients given in \eqref{freeOPE}, at which point these long multiplets become conserved current multiplets as expected for a free theory. }
   \es{Fshort}{
 \cF^{(0)}_\text{short}=&\frac{6 \left(-2 z \bar z \left(z^2+z \bar z+\bar z^2-4\right)+(z+\bar z) \left(z^2
   \bar z^2+z^2+\bar z^2-6\right)+4\right)}{(1-z)^2 z (1-\bar z)^2 \bar z}+\frac{24
   \log (1-z) \log (1-\bar z)}{z^2 \bar z^2}\\
   &+\frac{2 \left(z \left(2
   \bar z^4-\bar z^3+4 \bar z^2-18 \bar z+12\right)-3 \left(\bar z^4-6
   \bar z^2+4 \bar z\right)\right) \log (1-z)}{z^2 (1-\bar z)^2 \bar z
   (z-\bar z)}\\
   &+\frac{2 \left(3 \left(z^4-6 z^2+4 z\right)-\left(2 z^4-z^3+4 z^2-18
   z+12\right) \bar z\right) \log (1-\bar z)}{(1-z)^2 z \bar z^2 (z-\bar z)}\,,\\
 \cF^{(1)}_\text{short}=& \frac{36 \log (1-z) \log (1-\bar z)}{z^2 \bar z^2}-\frac{2 \left(\frac{9
   \bar z-18}{z^2 \bar z}+\frac{4}{z-\bar z}-\frac{4}{z}\right) \log
   (1-z)}{1-\bar z}\\
   &-\frac{2 \left(\frac{9 z-18}{z
   \bar z^2}-\frac{4}{z-\bar z}-\frac{4}{\bar z}\right) \log
   (1-\bar z)}{1-z}+\frac{18 \left(\frac{1}{(1-z) (1-\bar z)}+1\right)}{z \bar z}\,.
 }
All non-trivial interacting information in the block expansion \eqref{Gexp} of the correlator is then given by $\Delta$ and $\ell$ for the long multiplets. This data is constrained by the crossing equation derived by swapping $1\leftrightarrow3$ in \eqref{2222}, which in terms of the block expansion in \eqref{Gexp} leads to
\es{crossing}{
\sum_{\ell=0,2,\ldots}\sum_{\Delta\geq\ell+2}\lambda^2_{\Delta,\ell}F_{\Delta,\ell}(U,V)+F^{(0)}_\text{short}(U,V)+c^{-1}F^{(1)}_\text{short}(U,V)=0\,,
}
where we defined
\es{fdef}{
F_{\Delta,\ell}(U,V)&\equiv V^4G_{\Delta+4,\ell}(U,V)-U^4G_{\Delta+4,\ell}(V,U)\,,\\
F^{(0)}_\text{short}(U,V)&\equiv U^2V^4\cF^{(0)}_\text{short}(U,V)-U^4V^2\cF^{(0)}_\text{short}(V,U)+U^4V^2-V^4U^2\,,\\
F^{(1)}_\text{short}(U,V)&\equiv U^2V^4\cF^{(1)}_\text{short}(U,V)-U^4V^2\cF^{(1)}_\text{short}(V,U)+{U^2V^2(U-V) }\,.\\
} 

\subsection{Integrated constraints from supersymmetric localization}
\label{intCon}

Another source of nonperturbative constraints on the CFT data comes from the two integrated constraints discussed in the introduction, which relate certain integrals on $S^4$ of $\langle SSSS\rangle$ to derivatives of the mass-deformed sphere free energy $F(m,\tau,\bar\tau)$. These constraints can be written in terms of the function $\cT(U,V)$ defined in \eqref{T} as\cite{Binder:2019jwn,Chester:2020dja}
 \es{constraint1}{
\frac{1}{8c}\frac{\partial_m^2\partial_\tau\partial_{\bar\tau} F}{\partial_\tau\partial_{\bar\tau} F}\Big\vert_{m=0}&=I_2\left[ {\cal T}\right] \,,\\
  48 \zeta(3) c^{-1}+c^{-2}{\partial^4_m  F}\big\vert_{m=0}  &=I_4\left[ {\cal T}\right]\,.
 }
Here,\footnote{For later convenience, we defined $I_4$ with a minus sign relative to \cite{Chester:2020dja}.}
  \es{ints}{
   I_{2}[f]&\equiv -\frac{2}{\pi}\int dR\, d\theta\, \frac{R^3\sin^2\theta f(U,V)}{U^2}\bigg|_{\substack{U = 1 + R^2 - 2 R \cos \theta \\
    V = R^2}}\,,\\
       I_4[f]&\equiv-\frac{32}{\pi}  \int dR\, d\theta\, R^3 \sin^2 \theta \, (U^{-1}+U^{-2}V+U^{-2})\bar{D}_{1,1,1,1}(U,V) f(U, V) \bigg|_{\substack{U = 1 + R^2 - 2 R \cos \theta \\
    V = R^2}}\,,\\
 }
 where $\bar{D}_{1,1,1,1}(U,V)$ is a scalar 1-loop box integral that can be written explicitly as
 \es{Dbar}{
\bar{D}_{1,1,1,1}(U,V)= \frac{1}{z-\bar z}\left(\log(z \bar z)\log\frac{1-z}{1-\bar z}+2\text{Li}(z)-2\text{Li}(\bar z)\right)\,.
 }
Using \eqref{GtoT} and \eqref{Gexp}, the function $\cT$ appearing in \eqref{constraint1} takes the form 
 \es{calTBlocks}{
  {\cal T}(U, V) = \sum_{\Delta,\ell}\lambda^2_{\Delta,\ell}\frac{G_{\Delta+4,\ell}}{U^2}
   + \cT_\text{short}(U, V) \,, \quad
    \cT_\text{short}(U, V) \equiv \mathcal{F}^{(0)}_\text{short}+\frac{\mathcal{F}^{(1)}_\text{short}}{c}-\Big(1+\frac{1}{V^{2}}+\frac{1}{cV}\Big) \,.
 }

\begin{figure}
\centering
{\large $\grSU(2)$}\\[1em]%
\begin{subfigure}{0.48\linewidth}%
	\centering
	\includegraphics[width=0.99\linewidth]{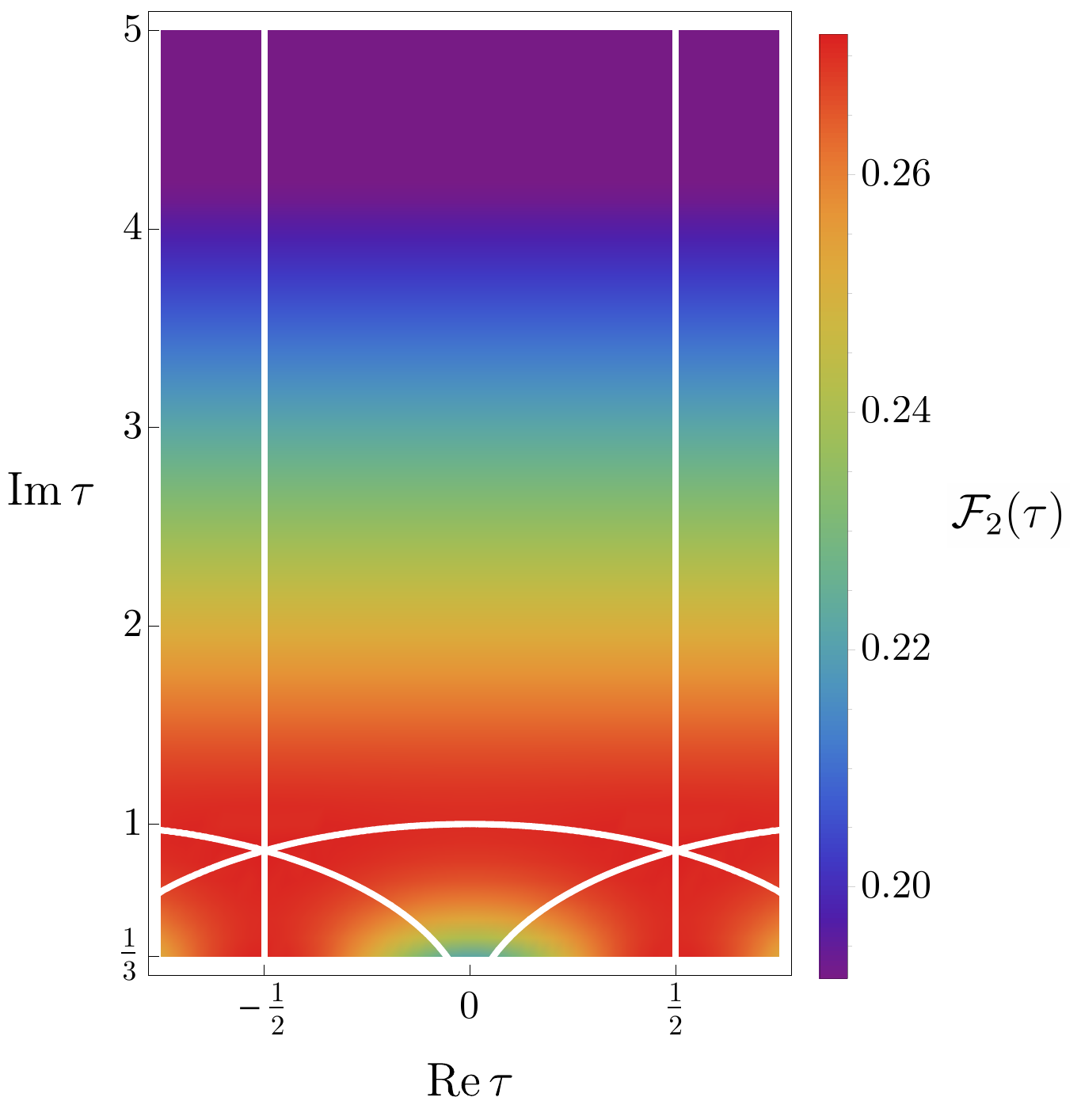}
\end{subfigure}%
\hspace{.04\linewidth}%
\begin{subfigure}{0.48\linewidth}%
	\centering
	\includegraphics[width=0.99\linewidth]{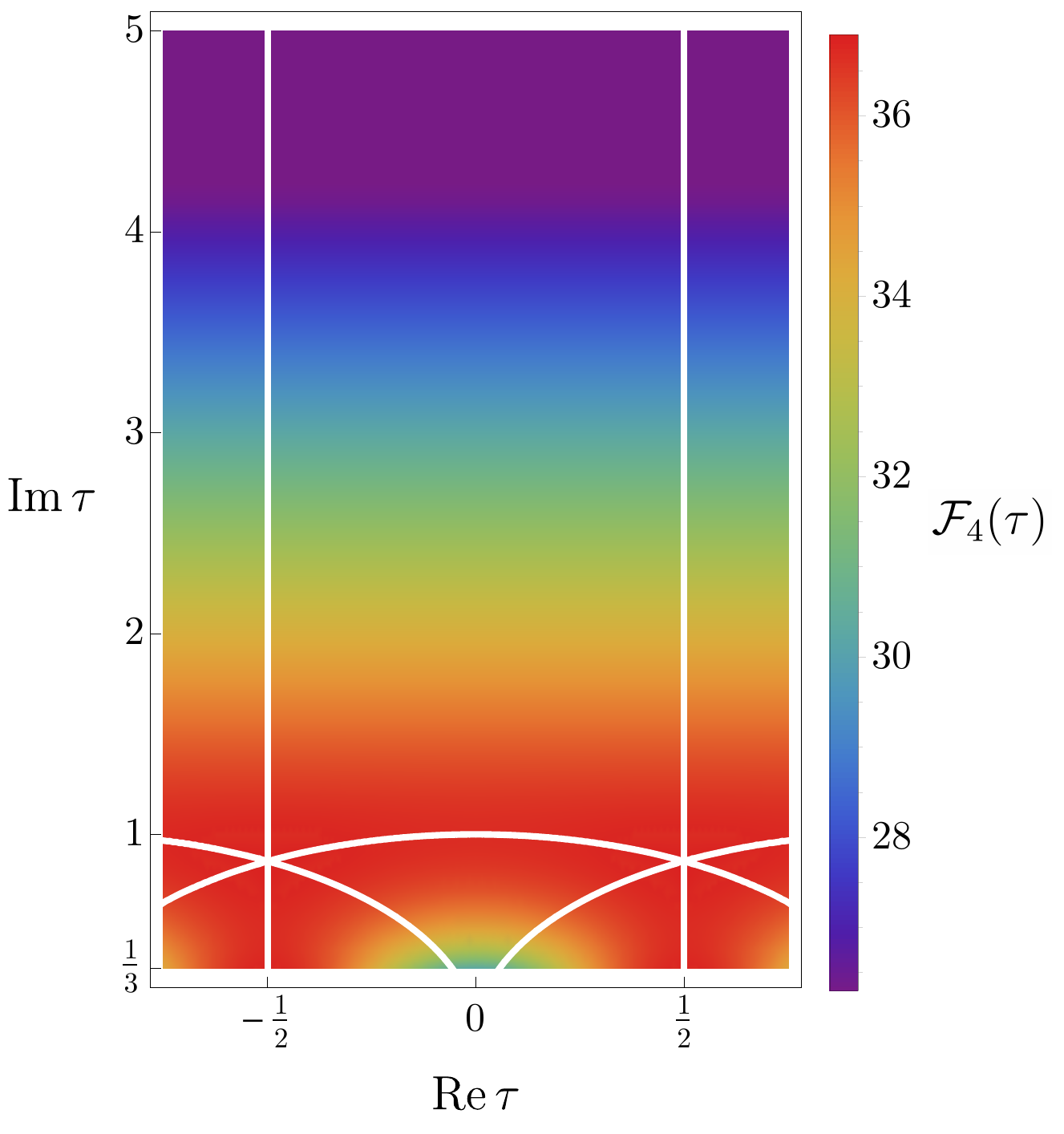}
\end{subfigure}\\[2cm]%
{\large $\grSU(3)$}\\[1em]
\begin{subfigure}{0.48\linewidth}%
	\centering
	\includegraphics[width=0.99\linewidth]{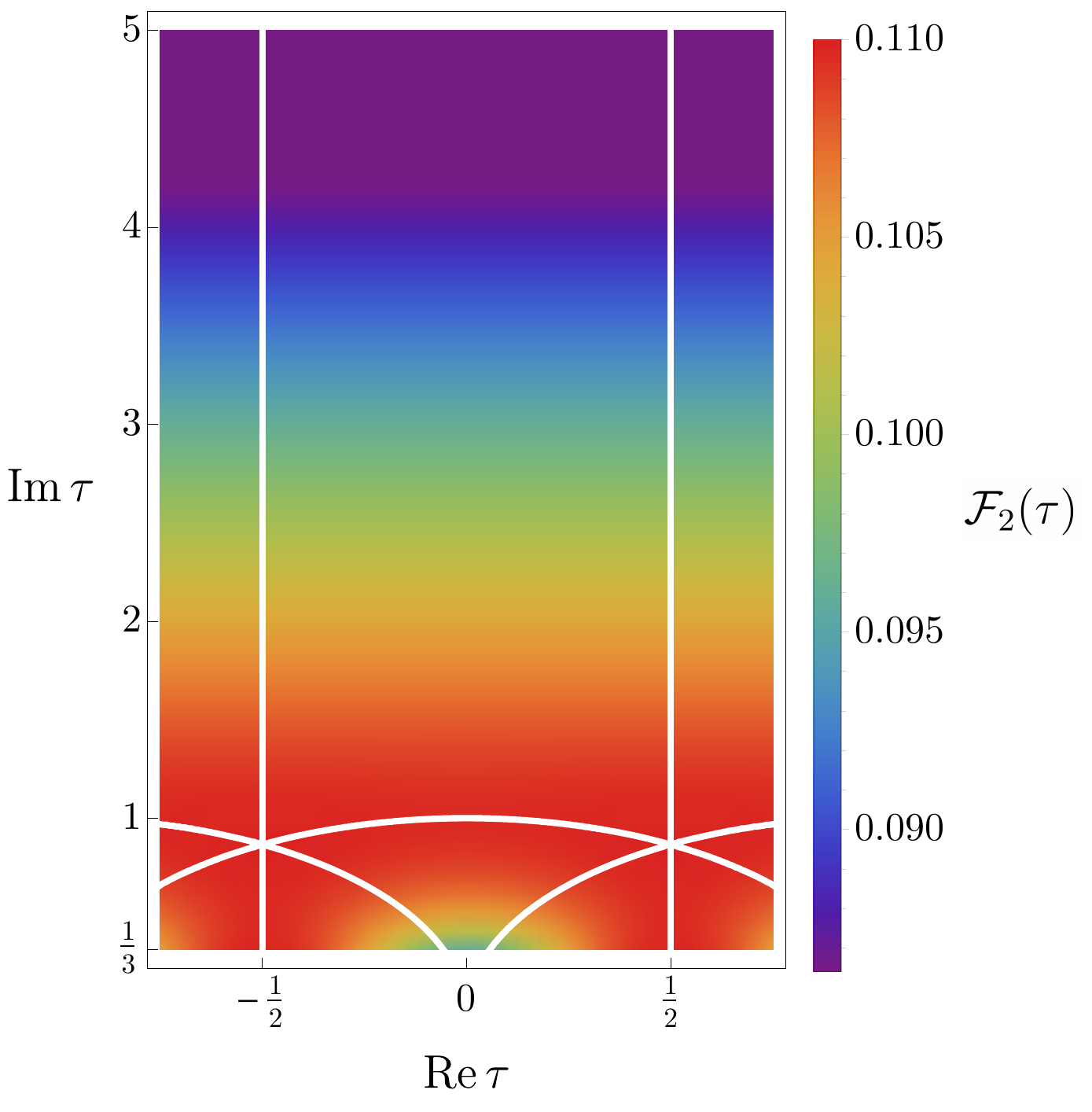}
\end{subfigure}%
\hspace{.04\linewidth}%
\begin{subfigure}{0.48\linewidth}%
	\centering
	\includegraphics[width=0.99\linewidth]{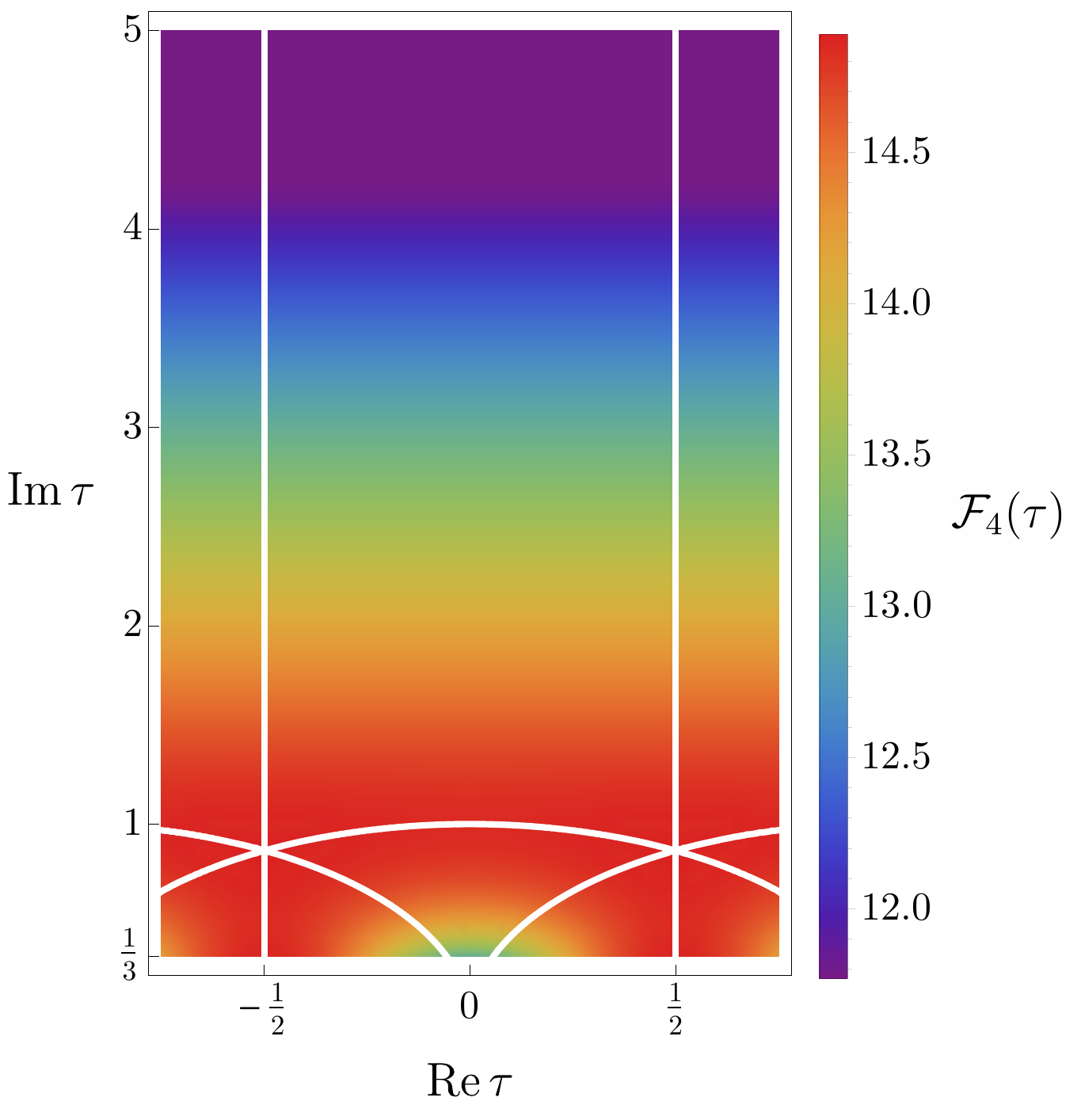}
\end{subfigure}%
 \caption{The localization inputs for $\grSU(2)$ and $\grSU(3)$ as a function of the complex coupling $\tau$. White lines show boundaries between some of the fundamental domains of $\grSL(2,\Z)$. We compute these within the standard fundamental domain $|z| > 1$, $|\Re z| \leq \frac{1}{2}$ using terms with up to 10 instantons in \eqref{ZInstSum} for $\grSU(2)$, and terms with up to 2 instantons for $\grSU(3)$, and extend by $\grSL(2,\Z)$ invariance to the remainder of the upper half-plane.}
\label{fig:localization_fd}
\end{figure}

\begin{figure}
\centering
\begin{subfigure}{0.48\linewidth}%
	\centering
	\includegraphics[width=0.99\linewidth]{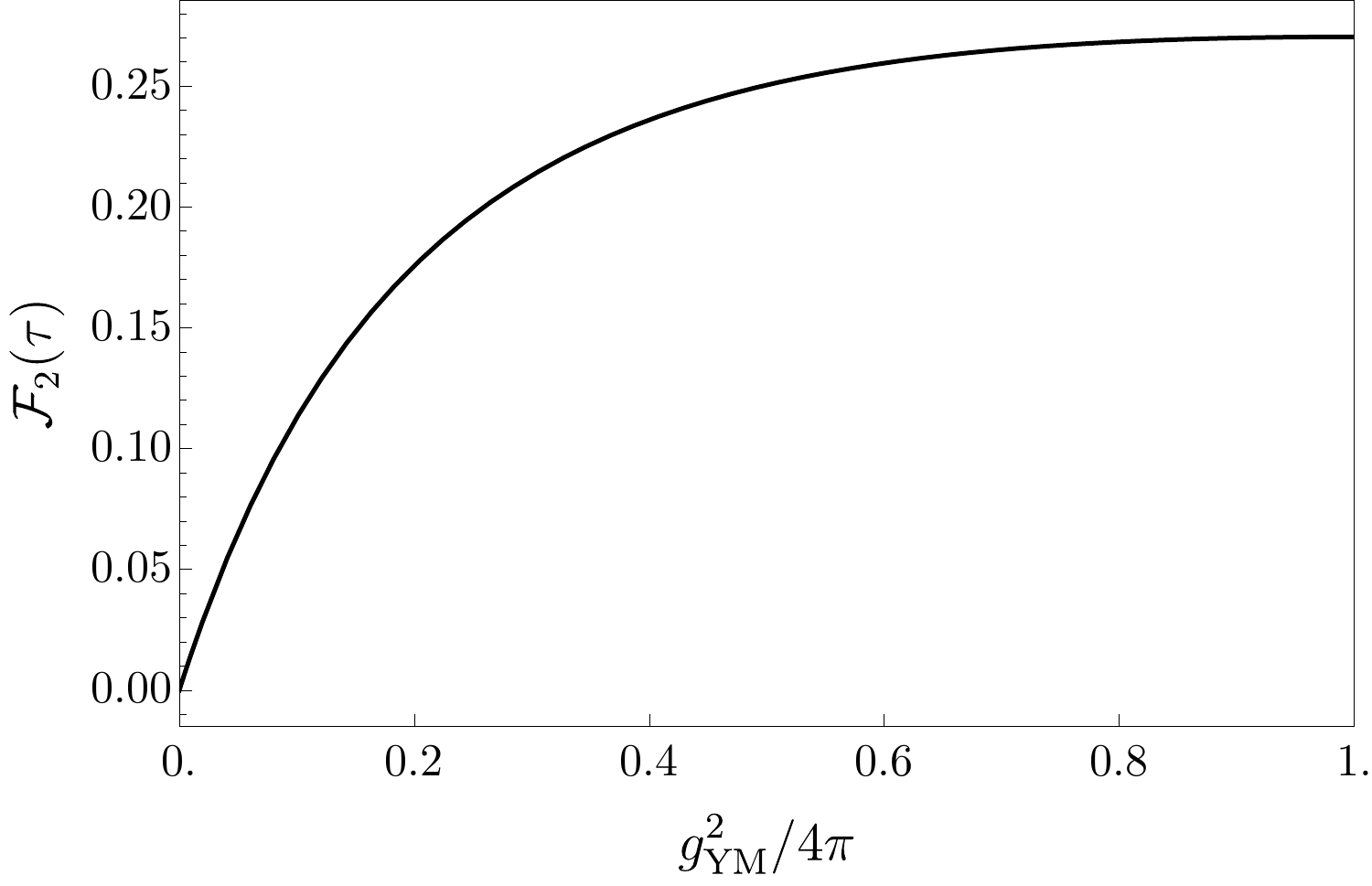}
\end{subfigure}%
\hspace{.04\linewidth}%
\begin{subfigure}{0.48\linewidth}%
	\centering
	\includegraphics[width=0.99\linewidth]{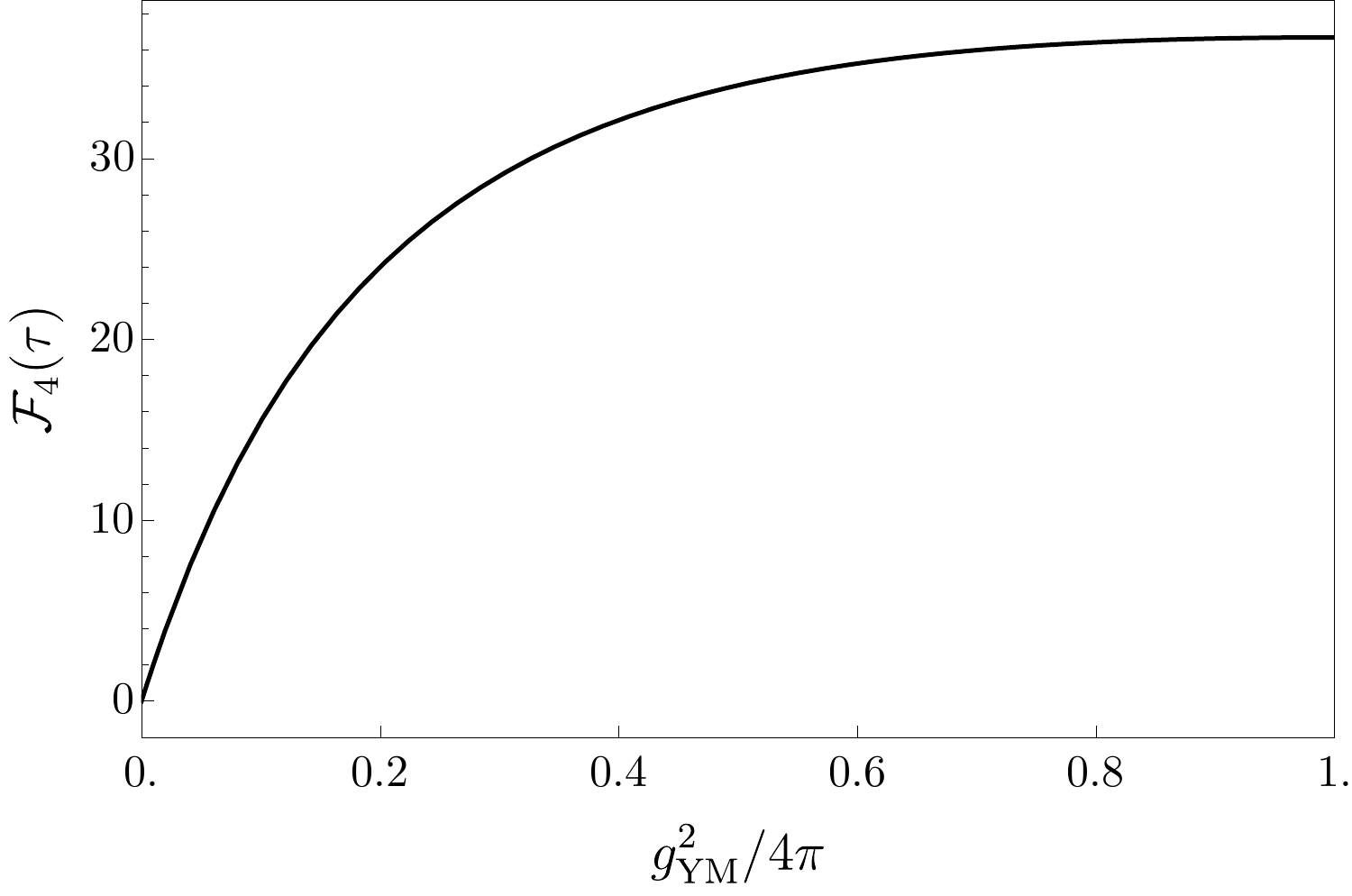}
\end{subfigure}\\[3cm]
\begin{subfigure}{0.48\linewidth}%
	\centering
	\includegraphics[width=0.99\linewidth]{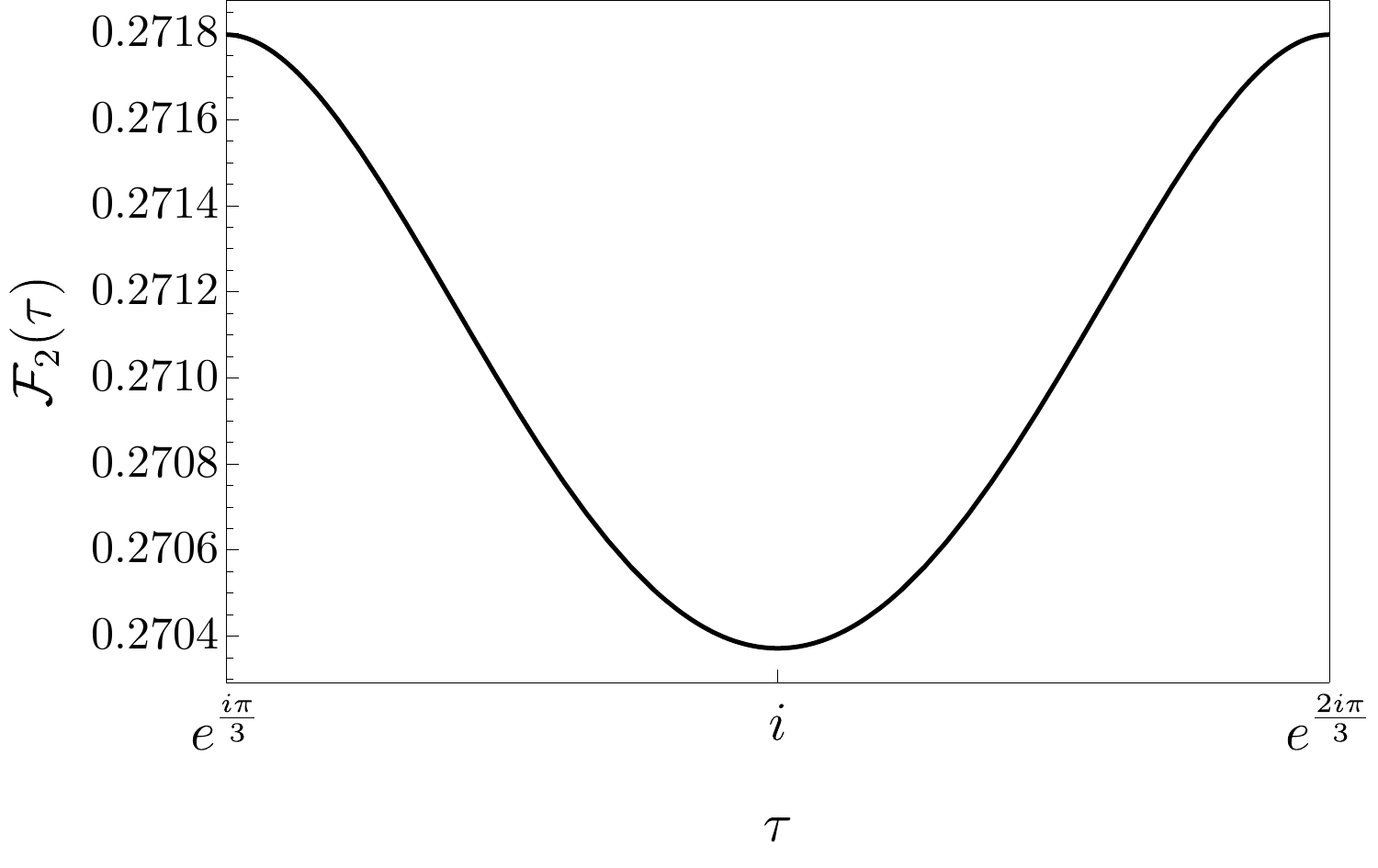}
\end{subfigure}%
\hspace{.04\linewidth}%
\begin{subfigure}{0.48\linewidth}%
	\centering
	\includegraphics[width=0.99\linewidth]{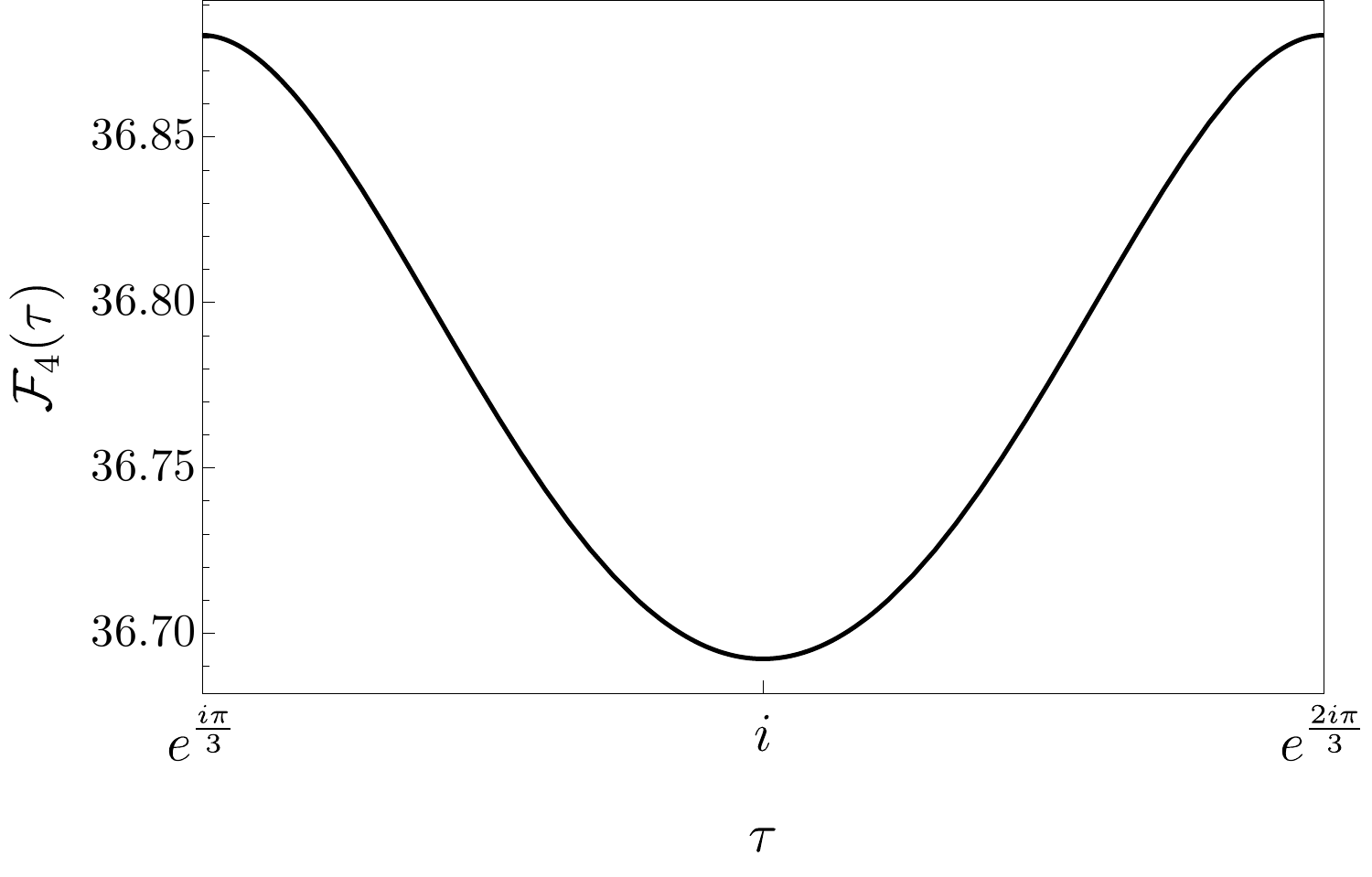}
\end{subfigure}%
 \caption{The localization inputs for $\grSU(2)$ for $\theta = 0$ as a function of $\gym^2$ (\textbf{above}), and as a function of $\tau$ along the arc from $e^{\pi i/3}$ to $e^{2\pi i/3}$ (\textbf{below}). We include terms with up to 10 instantons in \eqref{ZInstSum}, which as shown in Appendix \ref{loc} is more than sufficient for convergence inside the fundamental domain. Note that the slope of these inputs goes to zero at the $\Z_2$ self-dual point $\tau = i$ and at the $\Z_3$ self-dual point $\tau = e^{\pi i/3}$.}
\label{fig:localization_n2_cross_sections}
\end{figure}

\begin{figure}
\centering
\begin{subfigure}{0.48\linewidth}%
	\centering
	\includegraphics[width=0.99\linewidth]{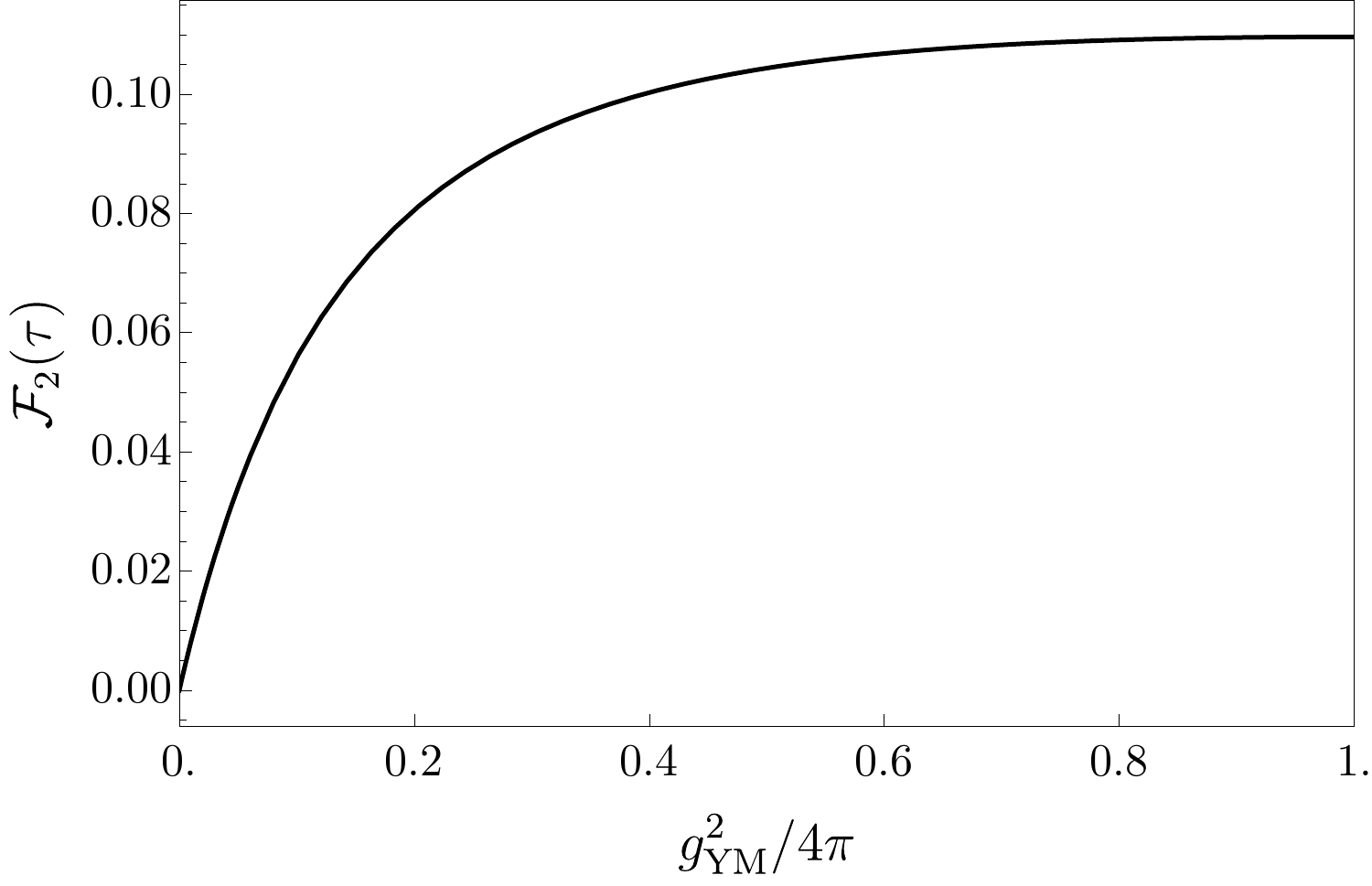}
\end{subfigure}%
\hspace{.04\linewidth}%
\begin{subfigure}{0.48\linewidth}%
	\centering
	\includegraphics[width=0.99\linewidth]{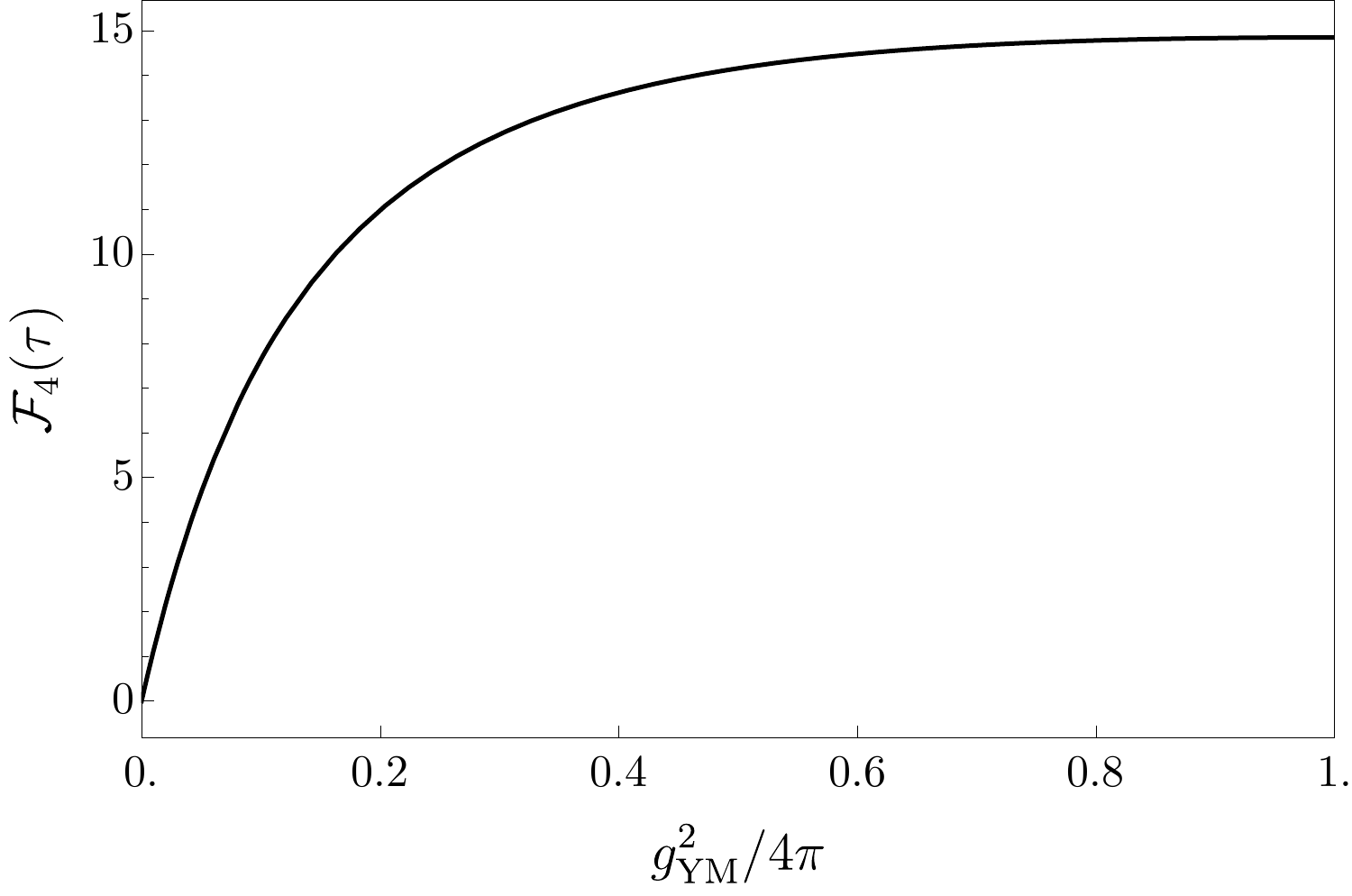}
\end{subfigure}\\[3cm]
\begin{subfigure}{0.48\linewidth}%
	\centering
	\includegraphics[width=0.99\linewidth]{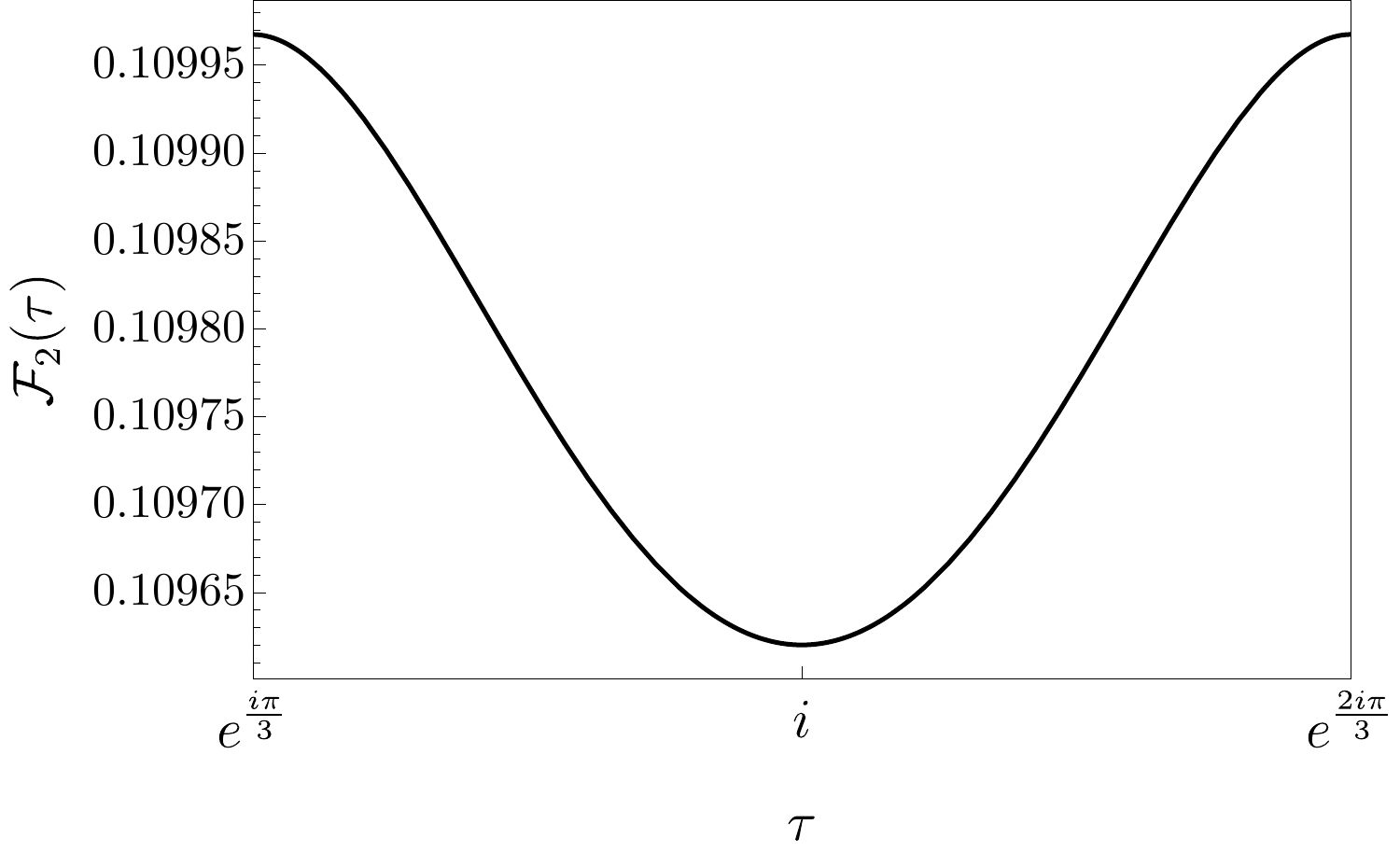}
\end{subfigure}%
\hspace{.04\linewidth}%
\begin{subfigure}{0.48\linewidth}%
	\centering
	\includegraphics[width=0.99\linewidth]{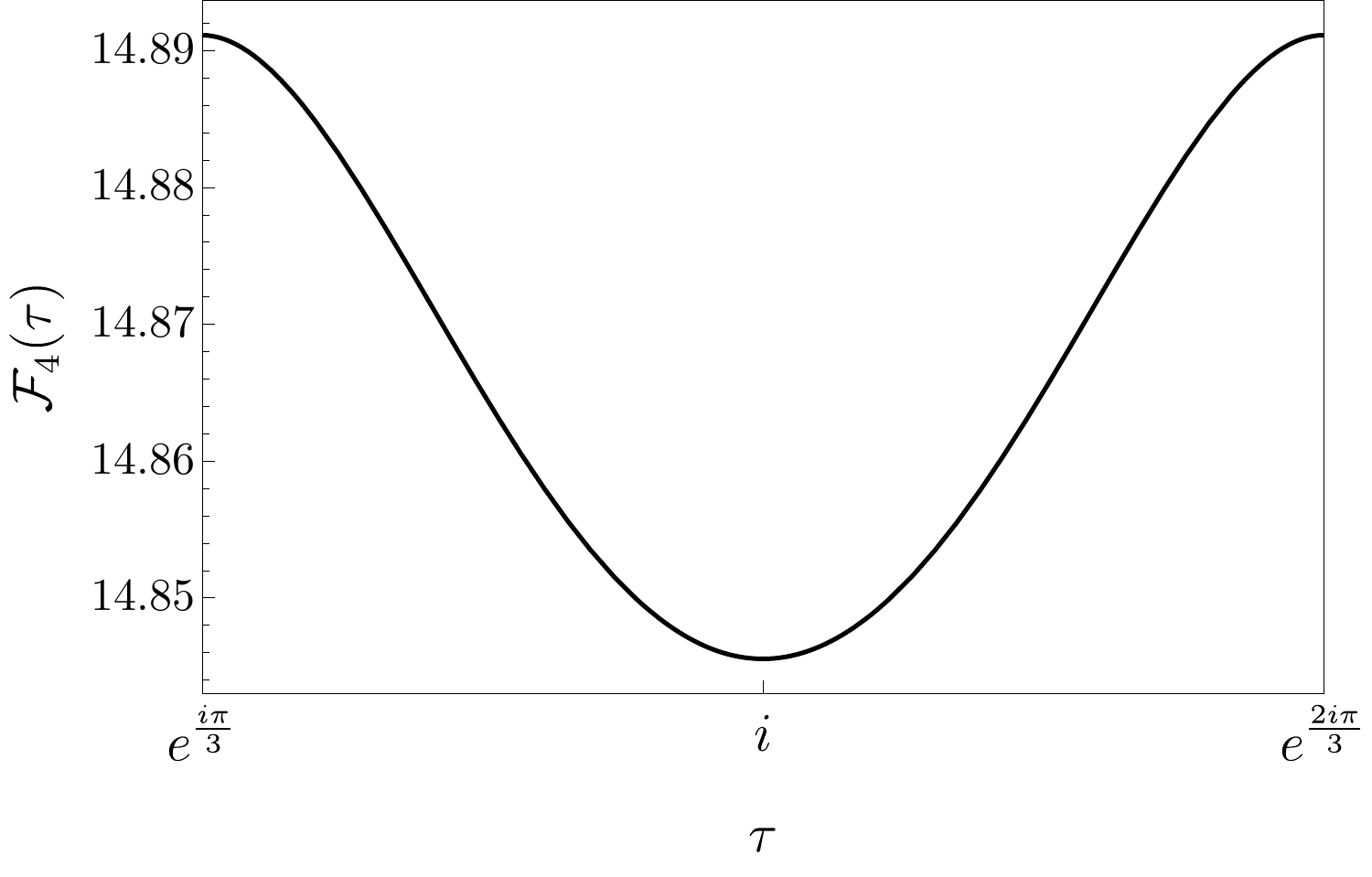}
\end{subfigure}%
 \caption{The localization inputs for $\grSU(3)$ at $\theta = 0$ as a function of $\gym^2$ (\textbf{above}), and as a function of $\tau$ along the arc from $e^{\pi i/3}$ to $e^{2\pi i/3}$ (\textbf{below}). We include terms with up to 2 instantons in \eqref{ZInstSum}, which as shown in Appendix \ref{loc} is sufficient for convergence inside the fundamental domain. Note that the slope of these inputs goes to zero at the $\Z_2$ self-dual point $\tau = i$ and at the $\Z_3$ self-dual point $\tau = e^{\pi i/3}$.}
\label{fig:localization_n3_cross_sections}
\end{figure}

The left-hand sides of the constraints \eqref{constraint1} are written in terms of the mass-deformed free energy $F(m, \tau, \bar \tau)$.  The partition function $Z(m,\tau,\bar\tau)\equiv \exp(-F(m,\tau,\bar\tau))$ was computed using supersymmetric localization in \cite{Pestun:2007rz} for $\mathcal{N}=4$ SYM with gauge group $G$ in terms of a $\text{rank}(G)$-dimensional integral.\footnote{If we ignore the instanton contribution, then the mass derivatives we consider can be computed exactly using the method of orthogonal polynomials \cite{Chester:2019pvm,Chester:2020dja}.} We are mostly interested in the $G=SU(N)$ expression,
  \es{ZFull}{
  Z(m, \tau, \bar \tau) = \int \frac{d^{N} a}{N!} \,  \frac{ \delta \left( \sum_i a_i \right) \prod_{i < j}a_{ij}^2 H^2(a_{ij})}{ H(m)^{N-1} \prod_{i \neq j} H(a_{ij}+ m)}e^{- \frac{8 \pi^2}{g_\text{YM}^2} \sum_i a_i^2} \abs{Z_\text{inst}(m, \tau, a_{ij})}^2 \,,
 }
where $a_{ij}\equiv a_i-a_j$.  Here, the integration is over $N$ real variables $a_i$, $i = 1, \ldots, N$, subject to the constraint $\sum_i a_i = 0$ imposed by the delta function, and $H(m) \equiv e^{-(1 + \gamma) m^2} G(1 + im) G(1 - im)$, where $G$ is the Barnes G-function and $\gamma$ is the Euler-Mascheroni constant. The Nekrasov partition function $Z_\text{inst}$ encodes the contribution from instantons localized at the north pole of $S^4$, and can be conveniently expanded as
 \es{ZInstSum}{
  Z_\text{inst}(m, \tau,  a_{ij}) = \sum_{k=0}^\infty e^{2 \pi i k \tau} Z_\text{inst}^{(k)} (m, a_{ij}) \,,
 } 
with $Z_\text{inst}^{(k)} (m, a_{ij})$ representing the contribution of the $k$-instanton sector and normalized such that  $Z_\text{inst}^{(0)} (m,  a_{ij}) = 1$. Each $Z_\text{inst}^{(k)} (m, a_{ij})$ can be explicitly computed using the expressions in \cite{Nekrasov:2002qd,Nekrasov:2003rj},\footnote{The instanton partition function for $\cN=2^*$ SYM was originally obtained for $U(N)$ gauge group \cite{Nekrasov:2002qd,Nekrasov:2003rj}. Later in \cite{Alday:2009aq,Alday:2010vg}, the $SU(N)$ instanton partition function was obtained by factoring out $U(1)$ contributions  $Z^\text{inst}_{U(1)}$ motivated by the AGT correspondence \cite{Alday:2009aq}. Since $Z^\text{inst}_{U(1)}$ is holomorphic in $\tau$ and independent of the $a_i$, this contribution is killed by the $m$ and $\tau$ derivatives that we consider, so we will simply use the results for $U(N)$ instanton partition functions in the following. } and the resulting integral of this expansion converges rapidly for any $\tau$ in the $SL(2,\mathbb{Z})$ fundamental domain
\es{tauDomain}{
|\tau|\geq1\,,\qquad |\Re(\tau)|\leq\frac12\,.
}
In Appendix \ref{loc} we provide more details about this calculation and explicit expressions for the $SU(2)$ and $SU(3)$ cases written as one- and two-dimensional integrals, respectively, that can be easily computed numerically for any $\tau$ in \eqref{tauDomain}. We plot the localization inputs
\es{Fs}{
\mathcal{F}_2(\tau)\equiv\frac{1}{8c}\frac{\partial_m^2\partial_\tau\partial_{\bar\tau} F}{\partial_\tau\partial_{\bar\tau} F}\Big\vert_{m=0}  \,,\qquad \mathcal{F}_4(\tau)\equiv 48 \zeta(3) c^{-1}+ c^{-2}{\partial^4_m  F}\big\vert_{m=0} 
}
in Figure~\ref{fig:localization_fd} as a function of $\tau$ for $SU(2)$ and $SU(3)$. Figures \ref{fig:localization_n2_cross_sections} and \ref{fig:localization_n3_cross_sections} additionally show cross sections of these inputs along the imaginary axis and on the arc from $e^{\pi i/3}$ to $e^{2\pi i/3}$.

\subsection{Weak coupling expansion}
\label{weak}

In the next section, we will find it useful to compare our results to the small $g_\text{YM}$ perturbative expansion, where instantons do not contribute. At $g_\text{YM}=0$, we have the free theory correlator \eqref{free}, which, when expanded in superconformal blocks, gives long multiplets of twist $t=2,4,6,\dots$ with OPE coefficients \cite{Dolan:2001tt}: 
\es{freeOPE}{
\lambda^2_{\ell+2,\ell}&=\frac{2 ((\ell+2)!)^2}{c (2 \ell+4)!}\,,\\
\lambda^2_{\ell+t,\ell}&=\frac{ \Big[(\ell+1) (\ell+t+2)+\frac{(-1)^{\frac t2}}{c}\Big]   \left(\frac{t}{2}!\right)^2 \left(\ell+\frac{t}{2}\right)!
   \left(\ell+\frac{t}{2}+1\right)!   }{t! (2 \ell+t+1)!}\quad\text{for}\quad t=4,6,8\dots\,.\\
}
For the $SU(N)$ theory with $c=\frac{N^2-1}{4}$, the correlator has been expanded to 4-loop order in $g_\text{YM}^2$ (at fixed $N$) in \cite{Fleury:2019ydf}, where non-planar corrections first appear at $O(g_\text{YM}^8)$. There is a unique twist two scalar operator called the Konishi, whose CFT data to 4-loop order is\footnote{The leading instanton correction has also been computed in \cite{Alday:2016tll,Alday:2016bkq,Alday:2016jeo}, and takes the form $\gamma_\text{1-insnt}=-\frac{9g_\text{YM}^4}{40 \pi ^4} e^{-\frac{8 \pi ^2}{g_\text{YM}^2}}  \cos (\theta )$. This correction is too small within the fundamental domain to meaningfully compare to our numerical results.} \cite{Velizhanin:2009gv,Eden:2012rr,Fleury:2019ydf,Eden:2016aqo,Goncalves:2016vir}
\es{konishiWeak}{
\Delta_{2,0}&=2+\frac{3 \lambda }{4 \pi ^2}-\frac{3 \lambda ^2}{16 \pi ^4}+\frac{21 \lambda ^3}{256 \pi ^6}+\frac{\lambda ^4
   \left(-1440 \left(\frac{12}{N^2}+1\right) \zeta (5)+576 \zeta (3)-2496\right)}{65536 \pi ^8} + O(\lambda^5)\,,\\
   (\lambda_{2,0})^2&=\frac1c\Bigg[\frac{1}{3}-\frac{\lambda }{4 \pi ^2}+\frac{\lambda ^2 (3 \zeta (3)+7)}{32 \pi
   ^4}-\frac{\lambda ^3 (8 \zeta (3)+25 \zeta (5)+48)}{256 \pi ^6}\\
   &+\frac{\lambda ^4 \left(  2488 + 328 \zeta(3) + 72 \zeta(3)^2 + 980 \zeta(5) + 1470 \zeta(7) +\frac{45}{N^2} (8 \zeta
   (5)+7 \zeta (7))\right)}{16384 \pi ^8 } \\
   &{}+ O(\lambda^5)\Bigg] 
   \,,
}
where the 't Hooft coupling is $\lambda\equiv g_\text{YM}^2N$. At twist 4, there are generically four degenerate operators, but when $N=2$ there are only two due to trace relations.  At one loop, the scaling dimensions of these twist 4 operators whose scaling dimension are written in terms of a quantity $\omega$ which obeys a quartic equation: \cite{Arutyunov:2002rs,Beisert:2003tq}
   \es{twist4}{
 \Delta_{4,0}=4+ \lambda \frac{\omega}{8\pi^2}\,,\quad \omega^4-25 \omega^3 + \left(188-\frac{160}{N^2}\right) \omega^2-\left(384-\frac{1760}{N^2}\right)
   \omega-\frac{7680}{N^2} = 0\,.
   }
Note that, for $N \geq 3$, only one of the four solutions to this quartic equations gives a negative anomalous dimension, while the other three solutions are real and positive.  For $N=2$, the quartic equation in \eqref{twist4} has only two real solutions, one positive and one negative, corresponding to the two operators that mix in this case.  The OPE coefficients have not been computed beyond the degenerate free theory result in \eqref{freeOPE}.

\section{Numerical bootstrap with integrated constraints}
\label{numBoot}

In this section we will describe how to combine the numerical bootstrap of the stress tensor correlator with the two integrated constraints. We will begin by applying the functionals in \eqref{ints} to each block $G_{\Delta,\ell}(U,V)$ in the expansion of the correlator for any spin $\ell$ and dimension $\Delta$. We will then describe the abstract bootstrap algorithms that can be used to bound the scaling dimensions and OPE coefficients of the unprotected operators that show up in the $S\times S$ OPE\@. We then discuss how to implement these algorithms in practice, and in particular why the conventional semidefinite programming approach \cite{Poland:2011ey} cannot efficiently impose both integrated constraints and the crossing equations, so instead we use an updated version of the original linear programming approach of \cite{Rattazzi:2008pe}. We conclude with non-perturbative bounds on low-lying CFT data coming from these constraints for finite $N$ and $\tau$, which we compare to the weak coupling predictions.  

\subsection{Integrated constraints for block expansion}
\label{blockInt}

\begin{figure}[]
\centering
	\begin{subfigure}{0.48\textwidth}%
		\includegraphics[width=\linewidth]{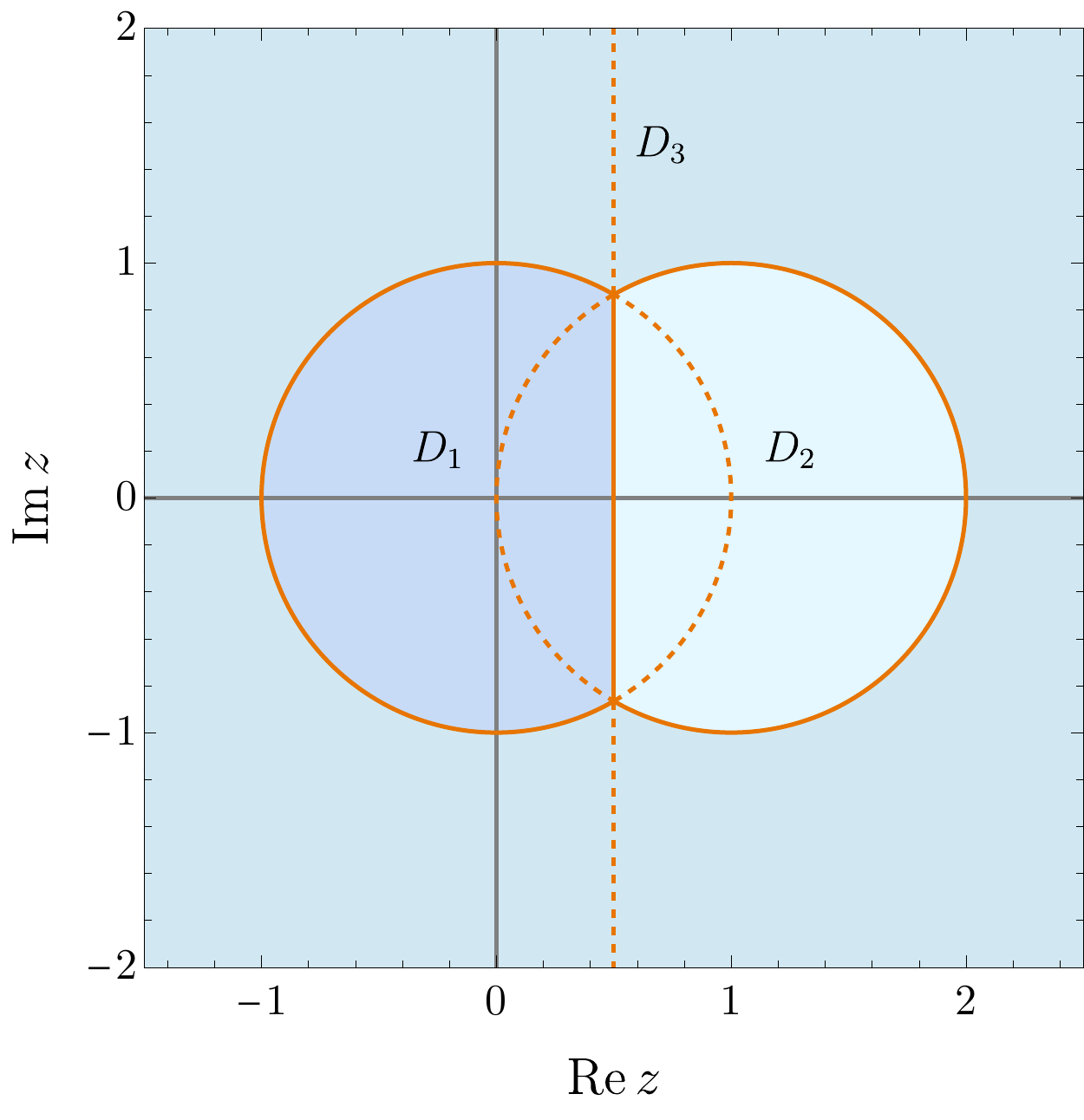}
	\end{subfigure}%
	\hspace{.04\textwidth}%
	\begin{subfigure}{0.48\textwidth}%
		\includegraphics[width=\linewidth]{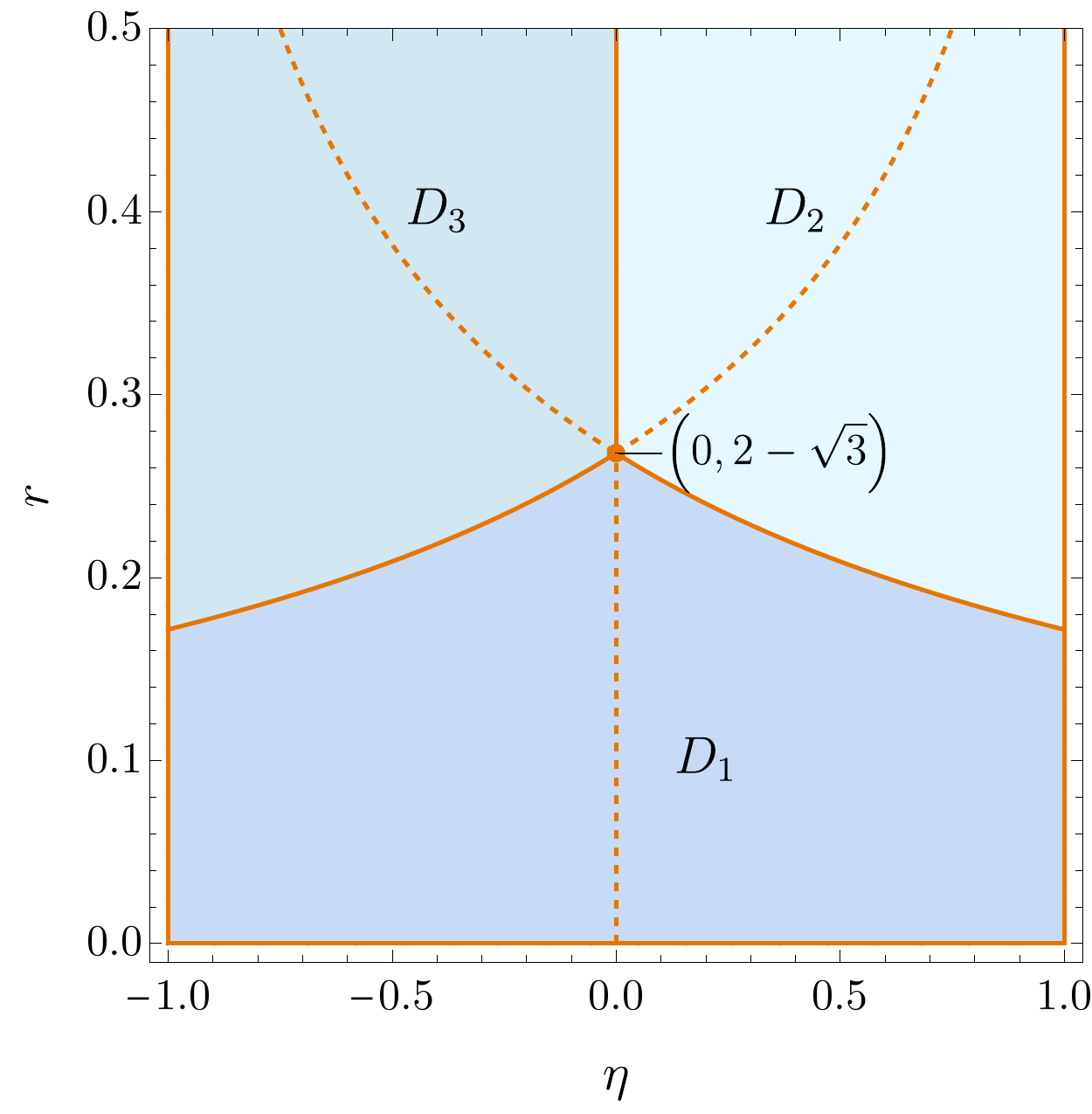}
	\end{subfigure}%
 \caption{The fundamental domains under the $S_3$ crossing symmetry in $z,\bar z$ (\textbf{left}) and $r,\eta$ coordinates (\textbf{right}). Dotted lines denote the boundaries between individual fundamental domains, and solid lines denote the boundaries between the unions of domains $D_1$, $D_2$, and $D_3$ discussed in the text. The $s$-channel block expansion converges in $D_1$.}
\label{domainsFigure}
\end{figure} 

The integrals in  \eqref{ints} are over $\R^2$ (written in polar coordinates $(R, \theta)$), while the block expansion of a four-point function has a finite radius of convergence. Thus, we divide $\R^2$ into domains that are permuted under the $S_3$ crossing symmetry of the four-point function, such that the conformal block expansion converges within each of these domains.  This division is more clearly seen  in the $(z, \bar z)$ coordinates used in the conformal block expansion, which are related to the polar coordinates $(R, \theta)$ used in the integrals in \eqref{ints} by
 \es{zToR}{
  z = 1 - R \cos \theta + i R \sin \theta \,, \qquad
   \bar z = 1 - R \cos \theta - i R \sin \theta \,.
 }
The $S_3$ crossing symmetry is generated by $z \to 1-z$ (coming from the interchange of the first and third operators in the four-point function) and by $z \to 1/z$ (coming from the interchange of the first and fourth operators in the four-point function), along with their complex conjugates.\footnote{Defining by $s_1$ the map $z \to 1-z$, by $s_2$ the map $z \to 1/z$, and by $e$ the identity transformation, it is clear that $s_1^2 = 1$ and $s_2^2 = 1$. One can also check $(s_1 s_2)^3 = 1$, which is the defining property of the permutation group $S_3$.} 
We define the three regions 
\begin{equation}\label{domains}
\begin{alignedat}{3}
	&D_1:\qquad &&|z| < 1\,,\qquad &&\Re z < \frac12\,,\\
	&D_2:\qquad &&|z-1| < 1\,,\qquad &&\Re z > \frac12\,,\\
	&D_3:\qquad &&|z| > 1\,,\qquad &&|1-z| > 1\,,\\
\end{alignedat}
\end{equation}
that are interchanged by the $S_3$ symmetry, and that are each marked with a different color in Figure~\ref{domainsFigure}.\footnote{Each of these regions is the union of two fundamental domains on which the $S_3$ permutation symmetry acts.}  Out of the three regions, the $s$-channel block expansion that we considered in \eqref{Gexp} converges in $D_1$ \cite{Dolan:2000ut}. 

For any crossing-invariant quantity, we can restrict the integration range in \eqref{ints} to one of these domains, or indeed any union of two of the fundamental domains under crossing symmetry, and then multiply by 3 to compensate. We will use a domain of integration denoted $D'(b)$, to be defined shortly. To motivate the definition, we first work out how to efficiently integrate the conformal blocks.

It is useful to expand the blocks such that one integral can be done analytically, leaving just one numerical integration.\footnote{We thank David Simmons-Duffin for suggesting this strategy, which follows Appendix C of \cite{Komargodski:2016auf}.} The block expansion is most easily expressed in terms of the radial coordinates $r,\eta$ defined in \cite{Hogervorst:2013sma} as
\es{retatozzb}{
U=\frac{16 r^2}{\left(r^2+2 \eta  r+1\right)^2}\,,\qquad V=\frac{\left(r^2-2 \eta  r+1\right)^2}{\left(r^2+2 \eta 
   r+1\right)^2}\,.
}
We can then use the recursion relation in Appendix \ref{blockExp}\footnote{This generalizes the recursion relation in \cite{Kos:2013tga} that works for odd spacetime dimensions to the case of 4d.} to expand the conformal blocks in a small $r$ expansion as
\es{4dblockNorm}{
G_{\Delta,\ell}(r,\eta)= (4r)^\Delta\sum_{n=0}^\infty \sum_sB_{n,s}(\Delta,\ell) {U_s(\eta)}r^n\,,
}
where $U_s(\eta)$ are Chebyshev polynomials, $B_{n,s}$ are numerical coefficients, and the block is normalized such that $B_{0,0}=1$. The integrals \eqref{ints} that act on the blocks $U^{-2}G_{\Delta+4,\ell}(U,V)$ that appear in the integrated constraints \eqref{constraint1} can then be written as the $r,\eta$ integrals
\es{intsreta}{
I_2\Big[\frac{G_{\Delta+4,\ell}}{U^2}\Big]&=-3\int_{D_1} drd\eta\frac{\sqrt{1-\eta ^2} \left(r^2-1\right)^2 \left(\left(2-4
   \eta ^2\right) r^2+r^4+1\right)}{128 \pi  r^5}G_{\Delta+4,\ell}(r,\eta)\,,\\
   I_4\Big[\frac{G_{\Delta+4,\ell}}{U^2}\Big]&=3\int_{D_1} drd\eta  \frac{\sqrt{1-\eta ^2} \left(r^2-1\right)^2 \left(2
   \eta  r-r^2-1\right) \left((4 \eta ^2+10)
   r^2+r^4+1\right)}{4 \pi  r^5 \left(r^2+2 \eta  r+1\right)} \bar{D}_{1,1,1,1}G_{\Delta+4,\ell}(r,\eta)\,,\\
}
where $\bar{D}_{1,1,1,1}$ can also be written in terms of $r,\eta$ using its explicit expression \eqref{Dbar} and the changes of variables from $(z, \bar z)$ to $(r, \eta)$ obtained by combining \eqref{Phis} and \eqref{retatozzb}. For a given $\Delta,\ell$, we can expand the blocks, the integration measures, and $\bar{D}_{1,1,1,1}$\footnote{The small $r$ expansion of $\bar{D}_{1,1,1,1}(U,V)$ also includes $\log r$ terms.}  all at small $r$ to some order $p$, perform the integrals in $r$ exactly, and then the remaining integral in $\eta$ numerically.

The error from this expansion goes like $r_\text{max}^p$, where $r_\text{max}$ is the maximum value of $r$ in our integration region $D'$. So long as $r_\text{max} < 1$, we see that the method converges quickly. For instance, in the region $D_1$ we have $r_\text{max} = 2-\sqrt{3} \approx 0.268$. 

However, this maximum value occurs at $\eta = 0$ where $U_s(0) = -(-1)^{s/2}$, so if we integrate over $D_1$ then the integrated constraints will oscillate with spin. This presents a problem for the bootstrap, because for large $\Delta$ the integrated constraints grow as $(4r_\text{max})^\Delta$ while the constraints from crossing symmetry grow as $(4(3-2\sqrt{2}))^\Delta$, where $3-2\sqrt{2}$ is the value of $r$ at the crossing-symmetric point. For $D_1$, we have $r_\text{max} > 3-2\sqrt{2}$, and so for large $\Delta$ the integrated constraints dominate. The sign oscillations in spin would then prevent us from including the integrated constraints in any positive functional, and so we could not use the integrated constraints to improve the rigorous bootstrap bounds\footnote{Despite the sign oscillations in $D_1$, one could apply the bootstrap using this region while only imposing positivity of functionals up to some scaling dimension $\Delta_\text{max}$. If we follow this approach, we arrive at bounds very similar to the ones obtained in this paper. Furthermore, we find that the range of scaling dimensions for which the functionals can be made positive increases with $n_\text{max}$, so that in the $n_\text{max}\to\infty$ limit we could obtain everywhere-positive functionals using $D_1$. By replacing $D_1$ with two non-oscillating regions, we make the bounds rigorous by achieving positivity at finite $n_\text{max}$.}.

\begin{figure}[h]
	\centering
	\includegraphics[width=0.7\linewidth]{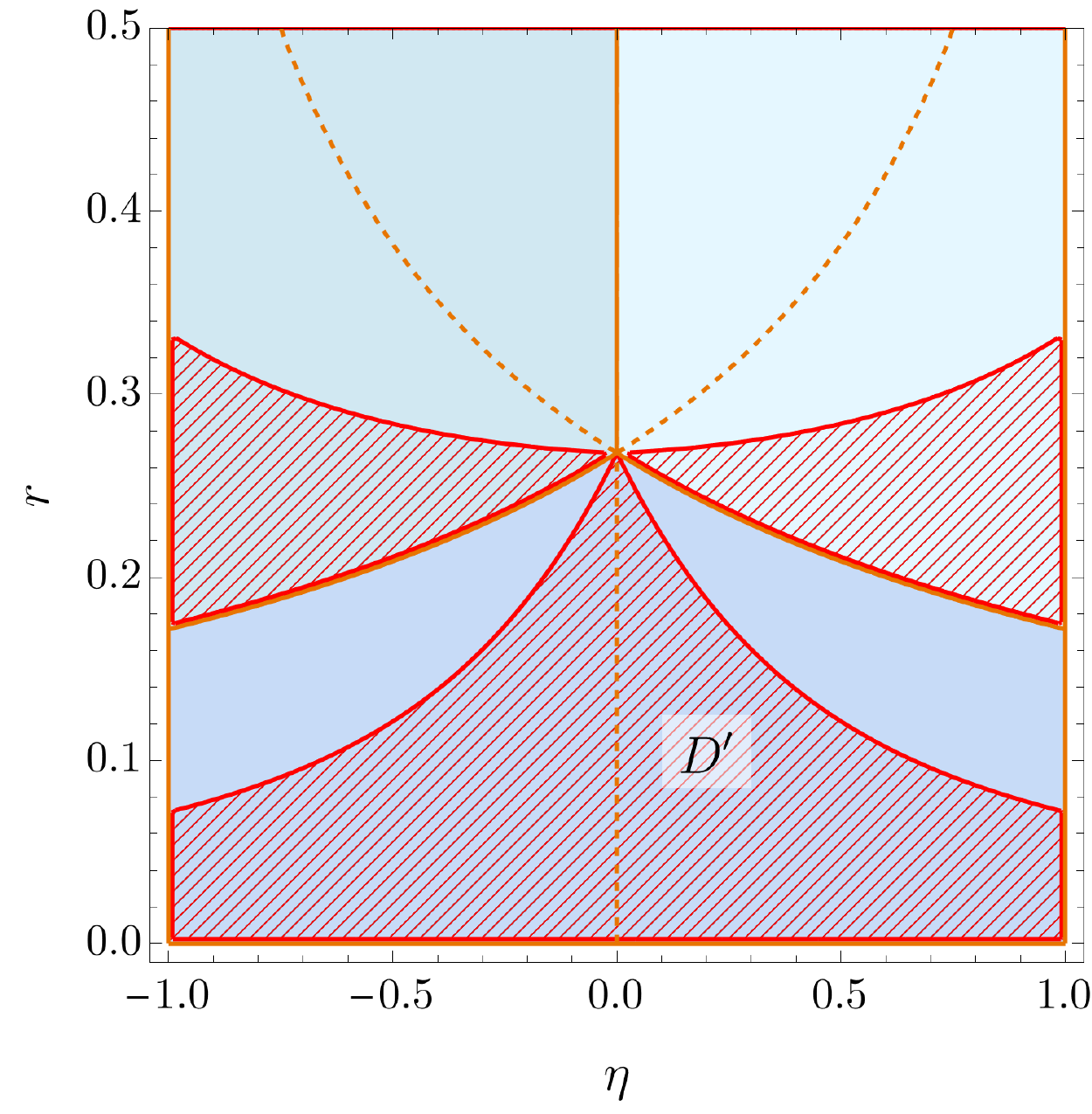}
	\caption{The integration region $D'$, shown in orange, defined by extending $D_1$ into $D_2\cup D_3$ (indicated with the same colors as on the right of Figure \ref{domainsFigure}) such that the maximal value of $r$ occurs at $\eta=\pm1$, and then subtracting off the image of the extension as mapped back into $D_1$ under crossing symmetry. This ensures that the integrated constraints will not exhibit spin oscillations.}
	\label{fig:nonosc_region}
\end{figure}

We will thus follow the approach of \cite{Lin:2015wcg} to modify the integration region such that the maximum value of $r$ occurs at $\eta = \pm 1$ instead of at 0, thus removing the sign oscillations. We first extend $D_1$ to form a region $D''(b)$ defined by
\begin{equation}\label{eq:nonosc}
	(z-\alpha)(\overline{z} - \alpha) = \beta,
\end{equation}
where $\alpha$ and $\beta$ are defined by requiring the boundary to intersect the points $(r,\eta) = (2-\sqrt 3, 0)$ and $(r,\eta) = (b, \pm 1)$. As long as $b > 2-\sqrt{3}$, the maximum value of $r$ in $D''(b)$ is then achieved at $\eta = \pm 1$ as desired. To correct for the expansion of the integration region, we should subtract off the image of $D''(b)\backslash D_1$ under $1\leftrightarrow 3$ crossing, which is bounded from below by the solution to
\begin{equation}\label{eq:nonosc_image}
	(1-z-\alpha)(1-\overline{z} - \alpha) = \beta
\end{equation}
and from above by the boundary of $D_1$, namely \eqref{eq:nonosc} with $\alpha = 0$ and $\beta = 1$. We define $D'(b)$ to be $D''(b)$ with this crossing image of $D''(b)\backslash D_1$ removed, as depicted in Figure \ref{fig:nonosc_region}.

In the $(r,\eta)$ coordinates, we can define this region by
\begin{equation}
	D'(b):\qquad -1 \leq \eta \leq 1, \qquad r_0(\eta) \leq r \leq r_1(\eta,b) \quad\text{or}\quad r \leq r_2(\eta,b),
\end{equation}
where
\begin{equation}
	r_0(\eta) = |\eta| + 2 - \sqrt{\eta^2 + 4|\eta| + 3}
\end{equation}
and $r_1(\eta, b)$ and $r_2(\eta, b)$ come from solving \eqref{eq:nonosc} and \eqref{eq:nonosc_image}, respectively.

\begin{figure}[]
	\centering
	\begin{subfigure}{0.48\textwidth}%
		\centering
		{\Large $b = \frac{1}{3}$}\\[1em]
		\includegraphics[width=\linewidth]{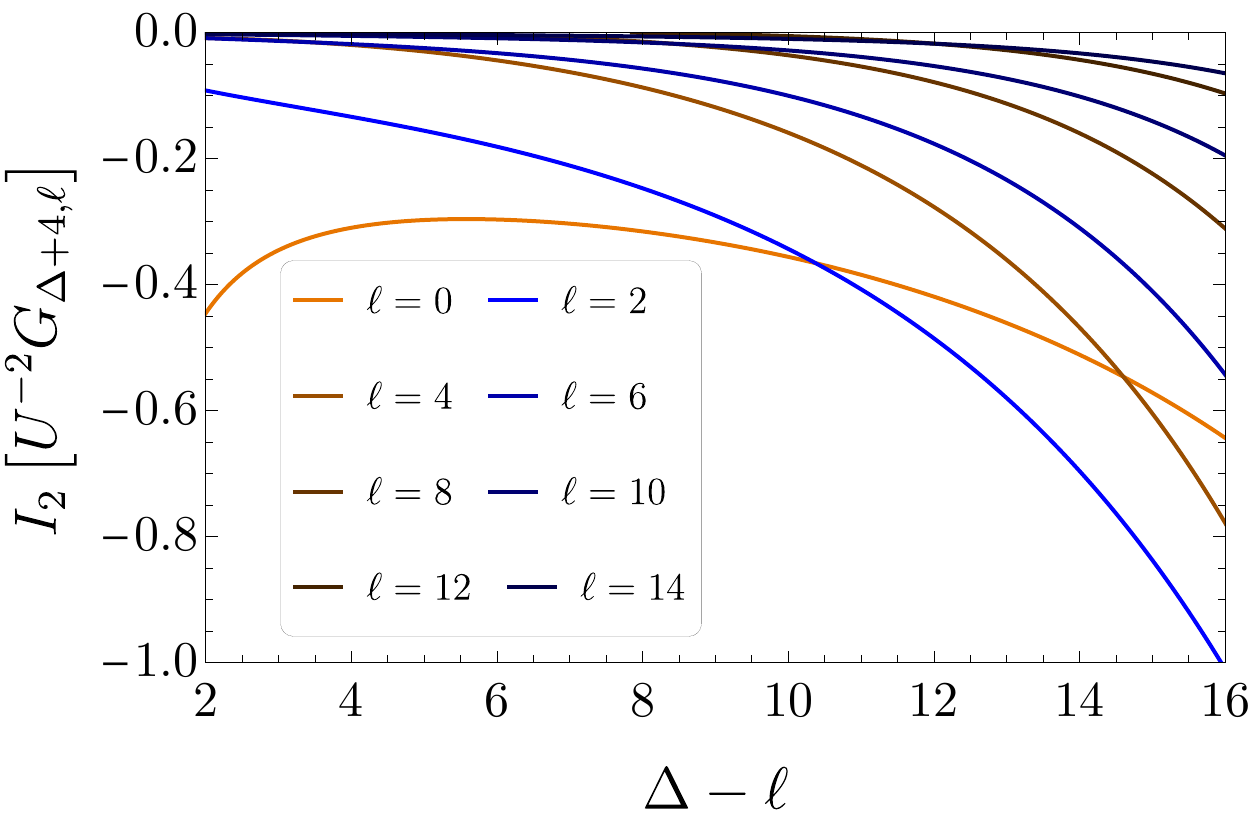}
	\end{subfigure}%
	\hspace{.02\textwidth}%
	\begin{subfigure}{0.48\textwidth}%
		\centering
		{\Large $b = \frac{7}{25}$}\\[1em]
		\includegraphics[width=\linewidth]{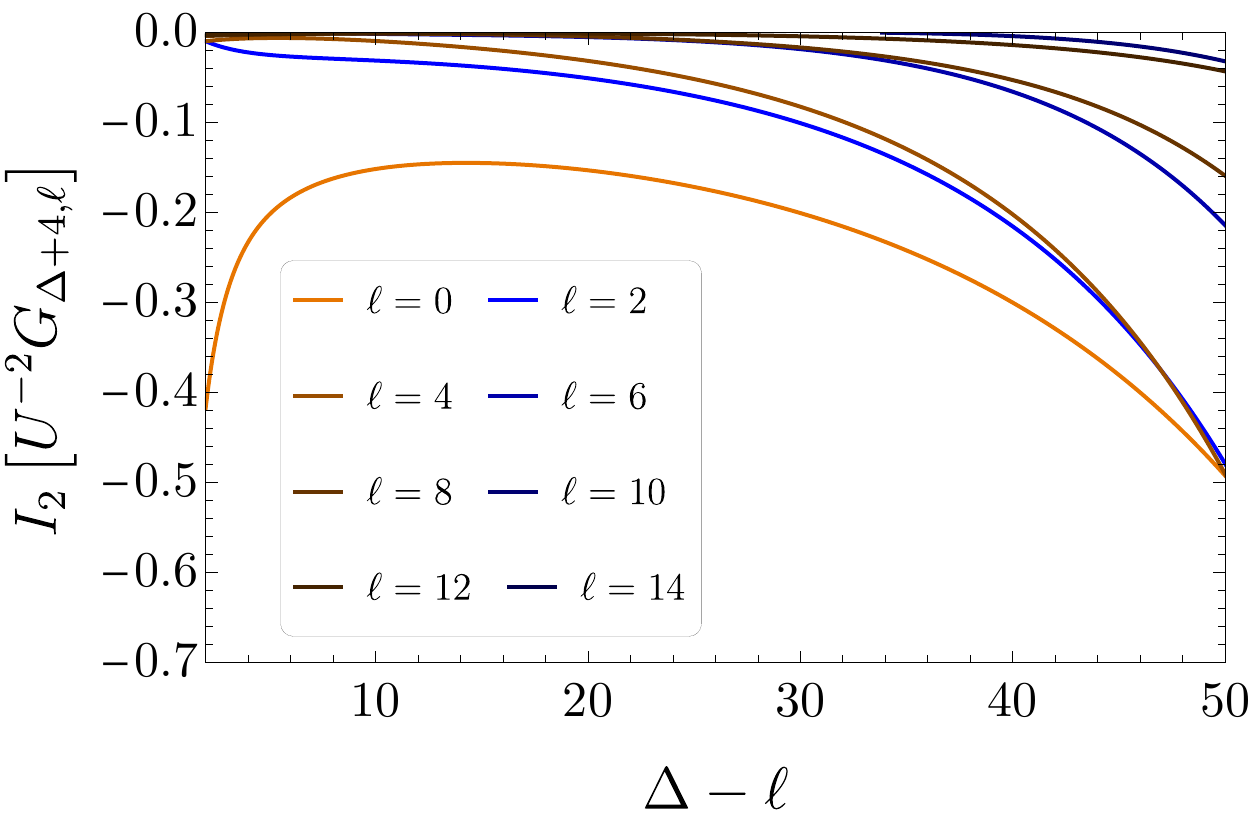}
	\end{subfigure}\\[1cm]
	\begin{subfigure}{0.48\textwidth}%
		\centering
		\includegraphics[width=\linewidth]{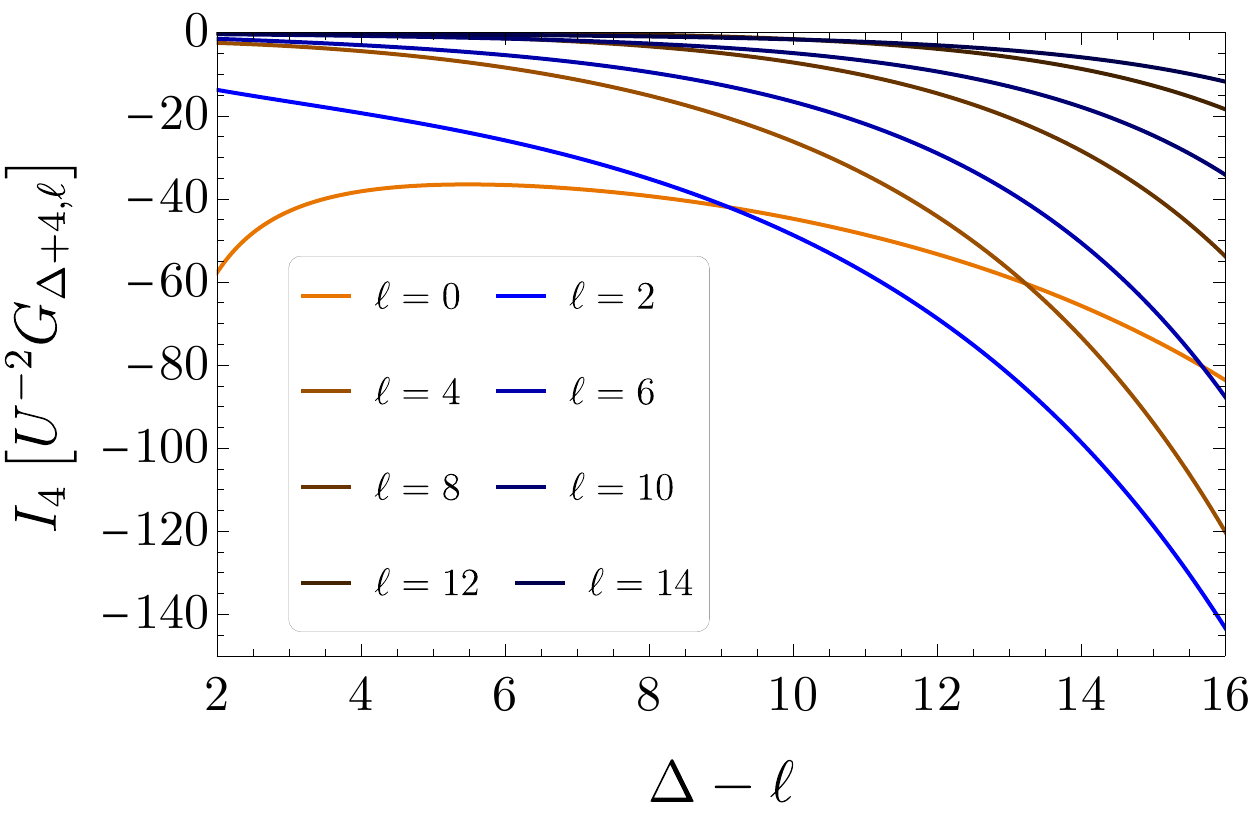}
	\end{subfigure}%
	\hspace{.02\textwidth}%
	\begin{subfigure}{0.48\textwidth}%
		\centering
		\includegraphics[width=\linewidth]{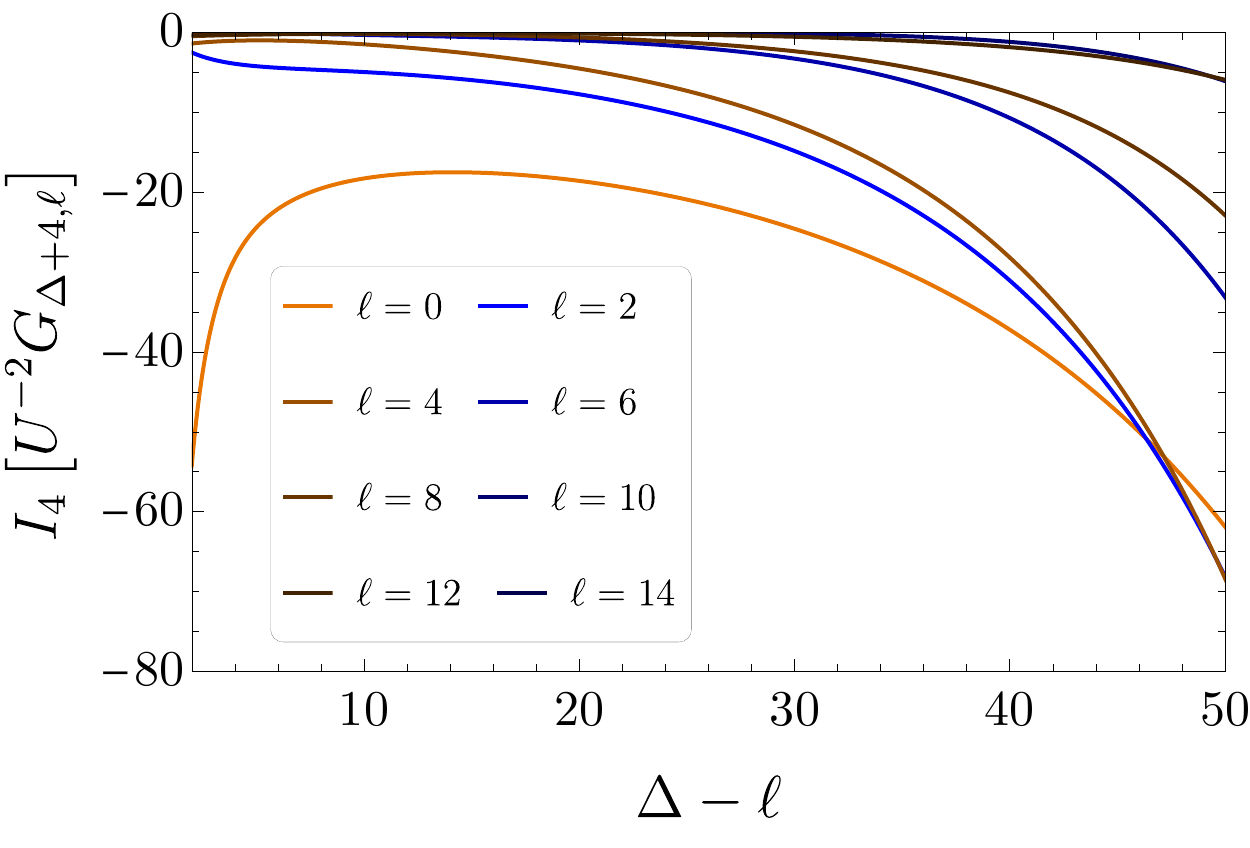}
	\end{subfigure}
 \caption{Integrated constraints $I_2$ and $I_4$ \eqref{ints} acting on blocks $U^{-2}G_{\Delta+4,\ell}$ as they appear in the block expansion \eqref{Gexp} for spins $\ell=0,2,\ldots,14$. For each spin, we begin plotting at the unitarity bound $\Delta = \ell + 2$. The blocks are computed with $r_\text{max}=20$ in the small $r$ expansion of the integrand in \eqref{intsreta}.}
\label{fig:block_integrals}
\end{figure}

Since the four-point function we are integrating is crossing-invariant, the choice of integration region should not affect the bounds we can achieve in the infinite-precision limit. However, as we have just mentioned, different regions can behave very differently for the finite truncations needed for the numerical bootstrap; this is why we choose a non-oscillating region like $D'(b)$ instead of $D_1$. Furthermore, although constraints coming from two different integration regions are redundant in the infinite-precision limit, for our numerical bootstrap approach we find that including constraints from two different regions significantly improves the bounds. We thus use both $D'(1/3)$ and $D'(7/25)$ in what follows. Both of these values of $b$ exceed $2-\sqrt{3}$, and thus do not exhibit the sign oscillations of $D_1$, so we can use them together to build positive functionals. The values of the integrated constraints computed using these regions are shown in Figure \ref{fig:block_integrals}.

In addition to the integrals of the conformal blocks, we need the integrals of the short contributions. We numerically integrate the closed-form expression for $\mathcal{T}_\text{short}$ given in \eqref{calTBlocks} for both of our integration regions, giving the following results:
\begin{equation}
\begin{split}
	b = \frac{1}{3} : \quad &\begin{cases} I_2[\mathcal{T}_\text{short}] \approx 0.1194001903 +  \frac{0.4681669312}{c} \\ I_4[\mathcal{T}_\text{short}] \approx 15.08334151 +  \frac{60.89159895}{c} \end{cases} \\
	b = \frac{7}{25} : \quad &\begin{cases} I_2[\mathcal{T}_\text{short}] \approx 0.07917567977 +  \frac{0.4287985056}{c} \\ I_4[\mathcal{T}_\text{short}] \approx 9.758836729 +  \frac{55.78193820}{c} \end{cases}
\end{split}
\end{equation}

\subsection{Numerical bootstrap algorithms}
\label{alg}

Let us now explain the numerical bootstrap algorithm that incorporates both the integrated constraints and the crossing equations.  For a given function $f$ of the cross ratios $(U, V)$, we consider the infinite-dimensional vector ${\bf v}_f$ by
\es{Vinf}{
 {\bf v}_f \equiv \begin{pmatrix}   I_{2,b=1/3}[f] \\
  I_{4,b=1/3}[f] \\
  I_{2,b=7/25}[f] \\
  I_{4,b=7/25}[f] \\
   U^2 V^4 f(U,V) - U^4 V^2 f(V, U) 
  \end{pmatrix}  \in {\bf V}_\infty\,.
}
The first four components of ${\bf v}_f$ are the functionals $I_2$ and $I_4$ defined in \eqref{ints} that appear in the integrated correlators, for both of the two non-oscillating integration regions we employ, acting on the function $f$.  The remaining components correspond to the function $U^2 V^4 f(U,V) - U^4 V^2 f(V, U)$.  We can denote by ${\bf V}_\infty$ the vector space consisting of the vectors ${\bf v}_f$ of the form \eqref{Vinf}.

 The constraints \eqref{crossing} and \eqref{constraint1} then impose that an infinite sum of vectors with positive coefficients $\lambda^2_{\Delta,\ell}$ plus the contribution from the protected terms must vanish:
 \es{SumRule}{
 \sum_{\Delta,\ell}\lambda^2_{\Delta,\ell} {\bf v}_{\Delta, \ell} + {\bf w} = 0 \,, \qquad
  {\bf w} \equiv 
  \begin{pmatrix}
    I_{2,b=1/3}\left[\cT_\text{short}\right]-\frac{1}{8c}\frac{\partial_m^2\partial_\tau\partial_{\bar\tau} F}{\partial_\tau\partial_{\bar\tau} F}\Big\vert_{m=0} \\
    I_{4,b=1/3}\left[\cT_\text{short}\right]- \frac{48 \zeta(3)}{c}- \frac{{\partial^4_m  F}\big\vert_{m=0}}{c^2} \\
    I_{2,b=7/25}\left[\cT_\text{short}\right]-\frac{1}{8c}\frac{\partial_m^2\partial_\tau\partial_{\bar\tau} F}{\partial_\tau\partial_{\bar\tau} F}\Big\vert_{m=0} \\
    I_{4,b=7/25}\left[\cT_\text{short}\right]- \frac{48 \zeta(3)}{c}- \frac{{\partial^4_m  F}\big\vert_{m=0}}{c^2} \\
    F^{(0)}_\text{short}(U,V)+ \frac{F^{(1)}_\text{short}(U,V)}{c} 
  \end{pmatrix}   \,,
}
where ${\bf v}_{\Delta, \ell}$ is a shorthand notation for ${\bf v}_f$ with $f(U, V) = F_{\Delta, \ell}(U, V)$.  To find instances in which the equation \eqref{SumRule} cannot be obeyed, consider a functional (written as an infinite-dimensional row vector)
 \es{alphaDef}{
   \alpha=\begin{pmatrix} \alpha_{2,1} & \alpha_{4,1} & \alpha_{2,2} & \alpha_{4,2} & \alpha_\infty \end{pmatrix}  \,,
  } 
where $\alpha_{2,i}$ and $\alpha_{4,i}$ are numbers, and $\alpha_\infty$ is a functional acting on functions of $(U, V)$.  

The functional $\alpha$ can be used to bound scaling dimensions $\Delta$ by the following algorithm:
\\
\\
{\bf Scaling dimension bound:}
\begin{enumerate}
\item Normalize $\alpha$ such that $\alpha [{\bf w}] =1$.\footnote{The normalization $\alpha[ {\bf w}] =1$ is a matter of convention.  One can more generally fix $\alpha [{\bf w}]$ to any positive real number.} 
\item Assume that the scaling dimensions $\Delta_\ell$ of all spin $\ell$ unprotected operators obey lower bounds $\Delta_\ell\geq\bar\Delta_\ell$, where the unitarity bound $\Delta_\ell\geq\ell+2$ provides a minimal choice.
\item Search for $\alpha$ obeying
\es{searchScal}{
 \alpha [{\bf v}_{\Delta, \ell}] \geq 0 
}
for all $\Delta \geq \bar \Delta_\ell$ and all $\ell$.
\item If such $\alpha$ exists, then by positivity of $\lambda^2_{{\Delta,\ell}}$ we have
 \es{alphScal}{
 \sum_{\Delta,\ell}\lambda^2_{\Delta,\ell} \alpha[{\bf v}_{\Delta, \ell}] + \alpha[{\bf w}] > 0 \,,
}
which contradicts Eq.~\eqref{SumRule}, and so our assumptions $\Delta\geq\bar\Delta_\ell$ must be false. If we cannot find such an $\alpha$, then we conclude nothing.
\end{enumerate}
For instance, we can set the lower bounds $\bar\Delta_\ell$ to their unitarity values $\bar \Delta_\ell = \ell+2$ for all $\ell$ except a certain $\ell'$, then by varying $\bar\Delta_{\ell'}$, $c$, and $\tau$, this algorithm can be used to find an upper bound on $\bar\Delta_{\ell'}$, i.e.~on the scaling dimension of the lowest dimension operator with spin $\ell'$, as a function of $c$ and $\tau$.

Without further assumptions we cannot do better than finding upper bonds on scaling dimensions. If we furthermore assume that only a few operators have dimensions below a certain value, then the allowed region as a function of the scaling dimensions of these operators could be more complicated.  For instance, consider adding the additional constraint on $\alpha$ to the algorithm above:
\es{gap}{
\text{\bf{Additional operator:}}\qquad \alpha[{\bf v}_{\tilde \Delta_\ell, \tilde \ell}] \geq 0 \,,
}
for some spin $\tilde\ell$ and dimension $\tilde\Delta_{\tilde \ell}$. One can then fix $\bar \Delta_{\tilde \ell}$ and vary $\tilde\Delta_{\tilde \ell}$. For instance, at weak coupling we saw in Section \ref{weak} that there are only two relevant scalar unprotected operators, while at large $N$ and large 't Hooft coupling $\lambda$ there is only one such operator.  Thus, we may assume that at any value of $N$ and $\tau$ there are at most two unprotected scalar operators.  We can impose this in our bootstrap study by setting a gap $\bar\Delta_0=4$, inserting two operators $\tilde\Delta_0$ and $\tilde\Delta_0'$ below, and varying these latter scaling dimensions for given $c$ and $\tau$.
  
We can also get upper bounds on a certain $\lambda^2_{{\Delta',\ell'}}$, and also lower bounds if we assume a gap, by a slight variation of the scaling dimension algorithm:
\\
\\
{\bf OPE coefficient bound:}
\begin{enumerate}
\item Normalize $\alpha$ such that 
\es{normOPE}{
s= \alpha [ {\bf v}_{\Delta', \ell'} ] 
\,,
}
where $s=\pm1$ for upper/lower bounds.
\item Assume that the scaling dimensions $\Delta_\ell$ of all unprotected operators other than the one with spin $\ell'$ and dimension $\Delta'$ obey lower bounds $\Delta_\ell\geq\bar\Delta_\ell$ (e.g.~unitarity).
\item Require that $\alpha$ satisfy \eqref{searchScal}. 
\item Maximize $\alpha[{\bf w}]$, for ${\bf w}$ as defined in \eqref{SumRule}, to get the upper/lower bounds
\es{OPEbound}{
&\text{Upper}:\qquad\qquad\lambda^2_{{\Delta',\ell'}}\leq -\alpha[{\bf w}] \,,\\
&\text{Lower}:\qquad\qquad\lambda^2_{{\Delta',\ell'}}\geq \alpha [ {\bf w} ] \,,
}
which follows from \eqref{crossing}, \eqref{constraint1}, positivity of $\lambda^2_{{\Delta,\ell}}$, and steps 1 and 3.
\end{enumerate}
If we just set $\bar\Delta_\ell$ to the unitarity bounds, then we can get upper bounds but not lower bounds, because the potential existence of operators with $\Delta$ arbitrarily close to $\Delta'$ makes step 1 inconsistent with step 3. To avoid this we must set a gap above unitarity for $\bar\Delta_\ell$ and below this insert the operator with $\Delta',\ell'$ as in \eqref{gap}. Note that this algorithm involved maximizing $\alpha[{\bf w}]$, whereas the scaling dimension algorithm only involved the existence of $\alpha$ for fixed $\alpha [{\bf w}]$.

\subsection{Linear versus semidefinite programming}
\label{program}

To implement the above algorithms numerically, we must perform three truncations / discretizations. The first truncation is the number of spins, which we can achieve by imposing a simple cap $\ell_\text{max}$ on the range of spins we consider. We then check afterwards that the functionals we find are positive for all spins. The second truncation is on the space of functions of $(U,V)$, or equivalently on the size of the functionals used to probe this space.  As is customary in conformal bootstrap studies, instead of functions of $(U, V)$, we consider a finite number of derivatives at the crossing symmetric point $z=\bar z=\frac12$.  Thus, the vector space ${\bf V}_\infty$ is replaced with a finite-dimensional vector space ${\bf V}_\Lambda$ defined by replacing a general vector
\begin{equation}\label{Vfinite}
	{\bf v} = \begin{pmatrix} v_2 \\ v_4 \\ v_2' \\ v_4' \\ v(U, V) \end{pmatrix} \qquad \text{width} \qquad {\bf v} = \begin{pmatrix} v_2 \\ v_4 \\ v_2' \\ v_4' \\ \partial_z^m \partial_{\bar z}^n v(U, V) \end{pmatrix}
\end{equation}
with $m+n = 1,3,5,\ldots,\Lambda$ and $m\leq n$.

Here, we restrict to odd derivatives with $m\leq n$ because $F_{\Delta,\ell}(U,V)$ \eqref{fdef} is symmetric in $z\leftrightarrow \bar z$ and odd under crossing.\footnote{We can efficiently compute $\partial_z^m\partial_{\bar z}^nF_{\Delta,\ell}(U,V)\vert_{z=\bar z=\frac12}$ using the \texttt{scalar\_blocks} code, available online from the bootstrap collaboration \cite{Kos:2013tga,Kos:2014bka,ElShowk:2012ht}. Note that their blocks differ from our conventions as $G^\text{ours}_{\Delta,\ell}(U,V)=(\ell+1)G^\text{theirs}_{\Delta,\ell}(U,V)$.} For sufficiently high $\ell_\text{max}$, the numerical bounds will monotonically improve with increasing $\Lambda$, which is what makes the numerical bootstrap rigorous.  Lastly, one should discretize the continuum of $\Delta_\ell\geq\bar\Delta_\ell$ in the space of allowed operators. 

Without integrated constraints, the most efficient way of dealing with the continuum in $\Delta$ is to use the semidefinite programming approach of \cite{Poland:2011ey}. The strategy is to use an approximation to the conformal blocks that makes the crossing equations  polynomial in $\Delta$.  The positivity constraints $\alpha [ {\bf v}_{\Delta, \ell}] \geq 0$ then reduce to the positivity of certain polynomials, which can be implemented using semidefinite programming software such as \texttt{SDPB} \cite{Simmons-Duffin:2015qma}. To write \eqref{crossing} as a polynomial in $\Delta$, one uses the small $r$ expansion of the blocks in \eqref{4dblockNorm} (truncated at some order $\tilde p$) to write the $\Delta$-dependence of $\partial_z^m\partial_{\bar z}^nF_{\Delta,\ell}(U,V)\vert_{z=\bar z=\frac12}$ as
\es{deltaSemi}{
\partial_z^m\partial_{\bar z}^nF_{\Delta,\ell}(U,V)\vert_{z=\bar z=\frac12}\sim (4(3-2\sqrt{2}))^\Delta \times(\text{meromorphic in $\Delta$})\,,
}  
where we used the fact that $r=3-2\sqrt{2}$ at the crossing symmetric point and the $B_{n,s}(\Delta,\ell)$ are meromorphic functions of $\Delta$. We can then multiply the crossing equations by $(4(3-2\sqrt{2}))^{-\Delta}$ as well as by the finite product of all positive factors that appear in the denominators of $B_{n,s}(\Delta,\ell)$ to get a polynomial in $\Delta$. 

The problem with using a similar approach here is that, as discussed in Section \ref{blockInt}, the integrated constraints scale at large $\Delta$ as
\es{largeDelta}{
I_{2,b}(G_{\Delta,\ell})\sim I_{4,b}(G_{\Delta,\ell})\sim (4b)^\Delta (-1)^{\frac\ell2}\,,
}
because the blocks scale as $(4r)^\Delta U_\ell(\eta)$ and $b$ is the maximum value of $r$ in the integration region $D'(b)$. Since the asymptotic scaling of the integrated blocks and the crossing equations differ by an exponential factor, writing both as polynomials in $\Delta$ would require approximating an exponential by a polynomial.\footnote{One cannot make the crossing equations scale the same as the integrated constraints by expanding the former around a different point, because then the crossing equations at large $\Delta$ would oscillate with spin as $(-1)^{\ell/2}$, and cause a similar oscillation problem. We thank David Simmons-Duffin for discussion on this topic.} While approximating an exponential by a polynomial seemed to give reasonable results in \cite{Lin:2015wcg}, in our 4d case we found that such an approximation made the bootstrap insensitive to the integrated constraints.

We will thus avoid this problem with semi-definite programming by returning to the original linear programming method of \cite{Rattazzi:2008pe}. In this case, we truncate each $\Delta_\ell$ to some upper cutoff $\Delta_\text{max}$, and then discretize $\Delta$ with spacing $\Delta_\text{sp}$, so the bootstrap algorithms become a finite set of linear constraints. We can then solve  these linear constraints with linear programming software.  In practice, we will still use SDPB for a linear programming problem.  Since we have cut off $\Delta_\text{max}$, the exponential scaling and oscillation of the integrated constraints relative to the crossing constraints do not matter as long as we use enough numerical precision so that we are sensitive to all the constraints. Also, for sufficiently small $\Delta_\text{sp}$ and large $\Delta_\text{max}$, the numerical bootstrap will become insensitive to the precise values of $\Delta_\text{sp}$ and $\Delta_\text{max}$ we use, similar to the cutoff on spin. See Appendix \ref{bootApp} for more details on our numerical implementation.

\subsection{Results}
\label{results}

\begin{figure}
	\centering
	\begin{subfigure}{0.75\linewidth}%
		\centering
		{\Large $\grSU(2)$}\\[1em]
		\includegraphics[width=\linewidth]{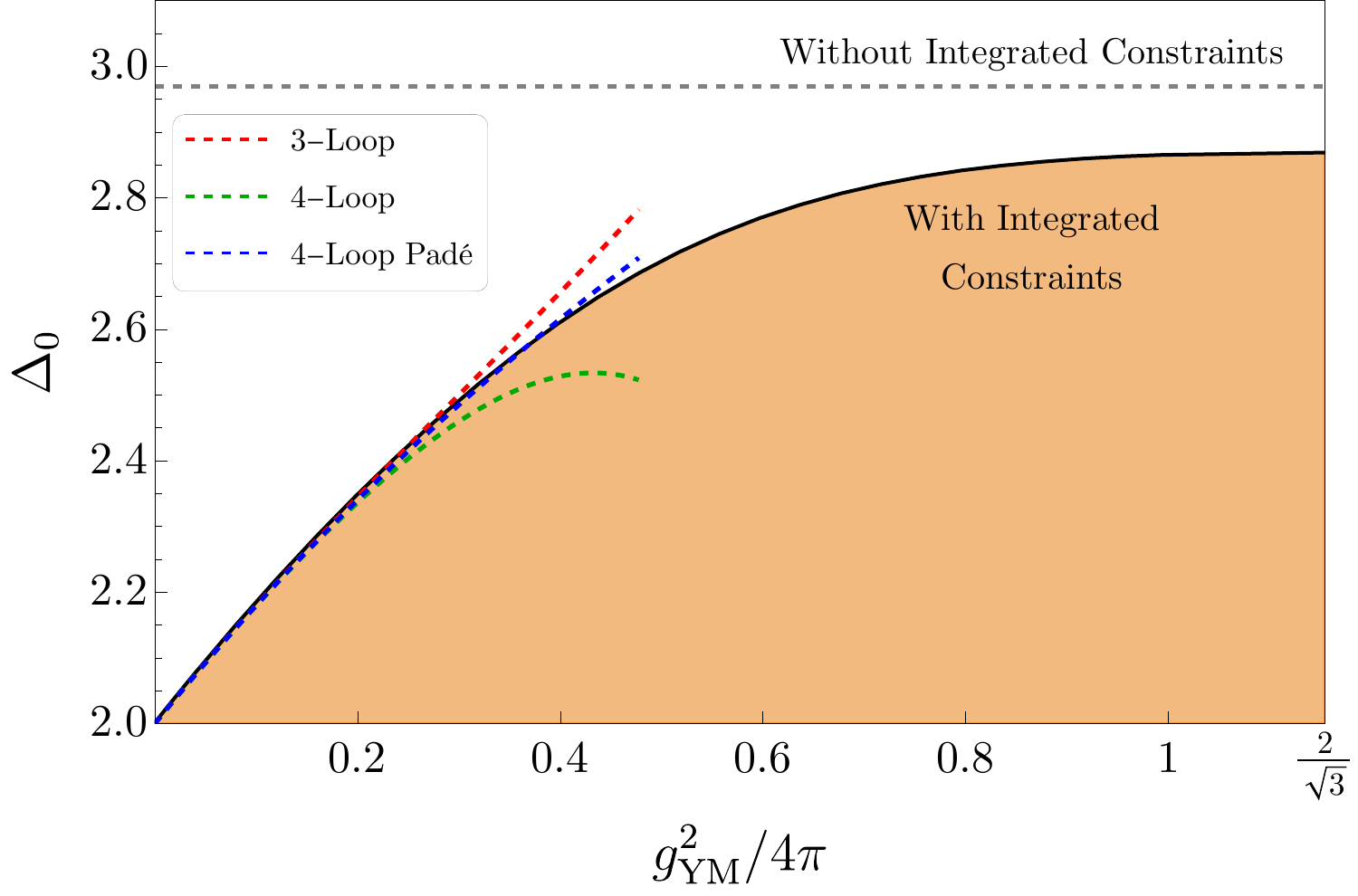}
	\end{subfigure}\\[2em]
	\begin{subfigure}{0.75\linewidth}%
		\centering
		{\Large $\grSU(3)$}\\[1em]
		\includegraphics[width=\linewidth]{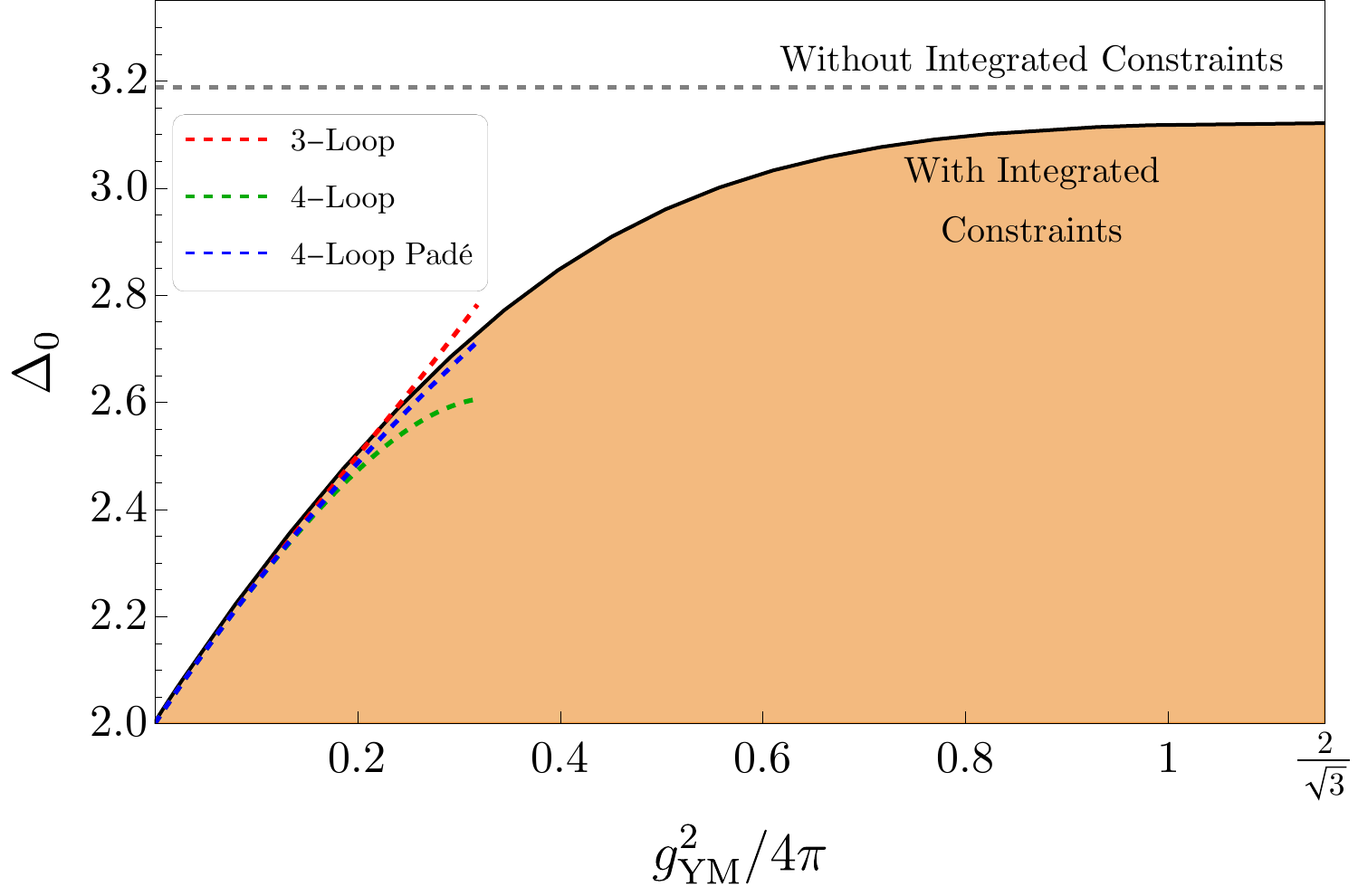}
	\end{subfigure}
	\caption{Upper bounds on the scaling dimension of the lowest scalar as a function of $\gym^2/4 \pi$ for $\grSU(2)$ and $\grSU(3)$, obtained with bootstrap resolution $\Lambda = 39$. At each value of $\gym^2/4 \pi$ from 0 to the bottom of the fundamental domain, we give the largest upper bound that we obtain for any $\theta$. The bounds are approximately saturated by the weak-coupling estimate \eqref{konishiWeak} in the regime where it is valid. We also show the upper bounds that would be obtained at the same bootstrap resolution without using the integrated constraints. Note that, like the localization inputs, the slope of these upper bounds go to zero at the $\Z_2$ self-dual point where $\gym^2 = 4\pi$.}
	\label{fig:scaling_dimension}
\end{figure}

Using the method outlined above, we can obtain an upper bound on the scaling dimension $\Delta_0$ of the lowest scalar as a function of complex coupling $\tau$.  Figure~\ref{fig:scaling_dimension} shows, for $\grSU(2)$ and $\grSU(3)$, the maximal upper bound on $\Delta_0$ that we obtain at any value of $\theta$, as a function of $g_\text{YM}$. For this plot and all others, we work at a bootstrap resolution of $\Lambda = 39$, as defined below \eqref{Vfinite}.

For $g^2_\text{YM} \ll 1$, our scaling dimension bounds are approximately saturated by the weak coupling result \eqref{konishiWeak}. In Figure~\ref{fig:scaling_dimension} we plot \eqref{konishiWeak} to three and four loops, and a Pad\'e approximant of order (2,2) to the four-loop expression. All of these expressions nearly coincide with our upper bound at very small coupling, and the Pad\'e approximant tracks our bound over a somewhat larger range of coupling values.

\begin{figure}
	\centering
	\begin{subfigure}{0.48\linewidth}%
		\centering
		{\large $\grSU(2)$}\\[1em]
		\includegraphics[width=\linewidth]{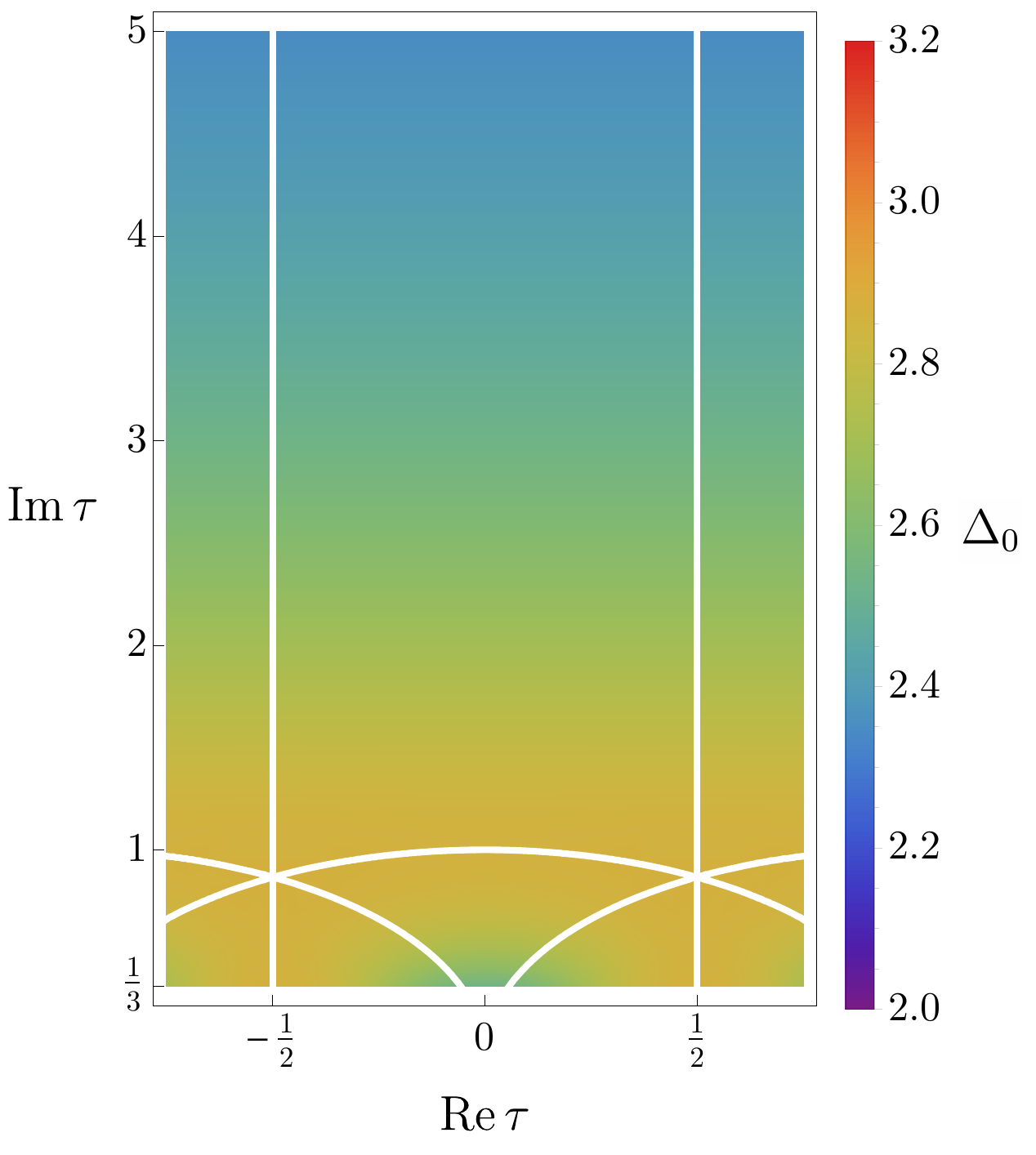}
	\end{subfigure}%
	\hspace{.04\linewidth}%
	\begin{subfigure}{0.48\linewidth}%
		\centering
		{\large $\grSU(3)$}\\[1em]
		\includegraphics[width=\linewidth]{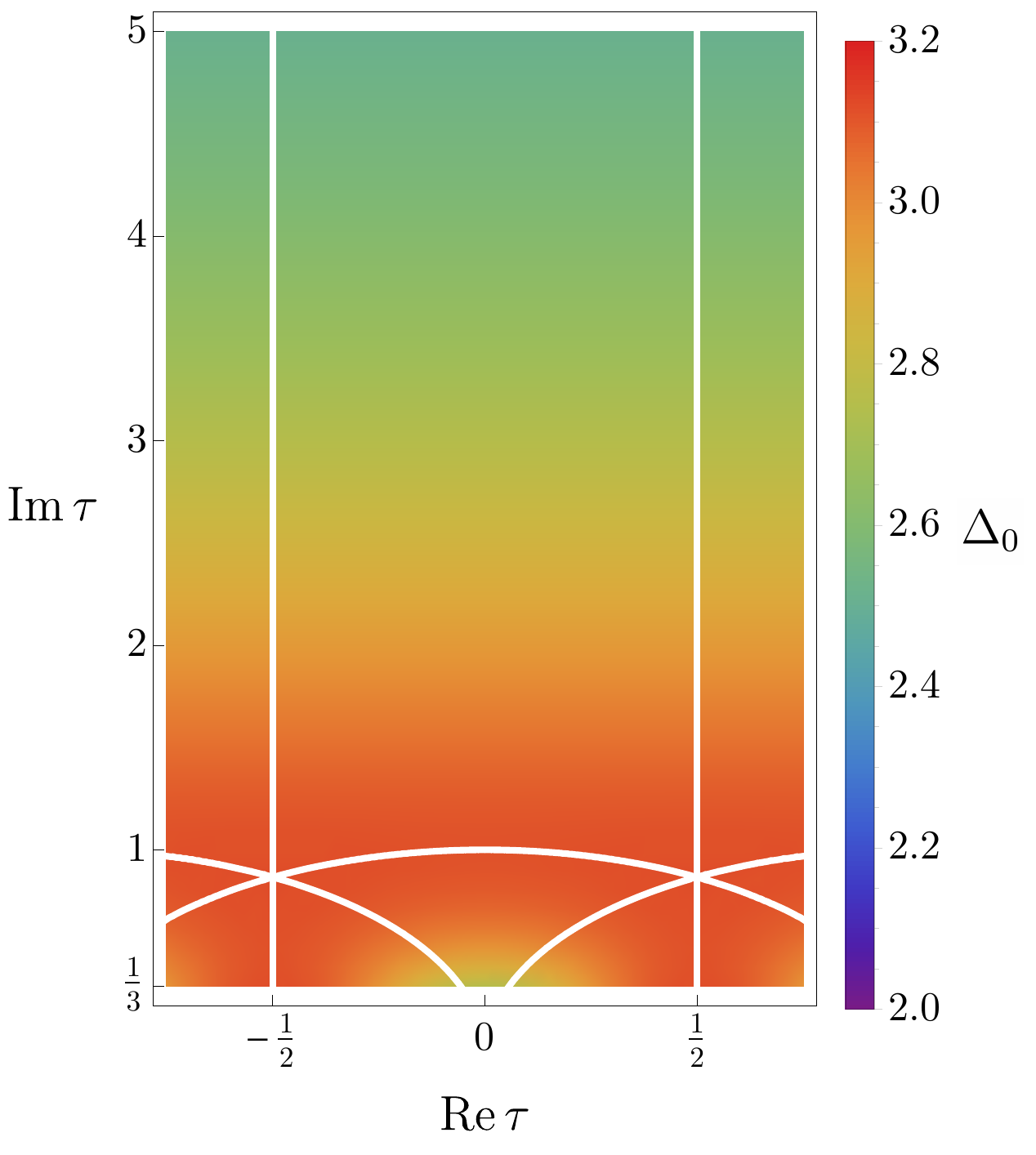}
	\end{subfigure}\\[.5cm]
	\begin{subfigure}{0.48\linewidth}%
		\centering
		\includegraphics[width=\linewidth]{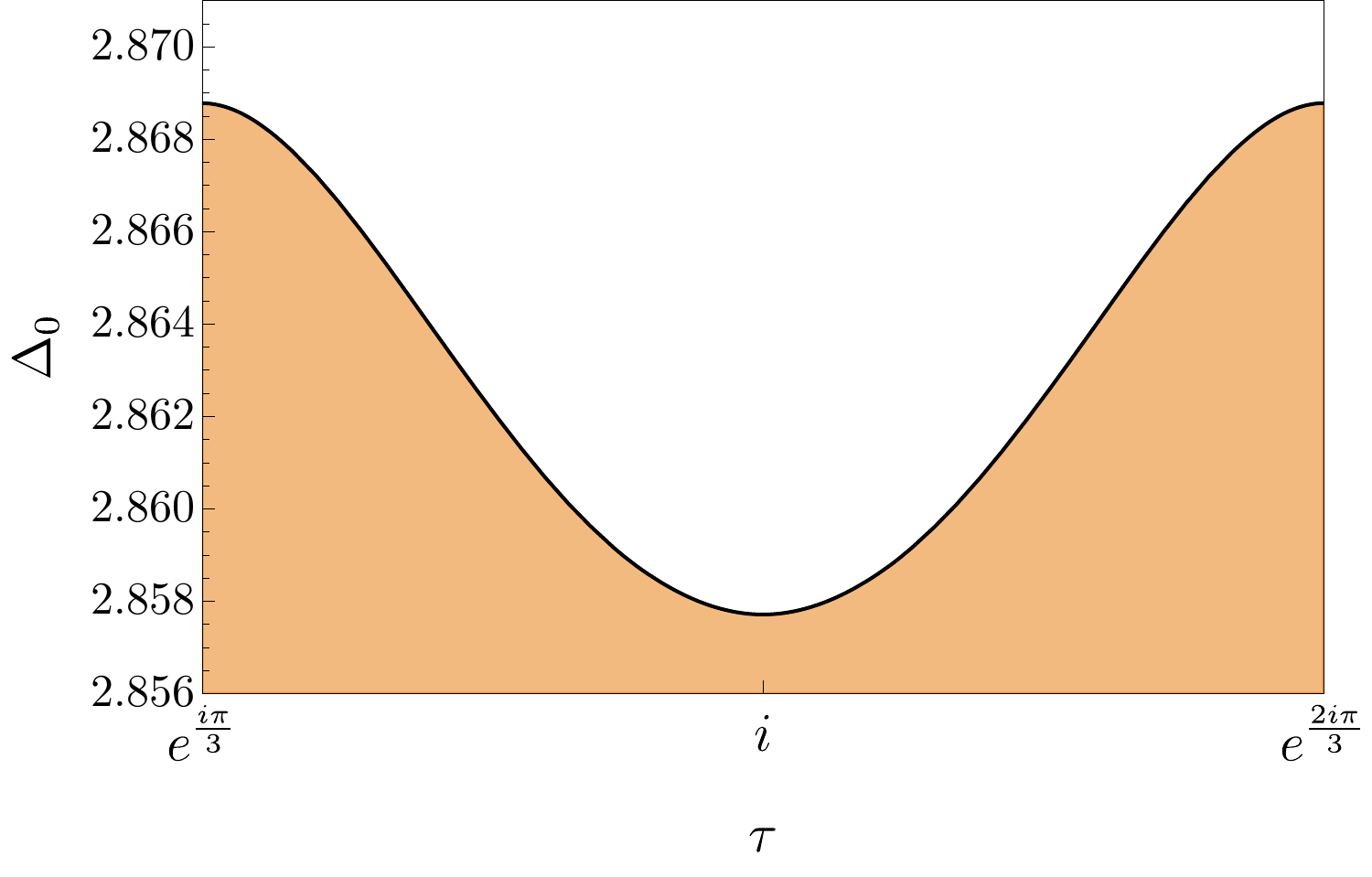}
	\end{subfigure}%
	\hspace{.04\linewidth}%
	\begin{subfigure}{0.48\linewidth}%
		\centering
		\includegraphics[width=\linewidth]{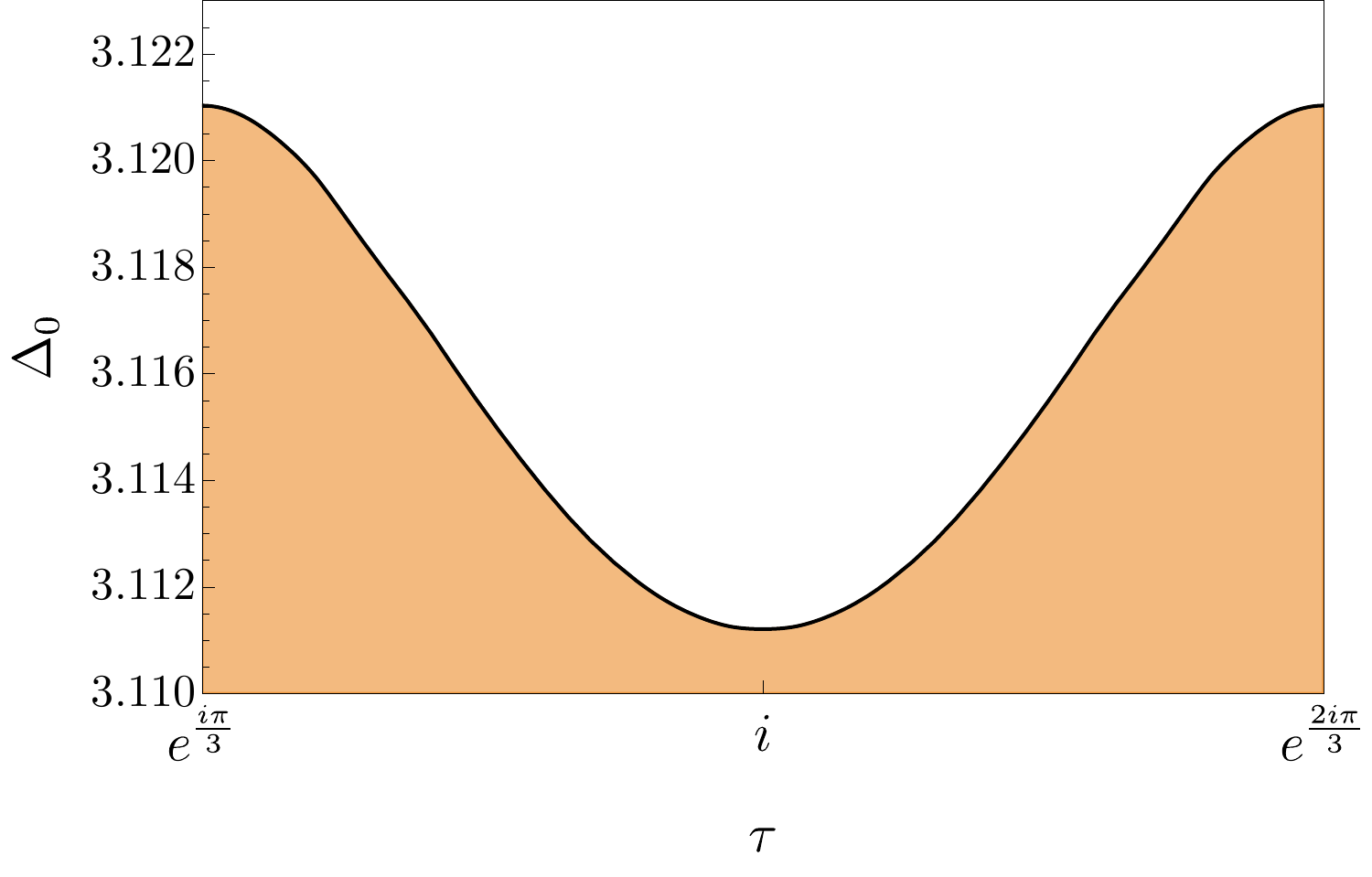}
	\end{subfigure}%
	\caption{Upper bounds on the scaling dimension of the lowest scalar as a function of complex coupling $\tau$ (\textbf{above}) and on the bottom boundary of the fundamental domain (\textbf{below}), obtained with bootstrap resolution $\Lambda = 39$. The dependence of the bounds on $\theta$ is weak within the fundamental domain, varying by less than 1\% between $\theta = 0$ and $\theta = \pi$. Note that, like the localization inputs, the slope of these upper bounds go to zero at the $\Z_2$ self-dual point $\tau = i$ and the $\Z_3$ self-dual point $\tau = e^{\pi i/3}$.}
	\label{fig:scaling_dimension_fd}
\end{figure}

In Figure~\ref{fig:scaling_dimension_fd} the upper bound on $\Delta_0$ is plotted across the standard fundamental domain \eqref{tauDomain} in terms of the complex coupling $\tau$, and extended via $\grSL(2,\Z)$ duality to the remainder of the upper half-plane. Within the standard fundamental domain, the dependence on $\theta$ is very weak. The bounds vary by less than 1\% between $\theta = 0$ and $\theta = \pi$. The bottom panels of Figure~\ref{fig:scaling_dimension_fd} show the bounds on the bottom boundary of the standard fundamental domain, namely along the arc of the unit circle between $e^{\pi i/3}$ and $e^{2\pi i/3}$.

\begin{figure}
	\centering
	\begin{subfigure}{0.75\linewidth}%
		\centering
		{\Large $\grSU(2)$}\\[1em]
		\includegraphics[width=\linewidth]{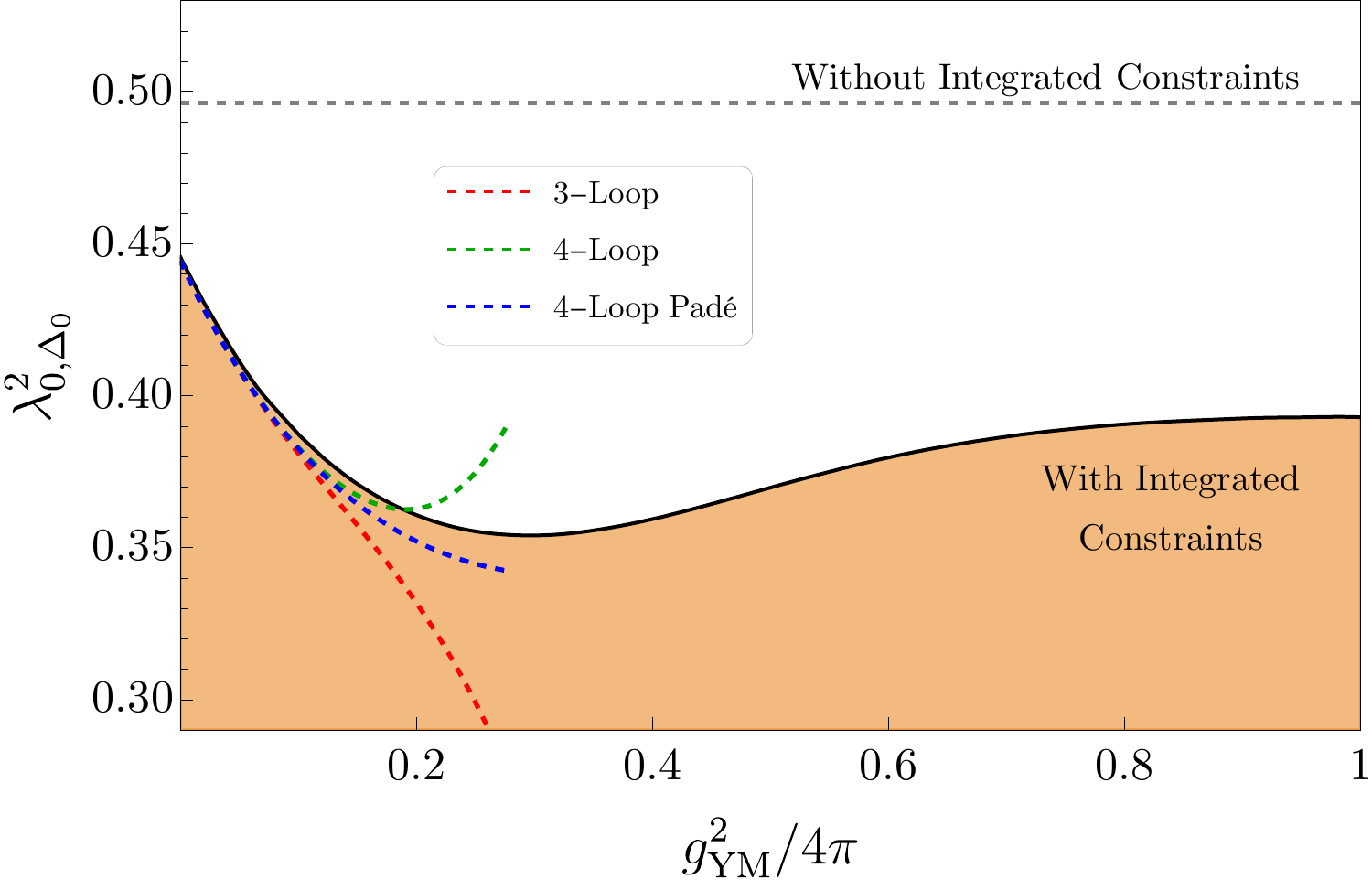}
	\end{subfigure}\\[2em]
	\begin{subfigure}{0.75\linewidth}%
		\centering
		{\Large $\grSU(3)$}\\[1em]
		\includegraphics[width=\linewidth]{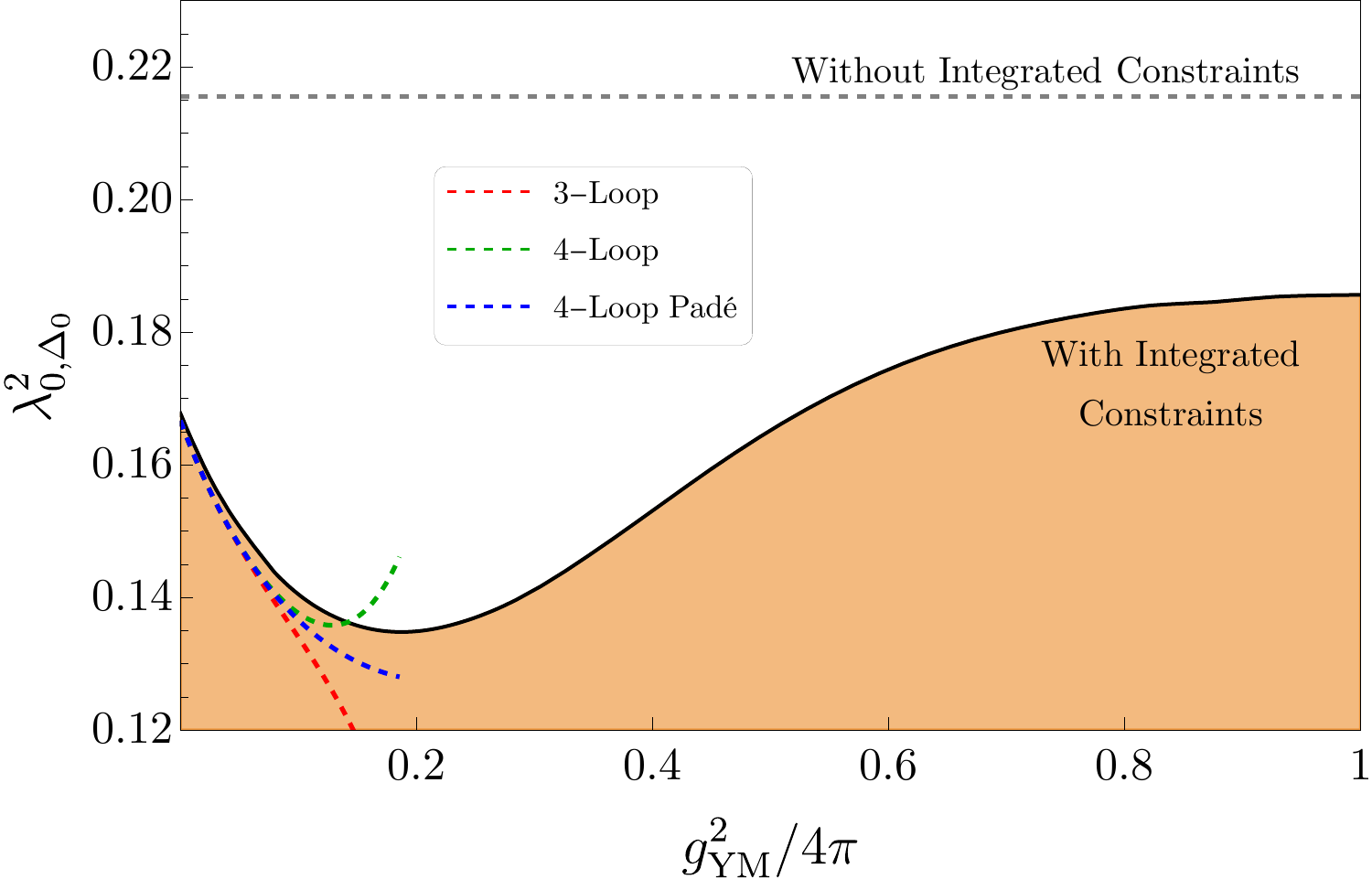}
	\end{subfigure}
	\caption{Upper bounds on the OPE coefficient of the lowest scalar for $\grSU(2)$ and $\grSU(3)$ as a function of $\gym^2$ at $\theta = 0$, obtained with bootstrap resolution $\Lambda = 39$. We plot from $\gym^2 = 0$ to the self-dual point at $\gym^2 = 4\pi$. The weak-coupling estimates for these coefficients are included, showing that our bounds are nearly saturated by these values at small $\gym^2$. We also show the upper bounds that would be obtained at the same bootstrap resolution without using the integrated constraints, and without using bounds on the scaling dimensions obtained with integrated constraints. Note that, like the localization inputs, the slope of these upper bounds go to zero at the $\Z_2$ self-dual point $\gym^2 = 1$.}
	\label{fig:ope_coefficient}
\end{figure}

As described above, we can also obtain bounds on the squared OPE coefficient of the lowest scalar operator, denoted by $\lambda_{0,\Delta_0}^2$. Figure~\ref{fig:ope_coefficient} shows the upper bounds we obtain on this OPE coefficient as a function of $\gym^2$ for $\theta = 0$. These bounds also appear close to being saturated by weak-coupling estimates. The agreement is less precise in this case because we have to input the scaling dimension of the lowest operator as estimated in Figure~\ref{fig:scaling_dimension}.

\begin{figure}
	\centering
	\begin{subfigure}{0.48\linewidth}%
		\centering
		{\large $\grSU(2)$}\\[1em]
		\includegraphics[width=\linewidth]{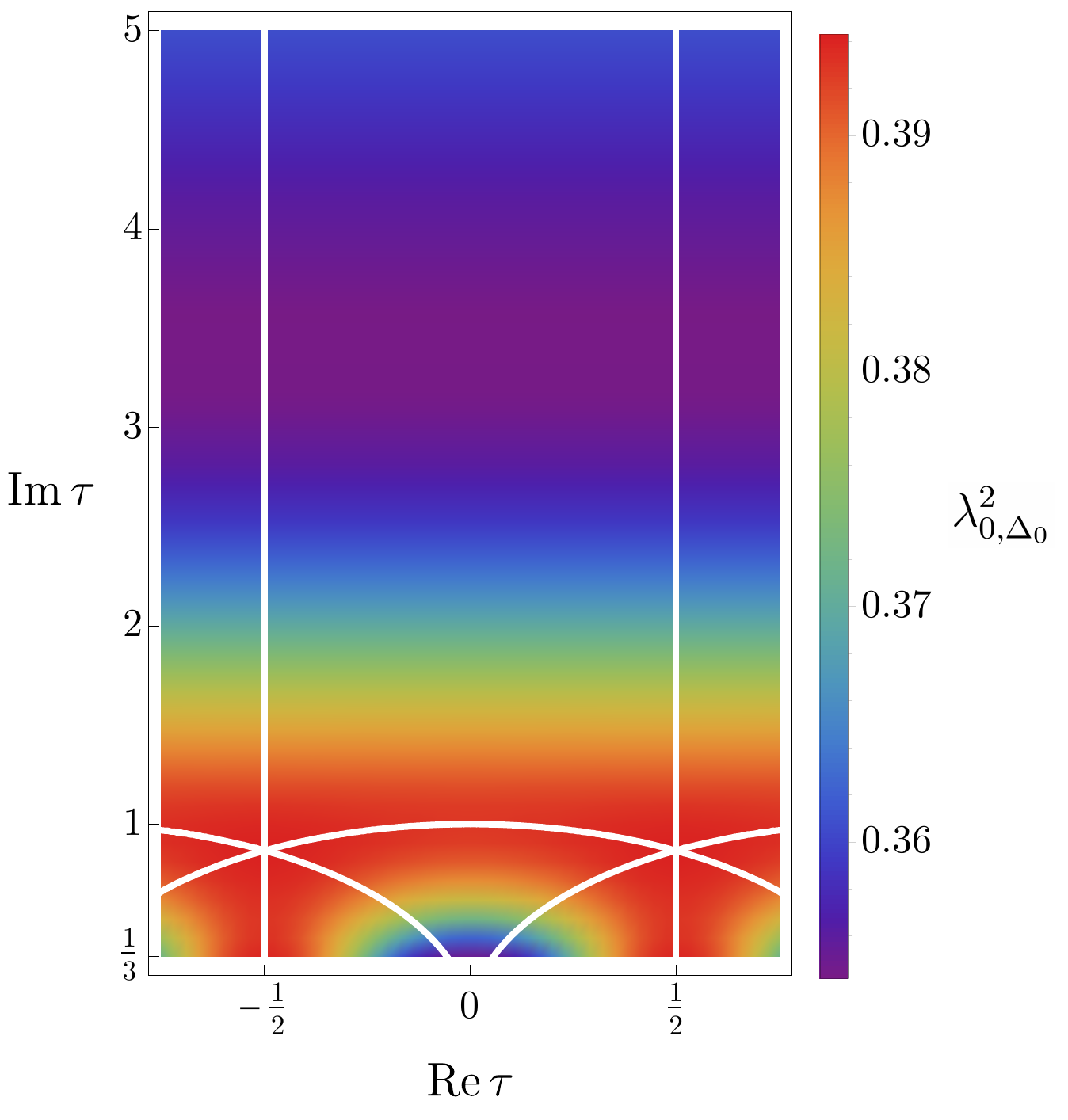}
	\end{subfigure}%
	\hspace{.04\linewidth}%
	\begin{subfigure}{0.48\linewidth}%
		\centering
		{\large $\grSU(3)$}\\[1em]
		\includegraphics[width=\linewidth]{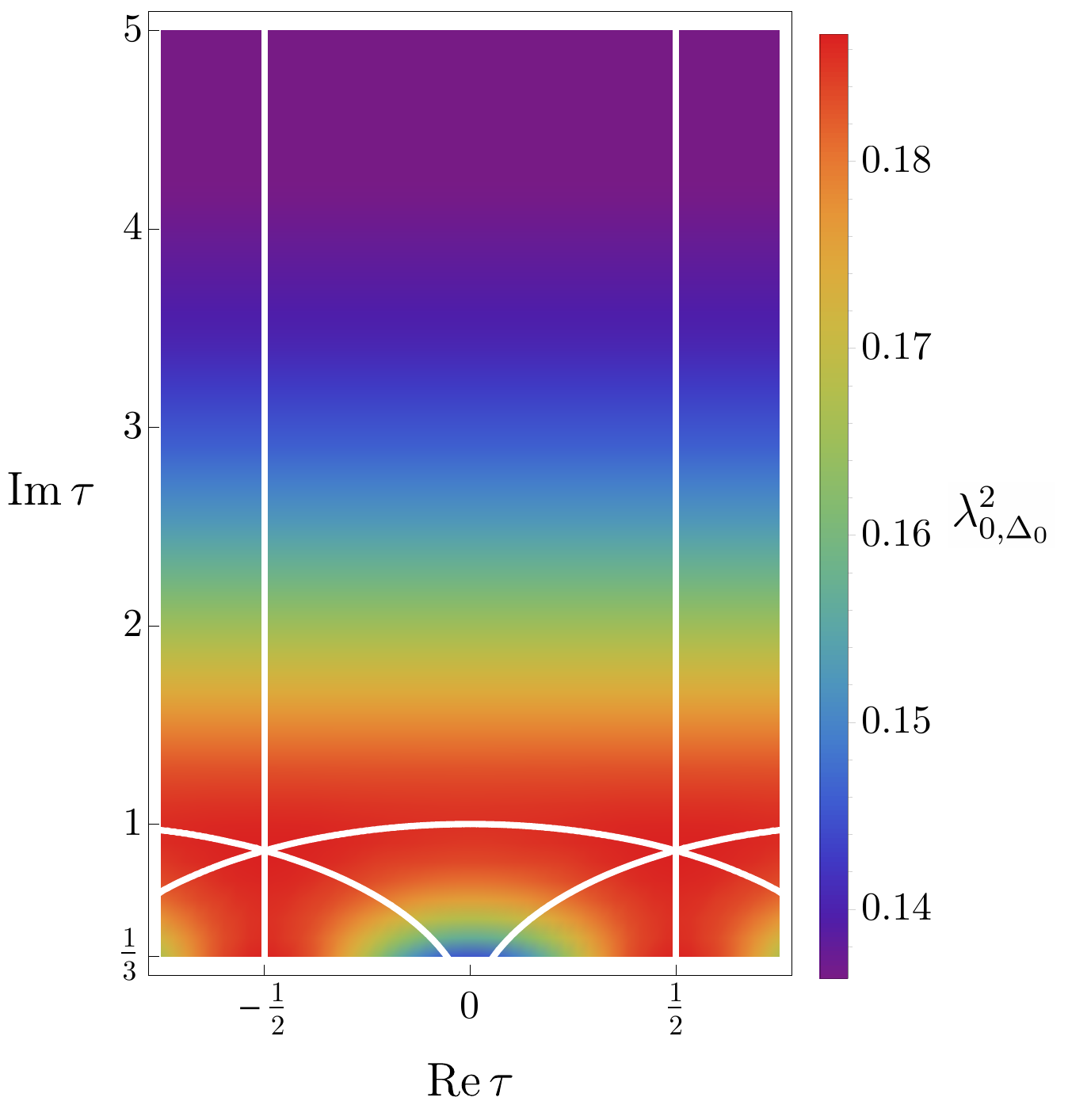}
	\end{subfigure}\\[.5cm]
	\begin{subfigure}{0.48\linewidth}%
		\centering
		\includegraphics[width=\linewidth]{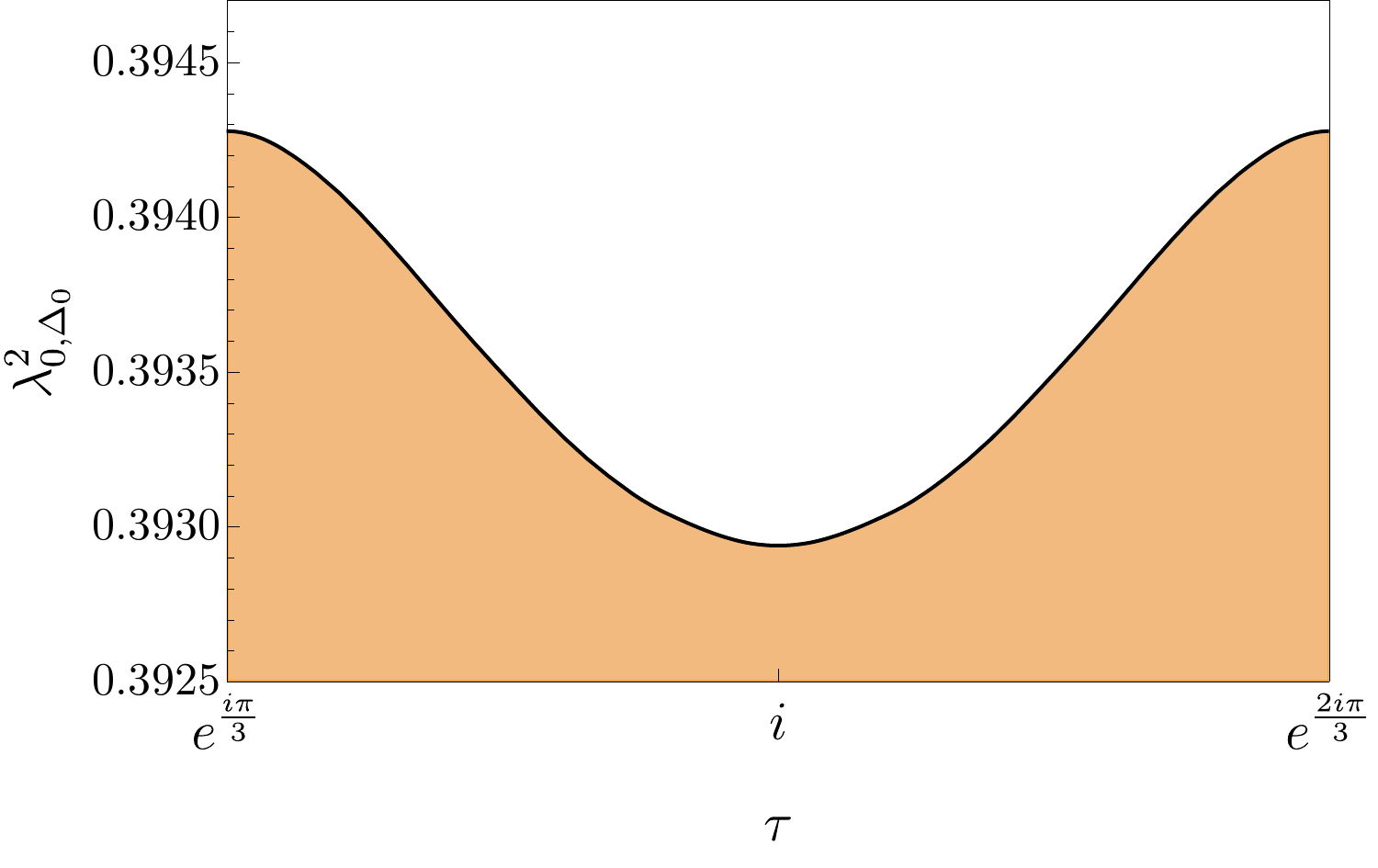}
	\end{subfigure}%
	\hspace{.04\linewidth}%
	\begin{subfigure}{0.48\linewidth}%
		\centering
		\includegraphics[width=\linewidth]{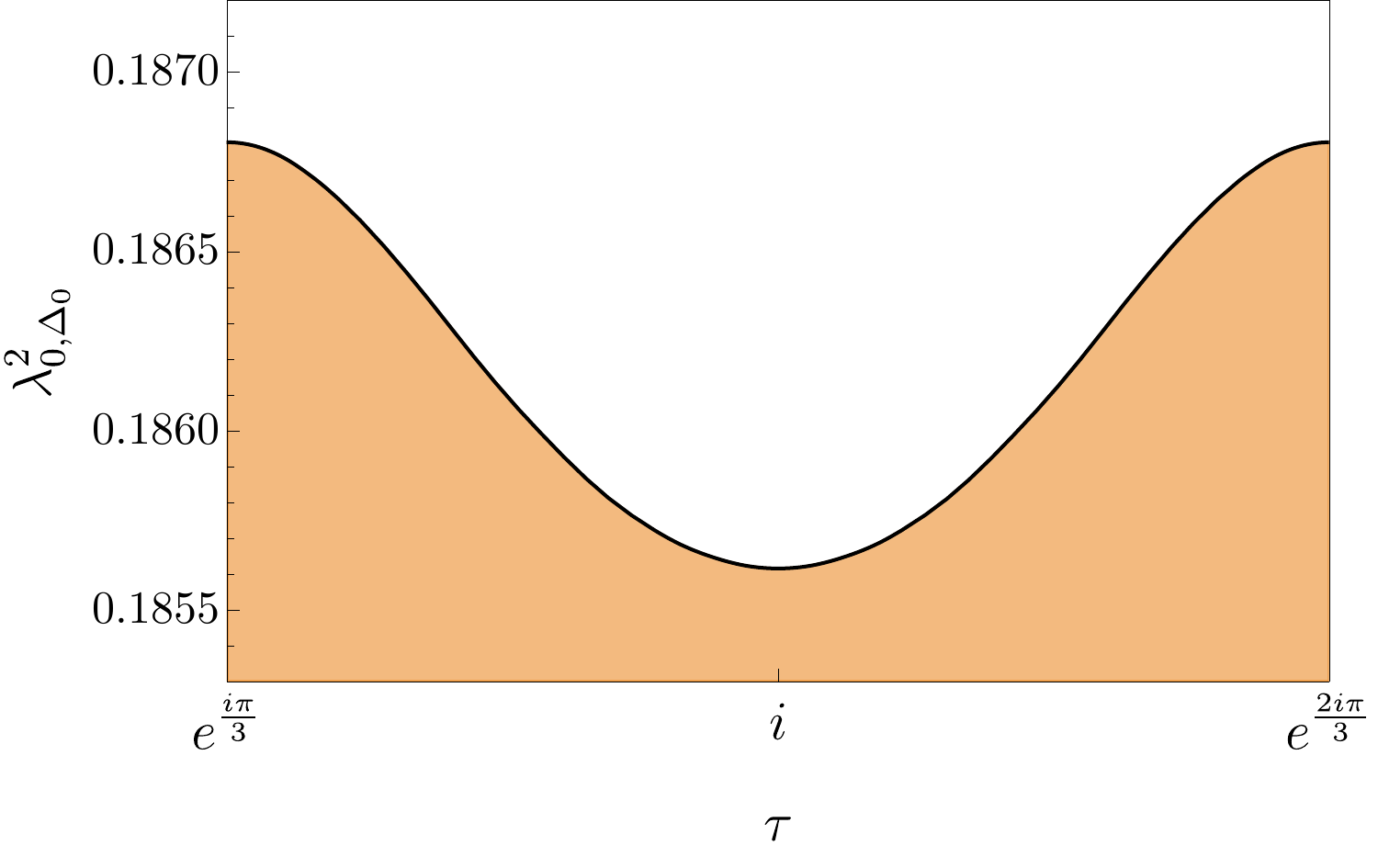}
	\end{subfigure}%
	\caption{The upper bounds on the OPE coefficient of the lowest scalar for $\grSU(2)$ and $\grSU(3)$ as a function of complex coupling $\tau$, obtained with bootstrap resolution $\Lambda = 39$. The bottom panels show how these bounds vary along the bottom boundary of the fundamental domain, the arc from $e^{\pi i/3}$ to $e^{2\pi i/3}$.  Note that, like the localization inputs, the slope of these upper bounds go to zero at the $\Z_2$ self-dual point $\tau = i$ and the $\Z_3$ self-dual point $\tau = e^{\pi i/3}$.}
	\label{fig:ope_coefficient_fd}
\end{figure}

We plot the bounds on the OPE coefficient as a function of $\tau$ in Figure~\ref{fig:ope_coefficient_fd}. Like in Figure~\ref{fig:scaling_dimension_fd}, our results are extended by $\grSL(2,\Z)$ duality to the remainder of the upper half-plane. We see again that the dependence on $\theta$ in the fundamental domain is weak. The bottom panels of Figure~\ref{fig:ope_coefficient_fd} show how the OPE coefficient bound varies on the arc between $e^{\pi i/3}$ and $e^{2\pi i/3}$.

Without imposing a gap, we are not yet able to obtain lower bounds on any of these pieces of CFT data. This is because, at the highest bootstrap resolution we have attempted ($\Lambda = 39$), theories with a single relevant scalar seem to still be allowed. Let $\Delta_0(\tau)$ denote an allowed scaling dimension for this scalar. If we then insert two relevant operators, the second operator could have any scaling dimension so long as the first operator has scaling dimension $\Delta_0(\tau)$. In particular, the second operator could have a dimension as low as the unitarity bound, thus preventing us from obtaining lower bounds.

\begin{figure}
	\centering
	\begin{subfigure}{0.33\textwidth}%
		\includegraphics[width=\linewidth]{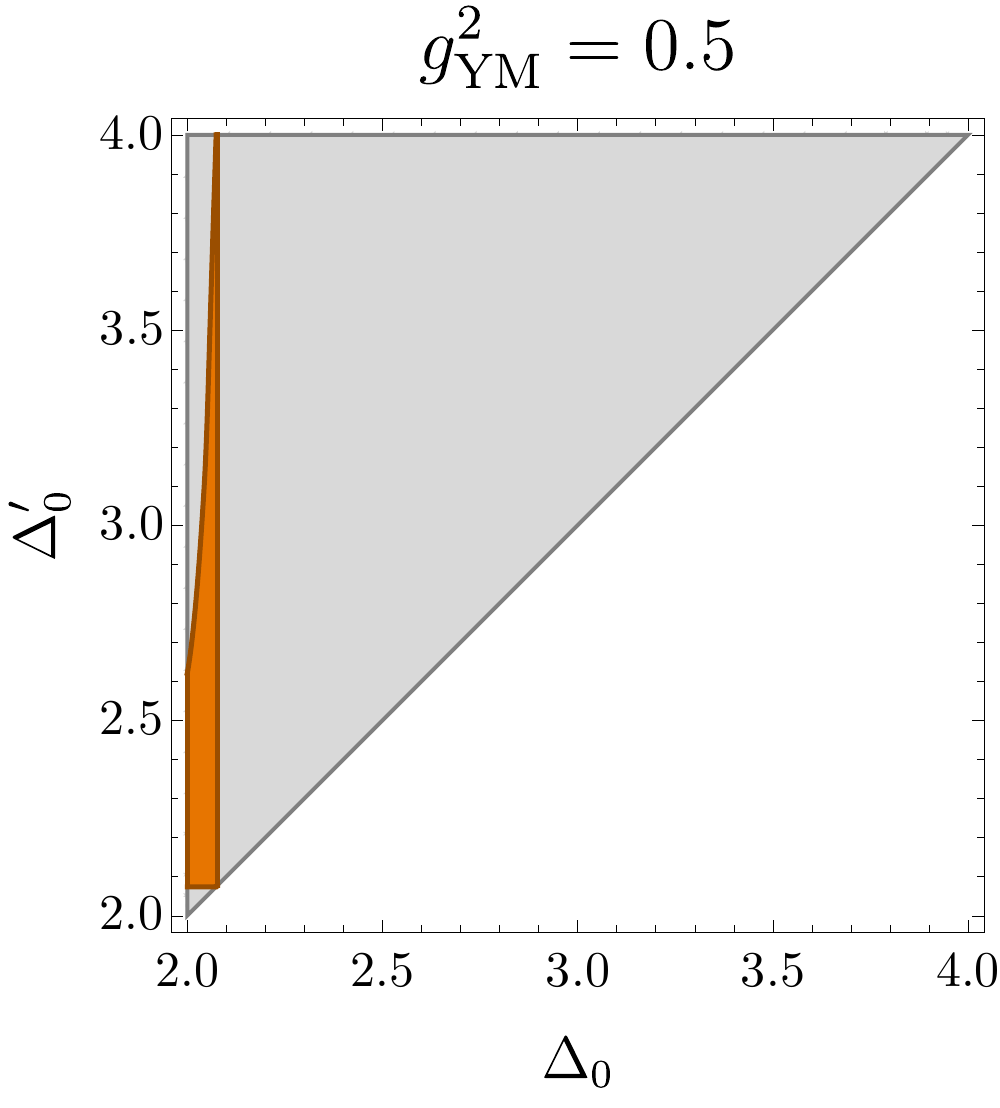}
	\end{subfigure}%
	\begin{subfigure}{0.33\textwidth}%
		\includegraphics[width=\linewidth]{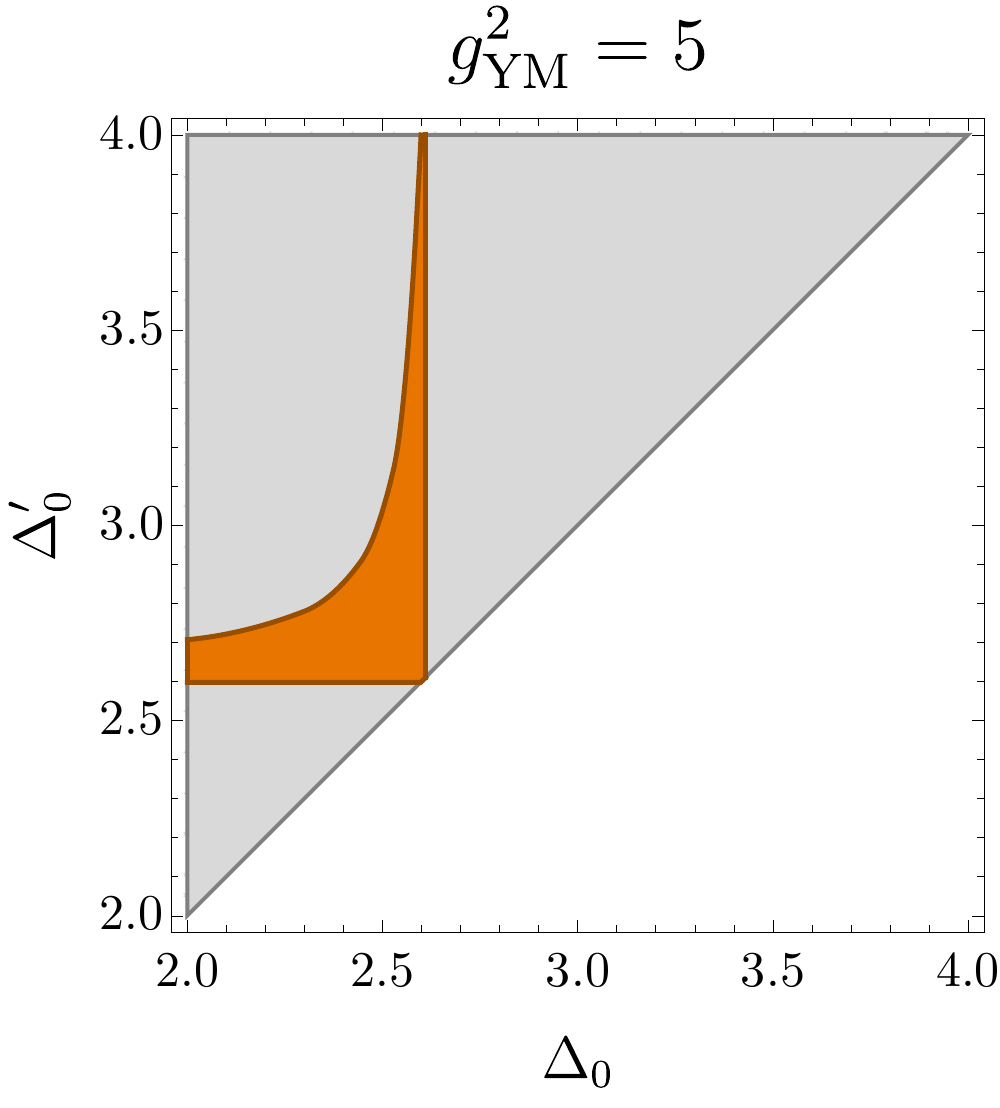}
	\end{subfigure}%
	\begin{subfigure}{0.33\textwidth}%
		\includegraphics[width=\linewidth]{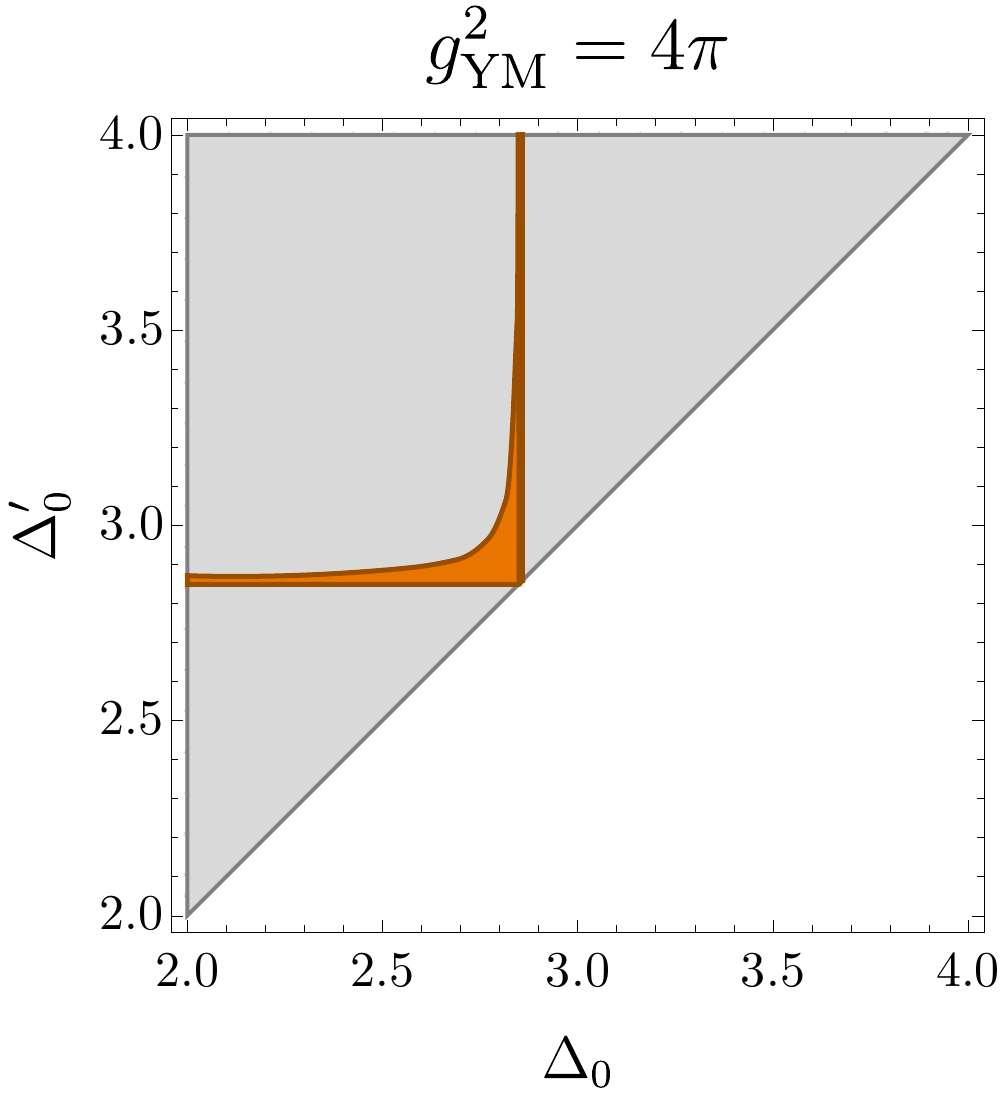}
	\end{subfigure}
	\caption{The allowed values (orange) for the two lowest scalar operator scaling dimensions, $\Delta_0$ and $\Delta_0'$, within the gray region in which $\Delta_0 < \Delta_0'$, obtained with bootstrap resolution $\Lambda = 39$ for  the $SU(2)$ theory with $\theta = 0$. Theories with $\Delta_0$ as low as the unitarity bound are allowed. However, if we assume some minimum value for $\Delta_0'$ then we can extract lower bounds, as shown in Figure~\ref{fig:lower_bounds}.}
	\label{fig:2d_feasible}
\end{figure}

This behavior is shown in Figure~\ref{fig:2d_feasible} in the $SU(2)$ case for $\theta = 0$. For three values of $\gym^2$, we plot the allowed region in the space of the lowest two scalar operator dimensions, denoted $\Delta_0$ and $\Delta_0'$, respectively. The gap on all other scalar operators in these plots is set to 4, so that the presence of allowed points with $\Delta_0' = 4$ indicates the feasibility of having a single relevant scalar. As a result of this fact, we see that there are also allowed points with $\Delta_0$ as low as 2, and so we cannot establish a lower bound.

Nevertheless, if we knew that $\Delta_0'$ is always above some threshold, then we could obtain a lower bound on $\Delta_0$. For instance, when $\gym^2 \ll 1$, we can use \eqref{twist4} to estimate the second-lowest scalar. It will be below 4, but barely so, and thus by imposing a gap slightly below 4 we should obtain lower bounds that include the weak-coupling result \eqref{konishiWeak}.

\begin{figure}
	\centering
	\begin{subfigure}{0.8\linewidth}%
		\centering
		\includegraphics[width=\linewidth]{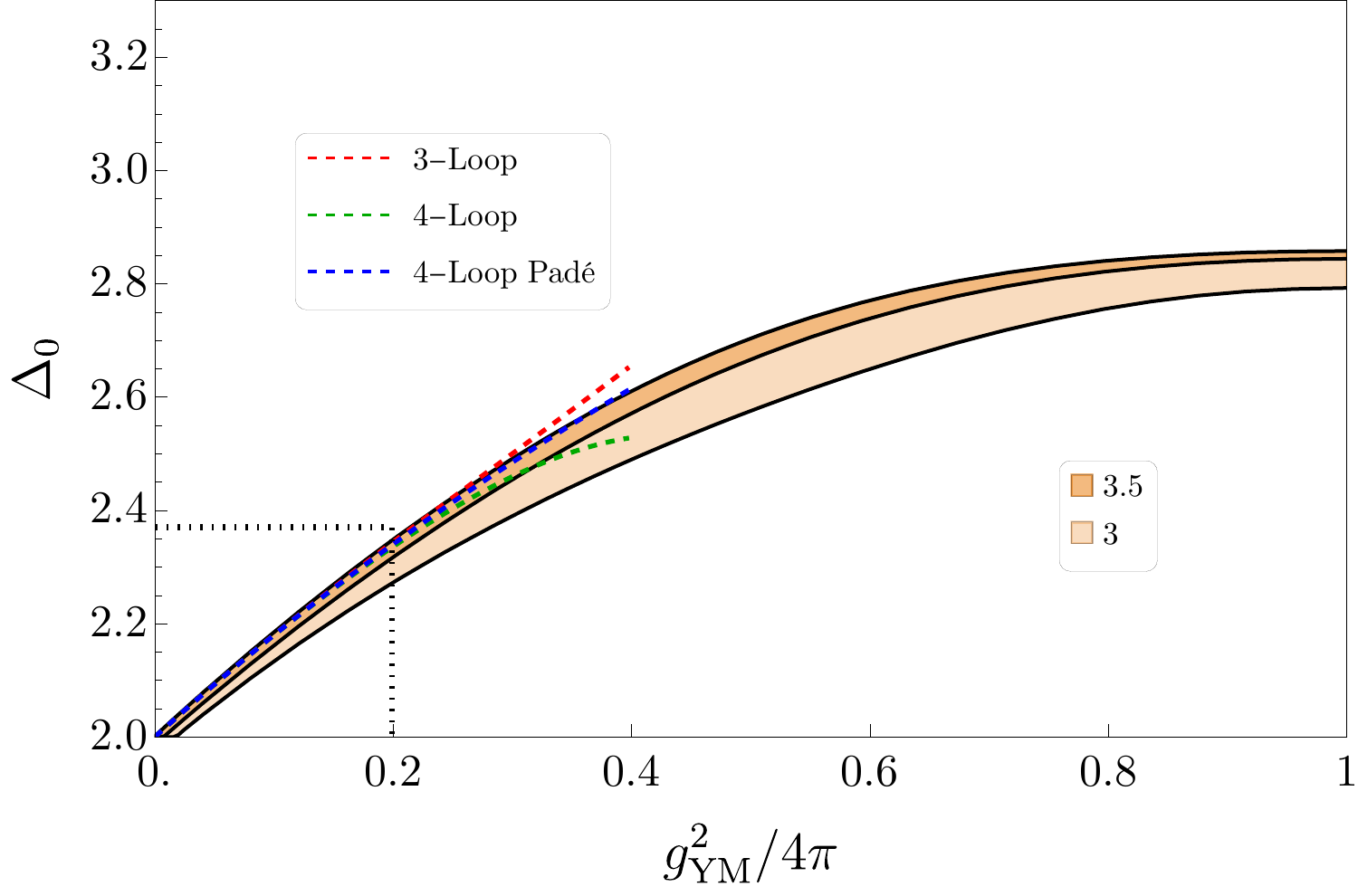}
	\end{subfigure}\\[2em]
	\begin{subfigure}{0.8\linewidth}%
		\centering
		\includegraphics[width=\linewidth]{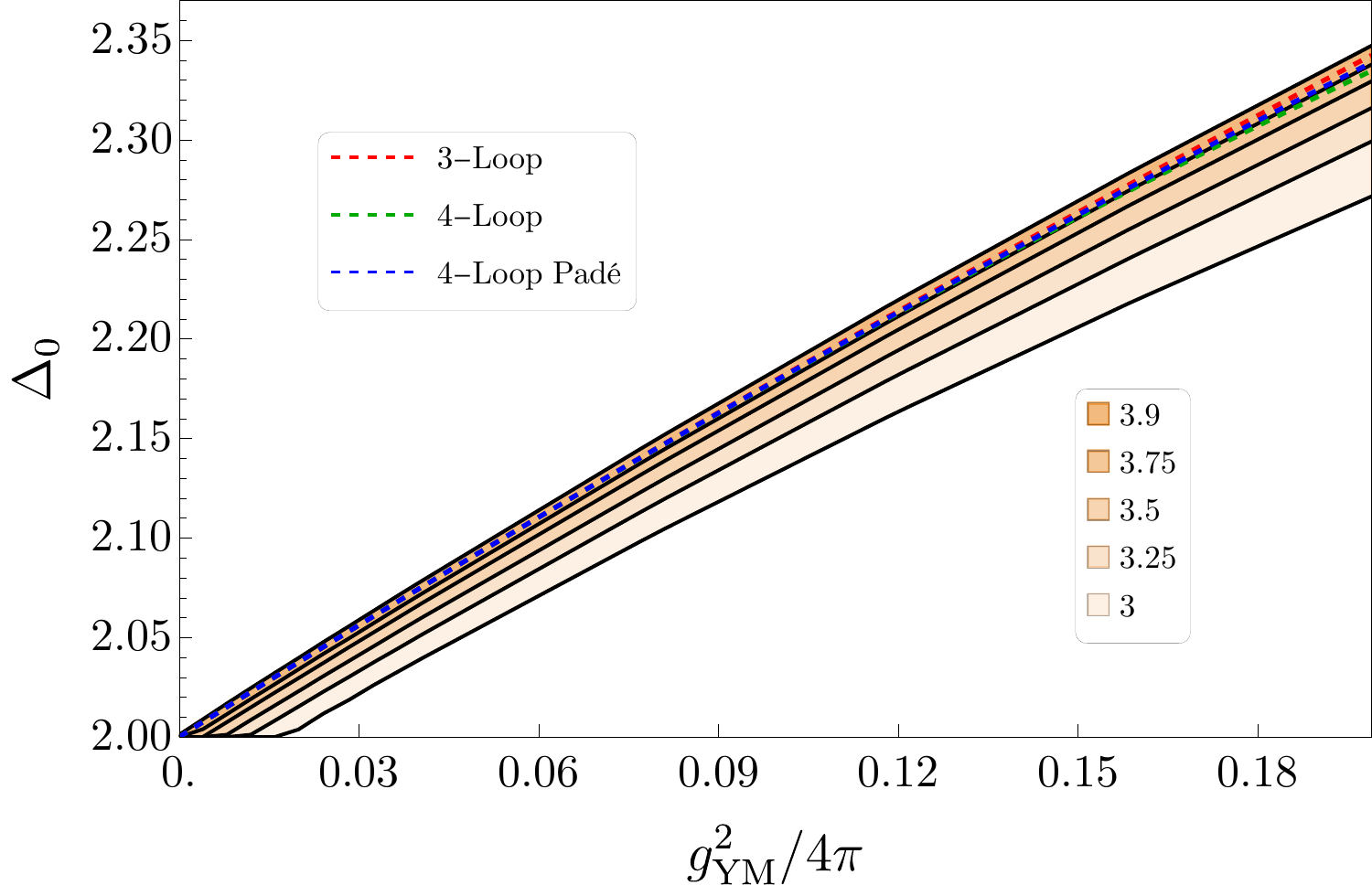}
	\end{subfigure}
	\caption{Lower bounds for the lowest scaling dimension obtained by assuming various gaps below 4, obtained with bootstrap resolution $\Lambda = 39$, for the $SU(2)$ theory at $\theta = 0$. The upper panel shows two of these lower bounds at $\theta = 0$ as a function of $\gym^2$, and the weak-coupling estimates of $\Delta_0$. The lower panel shows that the 4-loop Pad\'e approximant is within the bounds obtained with a gap at 3.9 up to $\gym^2/4\pi \sim 0.2$.}
	\label{fig:lower_bounds}
\end{figure}

Figure~\ref{fig:lower_bounds} shows the lower bounds on $\Delta_0$ we obtain for various gap assumptions for the $SU(2)$ theory at $\theta = 0$. If the second-lowest scalar is greater than one of these gaps for all coupling values, then the bounds corresponding to that gap are rigorous; otherwise, the bound applies only where the second-lowest scalar is above the gap value. In the bottom panel of Figure~\ref{fig:lower_bounds}, we zoom in on the weak-coupling region to show that the Pad\'e approximant that appears in Figure~\ref{fig:scaling_dimension} is indeed included within the lower bounds obtained by this method. 

\section{Discussion}
\label{disc}

In this paper we have shown how to combine two integrated constraints from supersymmetric localization with the infinitely many constraints from crossing symmetry and unitarity to numerically bootstrap the stress tensor multiplet correlator $\langle SSSS\rangle$ of $\mathcal{N}=4$ SYM as a function of the complexified coupling constant $\tau$.   Since the localization inputs are easier to compute for small rank gauge groups, we restricted our analysis to the cases $G=SU(2)$ and $SU(3)$.  For these cases, we determined upper bounds on the scaling dimension $\Delta_0$ and OPE coefficient of the lowest dimension scalar unprotected singlet operator. At weak coupling, we found that our upper bounds are close to being saturated by the four-loop weak coupling estimates \cite{Velizhanin:2009gv,Eden:2012rr,Fleury:2019ydf,Eden:2016aqo,Goncalves:2016vir} for a remarkably large range of couplings.  Interestingly, we found that the maximal values of our scaling dimension bounds occurs at the $\Z_3$ point $\tau=e^{i\pi/3}$, and that these maximal values are appreciably smaller than the conformal bootstrap bounds obtained without inputting the integrated constraints.   Lastly, we also found preliminary evidence for the entire range of $\tau$ that after imposing a gap $\Delta_0'$ above $\Delta_0$, a lower bound appears that meets the upper bound as $\Delta_0'$ is increased.

Looking ahead, our first goal is to upgrade these upper bounds to islands for every $G$ and $\tau$. As was the case for the $O(N)$ model bootstrap \cite{Kos:2014bka,Simmons-Duffin:2015qma,Kos:2015mba,Kos:2016ysd,Chester:2019ifh,Chester:2020iyt}, we expect that putting a gap to the spacetime dimension (4 in our case) and inserting the number of relevant operators below should give precise islands for the scaling dimensions of these operators. For the $\mathcal{N} = 4$ SYM theory, weak coupling and large $N$ estimates for $G=SU(N)$ suggest that there are at most two relevant operators for any $N$ and $\tau$. In order to get islands, the bootstrap must be able to accurately restrict the range of $\Delta_0'$ to an interval below 4 for all $\tau$, which we have not yet achieved. It is possible (but not likely) that we might accomplish this by simply increasing the parameter $\Lambda$ that measures the size of our functionals. Alternatively, we could consider a mixed correlator between $S$ and the dimension 3 half-BPS primary $S_3$, as was considered without integrated constraints in \cite{Bissi:2020jve,Alday:2021peq}. Not only will mixed correlators improve the bootstrap accuracy by allowing us to impose the uniqueness of the relevant operators as they appear in multiple correlators, which proved very effective for the $O(N)$ bootstrap \cite{Kos:2016ysd}, but this setup will also give us access to the integrated constraint on $\langle SSS_3S_3\rangle$ as derived in \cite{Binder:2019jwn}, where the localization input could be computed for any $G=SU(N)$ and $\tau$ following \cite{Losev:2003py,Nekrasov:2003rj,Nakajima:2003uh,Fucito:2015ofa}. In fact, there are infinite possible integrated constraints that we could access by considering $\langle SSS_pS_p\rangle$ for each integer $p$, where $S_p$ is the dimension $p$ half-BPS operator considered in \cite{Binder:2019jwn}.

In order to efficiently apply our algorithm to these more complicated correlators, we will need to improve the numerical implementation. Instead of discretizing $\Delta$ up to some $\Delta_\text{max}$ and applying linear programming, one could directly apply semidefinite programming to the continuous interval of $\Delta$ between the unitarity bound and $\Delta_\text{max}$, as was previously explored in \cite{Collier:2017shs}.\footnote{We thank Petr Kravchuk for suggesting this.} This will allow us to use the full power of \text{SDPB}, instead of treating it as a linear programming software in our current approach.

We would also like to apply our algorithm to any gauge group $G$, not just for $G=SU(2)$ and $SU(3)$. Since brute force evaluation of the matrix model integral for the localization input involves $\text{rank}(G)$ integrals, we will need more compact expressions. The $\partial_\tau\partial_{\bar\tau}\partial_m^2F\vert_{m=0}$ input was already computed for $G=SU(N)$ at finite $N$ and $\tau$ in \cite{Dorigoni:2021guq}, but we will also need a finite $N,\tau$ expression for the other input $\partial_m^4F\vert_{m=0}$. It would also be nice to generalize these finite $N,\tau$ expressions to general gauge group along the lines of \cite{Alday:2021peq}, and to the localization inputs that apply to $\langle SSS_pS_p\rangle$ in \cite{Binder:2019jwn}.

Finally, we would like to emphasize the broad applicability of our new method of combining numerical bootstrap with integrated constraints from localization to study SCFTs. This method should apply to any SCFT where the sphere free energy is computable from supersymmetric localization, which includes 3d $\mathcal{N}=2$ \cite{Kapustin:2009kz,Hama:2011ea}, 4d $\mathcal{N}=2$ \cite{Pestun:2007rz,Hama:2012bg}, and 5d $\mathcal{N}=1$ \cite{Imamura:2012xg,Imamura:2012bm} CFTs. In particular, 4d $\mathcal{N}=2$ SQCD also has a conformal manifold with one complex parameter, and we expect to be able to derive integrated constraints on correlation functions of some of the half-BPS operators. Another theory that would be interesting to study is 3d $\mathcal{N}=6$ ABJ(M) \cite{Aharony:2008ug,Aharony:2008gk} with gauge group $U(N)_k\times U(N+M)_{-k}$ and Chern-Simons level $k$, which has dual descriptions in either M-theory, string theory, or higher spin gravity depending on the values of $N$, $M$, and $k$. In \cite{Binder:2020ckj}, we combined the numerical conformal bootstrap with two protected OPE coefficients, which could be computed from supersymmetric localization \cite{Agmon:2017xes,Binder:2019mpb}, to fix a 2-parameter subspace of $\{k,N,M\}$. To fully fix the theory, we would need a third constraint, which could come from the integrated constraints derived in \cite{Binder:2018yvd,Binder:2021cif}. This would allow us to compute small precise islands for each $N,M,k$, as was already done for the $\mathcal{N}=8$ case with $k,M=1,2$ in \cite{Agmon:2019imm}. For the $\mathcal{N}=8$ case itself, adding this integrated constraint as well as another possible constraint coming from the squashed sphere free energy \cite{Chester:2020jay,Chester:2021gdw} could make the islands so precise that we could read off large $N$ corrections to the correlator, as initiated in \cite{Chester:2018lbz,Alday:2021ymb}, which would allow us to derive the M-theory S-matrix following \cite{Chester:2018aca}. We have truly only begun to explore the power of combining the conformal bootstrap with supersymmetric localization.

\section*{Acknowledgments} 

We thank Ofer Aharony, Alba Grassi, David Simmons-Duffin, Zohar Komargodski, Petr Kravchuk, Eric Perlmutter, and Leonardo Rastelli for useful discussions, and David Simmons-Duffin for collaboration at an early stage of this project. SSP and RD are supported in part by the Simons Foundation Grant No.~488653, and by the US NSF under Grant No.~2111977\@.   SMC is supported by the Zuckerman STEM Leadership Fellowship.

\appendix

\section{Supersymmetric localization}
\label{loc}

In this appendix, we will compute the mass derivatives of the $SU(N)$ $\mathcal{N}=2^*$ sphere free energy $F(m,\tau,\bar\tau)$ that appear in the integrated constraints \eqref{constraint1} as defined in \eqref{Fs}. We start by considering the massless partition function, where the instanton term cancels so that we simply get 
  \es{Z0}{
  Z(0, g_\text{YM}) &= \int \frac{d^N a}{N!} \, \delta \left( \sum_i a_i \right) \prod_{i < j}a_{ij}^2e^{- \frac{8 \pi^2}{g_\text{YM}^2} \sum_i a_i^2} \\
  &=2^{\frac{1}{2} \left(-4 N^2+N+3\right)} \sqrt{\frac{1}{N}} \pi ^{\frac{1}{2} \left(-2
   N^2+N+1\right)} G(N+1) g_\text{YM}^{N^2-1}\,,
 }
 where the second line was evaluated using orthogonal polynomials in \cite{Chester:2019pvm} and $G(x)$ is the Barnes-G function. We can then write $\mathcal{F}_2$ as an $(N-1)$-dimensional integral:
 \es{F2}{
 \cF_2(\tau)=\frac{4\pi^2}{g^4 c^2}\partial_\tau\partial_{\bar\tau}\Big[\sum_{i,j}\langle K'(a_{ij})\rangle+ \left \langle   \partial_m^2 (Z_\text{inst} + \bar Z_\text{inst})  \right \rangle \Big] \bigg|_{m=0} \,,
 }
where $K(z)\equiv -\frac{H'(z)}{H(z)}$ and the expectation value with respect to the unit-normalized massless partition function is
\es{Z0Exp}{
\langle f(a)\rangle\equiv  \int \frac{d^N a\, \delta \left( \sum_i a_i \right) }{N!  Z(0, g_\text{YM}) } f(a) \Big[\prod_{i < j}a_{ij}^2\Big]e^{- \frac{8 \pi^2}{g_\text{YM}^2} \sum_i a_i^2}\,.
} 
We can similarly write $\mathcal{F}_4$ as
\es{F4}{
c^2\cF_4(\tau)=&12\zeta(3)(1+4c)-\sum_{i, j}\langle K'''(a_{ij})\rangle-3\sum_{i,j,k,l}\left[\langle K'(a_{ij})K'(a_{kl})\rangle-\langle K'(a_{ij})\rangle\langle K'(a_{kl})\rangle\right]\\
&- \left \langle   \partial_m^4 (Z_\text{inst} + \bar Z_\text{inst})  \right \rangle- 6\langle  \partial_m^2Z_\text{inst}  \partial_m^2\bar Z_\text{inst}\rangle+6\langle  \partial_m^2Z_\text{inst} \rangle\langle \partial_m^2\bar Z_\text{inst}\rangle+3\langle \partial_m^2Z_\text{inst}\rangle^2+3\langle  \partial_m^2\bar Z_\text{inst}\rangle^2 \\
		&-6\langle K'(a_{ij})(\partial_m^2Z_\text{inst} +\partial_m^2\bar Z_\text{inst} )\rangle+6\langle K'(a_{ij})\rangle\langle(\partial_m^2Z_\text{inst} +\partial_m^2\bar Z_\text{inst} )\rangle\,.
}

To compute the mass derivatives of the Nekrasov partition function $Z_\text{inst}(m,\tau,a_{ij})$, we use the explicit expression for the $ Z_\text{inst}^{(k)} (m, a_{ij})$ in \eqref{ZInstSum}, which as shown in \cite{Nekrasov:2002qd,Nekrasov:2003rj} is given by a sum over $N$-tuples of Young diagrams $\vec Y=(Y_1, Y_2, \ldots, Y_N)$ with $k$ boxes in total:
\es{genNek}{
	& Z^{(k)}_{\rm inst}(m,a_{ij})= 
	\lim_{\epsilon_{1,2}\to1}\sum_{|\vec Y|=k} Z_{\vec Y}
	\\ 
	&	Z_{\vec{Y}}  \equiv 
 \prod_{i,j=1}^N	 	\frac{
		\prod_{s\in Y_i} (E(a_{ij},Y_i,Y_j,s)-i m -\epsilon_+/2)\prod_{t\in Y_j} (-E(a_{ji},Y_j,Y_i,t)-i m +\epsilon_+/2)
	}{
	\prod_{s\in Y_i}  E(a_{ij},Y_i,Y_j,s) \prod_{t\in Y_j} (\epsilon_+-E(a_{ji},Y_j,Y_i,t)) } \,,
}
where $\epsilon_\pm \equiv \epsilon_1\pm \epsilon_2$ and
\ie
E(a_{ij},Y_i,Y_j,s) \equiv i a_{ji}-\epsilon_1 h_j(s)+\epsilon_2 (v_i(s)+1) \,.
\fe
Here, $s$ labels a box $(\A,\B)$ ($\A$-th column and $\B$-th row) in a given Young diagram, while $h_i(s)$ and $v_i(s)$ denote the arm-length and leg-length, respectively, of the box $s$ in the diagram $Y_i$.  Each individual Young diagram $Y$ consists of columns of non-increasing heights $\lambda_1\geq \lambda_2 \geq \dots \geq \lambda_l$ with $\lambda_l\geq 1$. The transpose (conjugate) diagram $Y^T$ has columns of heights $\lambda^T_1\geq \lambda^T_2\geq \dots\geq \lambda^T_m$ with $\lambda^T_m\geq 1$.
Then the arm-length $h$ and leg-length $v$  of the box $s$ in $Y$ are given by
\ie
h(s)=\lambda^T_\B-\A \,, \qquad v(s)=\lambda_\A-\B.
\fe
Note that the definitions of $h$ and $v$ extend  beyond boxes in $Y$ to the entire quadrant $(\A,\B)\in \mZ_+^2$ in the obvious way, so they can be negative (e.g.~when $Y$ is empty).

The integrals in these expectation values in \eqref{F2} and \eqref{F4} can then be computed numerically for low $N$, where recall that there are $N-1$ integrals because we set $a_N=-\sum_{i=1}^{N -1} a_i$ as required by the $SU(N)$ measure. For numerical stability, it is useful to change integration variables so that $a_i-a_j$ terms do not appear in the Vandermonde determinant, and also to rescale the variables by $g_\text{YM}$ so that it does not appear in the Gaussian factor. We will now show some explicit results for the $SU(2)$ and $SU(3)$ cases that we consider in the main text.

\subsection{$SU(2)$}
\label{su2}

For $SU(2)$, after setting $a_2=-a_1=g_\text{YM}x$, we find that $\partial_m^2Z_\text{inst}^{(k)} (m, a_{ij})\vert_{m=0}$ for the lowest couple $k$ are:
\es{SU2m2}{
&k=1:\qquad -\frac{4 g_\text{YM}^2 x^2+3}{g_\text{YM}^2 x^2+1}\,,\\
&k=2:\qquad -\frac{6 (16 g_\text{YM}^4 x^4+64 g_\text{YM}^2 x^2+67)}{(4 g_\text{YM}^2 x^2+9)^2}\,,\\
}
while for $\partial_m^4Z_\text{inst}^{(k)} (m, a_{ij})\vert_{m=0}$ we get
\es{SU2m4}{
&k=1:\qquad \frac{12}{g_\text{YM}^2 x^2+1}\,,\\
&k=2:\qquad\frac{6 (128 g_\text{YM}^6 x^6+736 g_\text{YM}^4 x^4+1160 g^2 x^2+477)}{(g_\text{YM}^2 x^2+1)
   (4 g_\text{YM}^2 x^2+9)^2}\,.\\
}
In both cases it is easy to compute higher values of $k$, but we found that the numerical integrals for $\tau$ in the fundamental domain \eqref{tauDomain} converged quickly even with just a few terms. For instance, at the $\mathbb{Z}_3$ self-dual value $\tau=e^{i\pi/3}$ we get 
\es{SU2Ex}{
&\text{including $k=0$}:\qquad \cF_2(e^{i\pi/3})=0.270657218\,,\quad \cF_4(e^{i\pi/3})=36.72992\,,\\
&\text{including $k\leq1$}:\qquad \cF_2(e^{i\pi/3})=0.271801186\,,\quad \cF_4(e^{i\pi/3})=36.88321\,,\\
&\text{including $k\leq2$}:\qquad \cF_2(e^{i\pi/3})=0.271797213\,,\quad \cF_4(e^{i\pi/3})=36.88070\,,\\
&\text{including $k\leq3$}:\qquad \cF_2(e^{i\pi/3})=0.271797214\,,\quad \cF_4(e^{i\pi/3})=  36.88073\,,\\
}
while at the $\mathbb{Z}_2$ self-dual value $\tau=i$ we get
\es{SU2Ex2}{
&\text{including $k=0$}:\qquad \cF_2(e^{i\pi/3})=0.2708787623\,,\quad \cF_4(e^{i\pi/3})=36.759298\,,\\
&\text{including $k\leq1$}:\qquad \cF_2(e^{i\pi/3})=0.2703720954\,,\quad \cF_4(e^{i\pi/3})=36.692697\,,\\
&\text{including $k\leq2$}:\qquad \cF_2(e^{i\pi/3})=0.2703713536\,,\quad \cF_4(e^{i\pi/3})=36.692241\,,\\
&\text{including $k\leq3$}:\qquad \cF_2(e^{i\pi/3})=0.2703713535\,,\quad \cF_4(e^{i\pi/3})=  36.692238\,.\\
}
As a further consistency check, we can check that these explicit expressions satisfy $SL(2,\mathbb{Z})$ invariance after including sufficient values of $k$ in the expansion. For instance, we can look at the difference between the dual values $\tau=3/4i$ and $\tau=4/3i$ to get
\es{SU2diff}{
&\text{including $k=0$}:\qquad \cF_2(3/4i)- \cF_2(4/3i)\sim10^{-3}\,,\quad \cF_4(3/4i)-\cF_4(4/3i)\sim10^{-1}\,,\\
&\text{including $k\leq1$}:\qquad \cF_2(3/4i)- \cF_2(4/3i)\sim10^{-5}\,,\quad \cF_4(3/4i)-\cF_4(4/3i)\sim10^{-2}\,,\\
&\text{including $k\leq2$}:\qquad \cF_2(3/4i)- \cF_2(4/3i)\sim10^{-8}\,,\quad \cF_4(3/4i)-\cF_4(4/3i)\sim10^{-4}\,,\\
&\text{including $k\leq3$}:\qquad \cF_2(3/4i)- \cF_2(4/3i)\sim10^{-10}\,,\quad \cF_4(3/4i)-\cF_4(4/3i)\sim10^{-6}\,.\\
}

\subsection{$SU(3)$}
\label{su3}

For $SU(3)$, after setting $a_3=-a_1-a_2$, $a_1=\frac{g_\text{YM}}{2}(y+x)$, and $a_2=\frac{g_\text{YM}}{2}(y-x)$, we find that $\partial_m^2Z_\text{inst}^{(k)} (m, a_{ij})\vert_{m=0}$ for the lowest couple $k$ are:
\es{SU3m2}{
k=1:\qquad& \frac{8 (2 g_\text{YM}^2 x^3+5 x+3 y)}{x (g_\text{YM}^2 x^2+4) (g_\text{YM}^2 x^2-6 g_\text{YM}^2 x
   y+9 g_\text{YM}^2 y^2+16)}\\
   &+\frac{8 (2 g_\text{YM}^2 x^3+5 x-3 y)}{x (g_\text{YM}^2
   x^2+4) (g_\text{YM}^2 x^2+6 g_\text{YM}^2 x y+9 g_\text{YM}^2 y^2+16)}-\frac{2 (3 g_\text{YM}^2
   x^2+10)}{g_\text{YM}^2 x^2+4}\,,\\
k=2:\qquad& \frac{48 (3 g_\text{YM}^2 x^2 y+g_\text{YM}^4 x^5+15 g_\text{YM}^2 x^3+60 x+21 y)}{x (g_\text{YM}^2
   x^2+9)^2 (g_\text{YM}^2 x^2-6 g_\text{YM}^2 x y+9 g_\text{YM}^2 y^2+36)}\\
   &+\frac{48 (-3 g_\text{YM}^2 x^2
   y+g_\text{YM}^4 x^5+15 g_\text{YM}^2 x^3+60 x-21 y)}{x (g_\text{YM}^2 x^2+9)^2 (g_\text{YM}^2 x^2+6 g_\text{YM}^2
   x y+9 g_\text{YM}^2 y^2+36)}\\
   &-\frac{384 (g_\text{YM}^2 x^3+6 x+3 y)}{x (g_\text{YM}^2
   x^2+9) (g_\text{YM}^2 x^2-6 g_\text{YM}^2 x y+9 g_\text{YM}^2 y^2+36)^2}\\
   &-\frac{384 (g_\text{YM}^2 x^3+6
   x-3 y)}{x (g_\text{YM}^2 x^2+9) (g_\text{YM}^2 x^2+6 g_\text{YM}^2 x y+9 g_\text{YM}^2
   y^2+36)^2}-\frac{3 (3 g_\text{YM}^4 x^4+50 g_\text{YM}^2 x^2+215)}{(g_\text{YM}^2
   x^2+9)^2}\,,\\
}
while for $\partial_m^4Z_\text{inst}^{(k)} (m, a_{ij})\vert_{m=0}$ we get
\es{SU3m4}{
 k=1:\qquad&\scriptstyle\frac{192 (g_\text{YM}^2 x^3+x+3 y)}{x (g_\text{YM}^2 x^2+4)
   (g_\text{YM}^2 x^2-6 g_\text{YM}^2 x y+9 g_\text{YM}^2 y^2+16)}+\frac{48}{g_\text{YM}^2 x^2+4}+ \frac{192 (g_\text{YM}^2 x^3+x-3 y)}{x (g_\text{YM}^2 x^2+4) (g_\text{YM}^2 x^2+6 g_\text{YM}^2 x y+9
   g_\text{YM}^2 y^2+16)}\,,\\
k=2:\qquad&\scriptstyle\frac{12 (9 g_\text{YM}^6 x^6+198 g_\text{YM}^4 x^4+1273 g_\text{YM}^2 x^2+2250)}{(g_\text{YM}^2 x^2+4)
   (g_\text{YM}^2 x^2+9)^2}-\frac{1152 (9 g_\text{YM}^6 x^7+87 g_\text{YM}^4 x^5+39 g_\text{YM}^4 y x^4+254 g_\text{YM}^2 x^3+183 g_\text{YM}^2
   y x^2+120 x+168 y)}{x (g_\text{YM}^2 x^2+1) (g_\text{YM}^2 x^2+4) (g_\text{YM}^2
   x^2+9) (x^2 g_\text{YM}^2+9 y^2 g_\text{YM}^2-6 x y g_\text{YM}^2+36)^2}\\
   &\scriptstyle-\frac{96 (5 g_\text{YM}^{10} x^{11}+102 g_\text{YM}^8 x^9+18 g_\text{YM}^8 y x^8+702
   g_\text{YM}^6 x^7+252 g_\text{YM}^6 y x^6+1794 g_\text{YM}^4 x^5+1026 g_\text{YM}^4 y x^4+1657 g_\text{YM}^2 x^3+1032 g_\text{YM}^2 y x^2+660
   x+432 y)}{x (g_\text{YM}^2 x^2+1)^2 (g_\text{YM}^2 x^2+4) (g_\text{YM}^2
   x^2+9)^2 (x^2 g_\text{YM}^2+9 y^2 g_\text{YM}^2-6 x y g_\text{YM}^2+16)}\\
   &\scriptstyle+\frac{96 (5 g_\text{YM}^{10}
   x^{11}+87 g_\text{YM}^8 x^9+33 g_\text{YM}^8 y x^8+624 g_\text{YM}^6 x^7+360 g_\text{YM}^6 y x^6+1881 g_\text{YM}^4 x^5+1269 g_\text{YM}^4 y
   x^4+1557 g_\text{YM}^2 x^3+1926 g_\text{YM}^2 y x^2+666 x+792 y)}{x (g_\text{YM}^2 x^2+1)^2
   (g_\text{YM}^2 x^2+4) (g_\text{YM}^2 x^2+9)^2 (x^2 g_\text{YM}^2+9 y^2 g_\text{YM}^2-6 x y
   g_\text{YM}^2+36)}\\
   &\scriptstyle-\frac{96 (5 g_\text{YM}^{10} x^{11}+102 g_\text{YM}^8 x^9-18 g_\text{YM}^8 y x^8+702 g_\text{YM}^6 x^7-252
   g_\text{YM}^6 y x^6+1794 g_\text{YM}^4 x^5-1026 g_\text{YM}^4 y x^4+1657 g_\text{YM}^2 x^3-1032 g_\text{YM}^2 y x^2+660 x-432
   y)}{x (g_\text{YM}^2 x^2+1)^2 (g_\text{YM}^2 x^2+4) (g_\text{YM}^2 x^2+9)^2
   (x^2 g_\text{YM}^2+9 y^2 g_\text{YM}^2+6 x y g_\text{YM}^2+16)}\\
   &\scriptstyle+\frac{96 (5 g_\text{YM}^{10} x^{11}+87 g_\text{YM}^8
   x^9-33 g_\text{YM}^8 y x^8+624 g_\text{YM}^6 x^7-360 g_\text{YM}^6 y x^6+1881 g_\text{YM}^4 x^5-1269 g_\text{YM}^4 y x^4+1557 g_\text{YM}^2
   x^3-1926 g_\text{YM}^2 y x^2+666 x-792 y)}{x (g_\text{YM}^2 x^2+1)^2 (g_\text{YM}^2
   x^2+4) (g_\text{YM}^2 x^2+9)^2 (x^2 g_\text{YM}^2+9 y^2 g_\text{YM}^2+6 x y
   g_\text{YM}^2+36)}\\
   &\scriptstyle-\frac{1152 (9 g_\text{YM}^6
   x^7+87 g_\text{YM}^4 x^5-39 g_\text{YM}^4 y x^4+254 g_\text{YM}^2 x^3-183 g_\text{YM}^2 y x^2+120 x-168 y)}{x (g_\text{YM}^2
   x^2+1) (g_\text{YM}^2 x^2+4) (g_\text{YM}^2 x^2+9) (x^2 g_\text{YM}^2+9 y^2 g_\text{YM}^2+6 x
   y g_\text{YM}^2+36)^2}\,.\\
}
At the $\mathbb{Z}_3$ self-dual value $\tau=e^{i\pi/3}$ we get
\es{SU3Ex}{
&\text{including $k=0$}:\qquad \cF_2(e^{i\pi/3})=0.1096871\,,\quad \cF_4(e^{i\pi/3})=14.85431\,,\\
&\text{including $k\leq1$}:\qquad \cF_2(e^{i\pi/3})=0.1099688\,,\quad \cF_4(e^{i\pi/3})=14.89174\,,\\
&\text{including $k\leq2$}:\qquad \cF_2(e^{i\pi/3})=0.1099673\,,\quad \cF_4(e^{i\pi/3})=14.89111\,,\\
&\text{including $k\leq3$}:\qquad \cF_2(e^{i\pi/3})=0.1099674\,,\quad \cF_4(e^{i\pi/3})= 14.89112 \,,\\
}
while at the $\mathbb{Z}_2$ self-dual value $\tau=i$ we get
\es{SU3Ex2}{
&\text{including $k = 0$}:\qquad \cF_2(e^{i\pi/3})=0.1097433000\,,\quad \cF_4(e^{i\pi/3})=14.8617013\,,\\
&\text{including $k\leq1$}:\qquad \cF_2(e^{i\pi/3})=0.1096203854\,,\quad \cF_4(e^{i\pi/3})=14.8456504\,,\\
&\text{including $k\leq2$}:\qquad \cF_2(e^{i\pi/3})=0.1096201216\,,\quad \cF_4(e^{i\pi/3})=14.8455356\,,\\
&\text{including $k\leq3$}:\qquad \cF_2(e^{i\pi/3})=0.1096201214\,,\quad \cF_4(e^{i\pi/3})= 14.8455351\,.\\
}
As a further consistency check, we can check that these explicit expressions satisfy $SL(2,\mathbb{Z})$ invariance after including sufficient values of $k$ in the expansion. For instance, we can look at the difference between the dual values $\tau=3/4i$ and $\tau=4/3i$ to get
\es{SU3diff}{
&\text{including $k=0$}:\qquad \cF_2(3/4i)- \cF_2(4/3i)\sim10^{-4}\,,\quad \cF_4(3/4i)-\cF_4(4/3i)\sim10^{-2}\,,\\
&\text{including $k\leq1$}:\qquad \cF_2(3/4i)- \cF_2(4/3i)\sim10^{-6}\,,\quad \cF_4(3/4i)-\cF_4(4/3i)\sim10^{-3}\,,\\
&\text{including $k\leq2$}:\qquad \cF_2(3/4i)- \cF_2(4/3i)\sim10^{-8}\,,\quad \cF_4(3/4i)-\cF_4(4/3i)\sim10^{-5}\,,\\
&\text{including $k\leq3$}:\qquad \cF_2(3/4i)- \cF_2(4/3i)\sim10^{-12}\,,\quad \cF_4(3/4i)-\cF_4(4/3i)\sim10^{-6}\,.\\
}

\section{Conformal block expansion}
\label{blockExp}

In this appendix we describe how to efficiently compute the small $r$ expansion of 4d conformal blocks $G_{\Delta,\ell}(U,V)$. We start by defining the variables
\es{rho}{
\rho=r(\eta+i\sqrt{1-\eta^2})\,,\qquad \bar\rho=r(\eta-i\sqrt{1-\eta^2})\,,
}
which are simply related to $z$ and $\bar z$ as
\es{ztorho}{
z=\frac{4\rho}{(1+\rho)^2}\,,\qquad \bar z=\frac{4\bar\rho}{(1+\bar\rho)^2}\,.
}
Expanding in small $\rho,\bar\rho$ thus coresponds to expanding at small $r$. 

Recall that the 4d conformal blocks are written in terms of the functions $k_\beta(z)$ defined in terms of hypergeometrics in \eqref{4dblock}. From the standard series representation of the hypergeometric, we can define a recursion relation for $k_\beta(z)$ in terms of $\rho$ as
 \es{RecFormula}{
  k_h=&(4\rho)^h u_h\,,\qquad u_h(\rho) = u_\infty(\rho) - \sum_{k=1}^\infty \frac{c_k  \rho^{2k}}{h + k-\frac 12} u_{k + \frac 12}(\rho) \,,
 }
where
 \es{uinfcDefs}{
  u_\infty(\rho) = \left(\frac{1}{1 - \rho^2}\right)^{\frac{1}{2}} \,, \qquad c_k = \frac{\Gamma^2\left( k + \frac 12 \right)}{ \pi \Gamma(k) \Gamma(k+1)} \,.
 }
We can then use this recursion relation along with the identity
\es{arhoidentity}{
\frac{z\bar z}{z-\bar z}=\frac{2r}{(1-r^2)\sqrt{\eta^2-1}}\,,
}
to efficiently expand each term in \eqref{4dblock} at small $r$ to get the block expansion \eqref{4dblockNorm}.

\section{Numerical bootstrap details}
\label{bootApp}

As described in Section \ref{alg}, our bootstrap calculations reduce to solving linear programs, following the setup in \cite{Rattazzi:2008pe}. We search through a space of functionals $\alpha$ parametrized by $\Lambda$, where
\begin{equation}
	\alpha_\infty \cdot F_{\Delta, \ell}(U, V) = \sum_{m + n \leq \Lambda} \alpha_{m, n}\left(\partial^m_z \partial^n_{\bar z} F_{\Delta, \ell}(U, V)\right)_{z=\bar z = 1/2}.
\end{equation}
The parameters $\alpha_{m,n}$, along with $\alpha_{2,1}$, $\alpha_{2,2}$, $\alpha_{4,1}$, and $\alpha_{4,2}$ which define the action of the functional on the integrated constraints, are the variables of the linear program.

The constraints of the linear program implement the positivity conditions \eqref{searchScal}. For each spin $\ell$, we wish to include constraints corresponding to a continuum of possible operators from some gap $\bar\Delta_\ell$ all the way to infinity. To make the problem finite, we consider operators with dimensions spaced by some separation $\Delta_\text{sp}$, and up to some cutoff $\Delta_\text{max}$. We also include one additional constraint for each spin at $\Delta = 500$, which approximates the asymptotic large-$\Delta$ behavior.

In practice, once we make $\Delta_\text{sp}$ sufficiently small and $\Delta_\text{max}$ sufficiently large, our set of constraints approximates a continuum well enough that the results are no longer sensitive to the values of these parameters. In all the plots shown in this paper, $\Delta_\text{sp} = 0.1$ is small enough to see this behavior, and $\Delta_\text{max} = 100$ is large enough.

\begin{figure}
	\centering
	\begin{subfigure}{\textwidth}%
	\centering
	\includegraphics[width=0.8\linewidth]{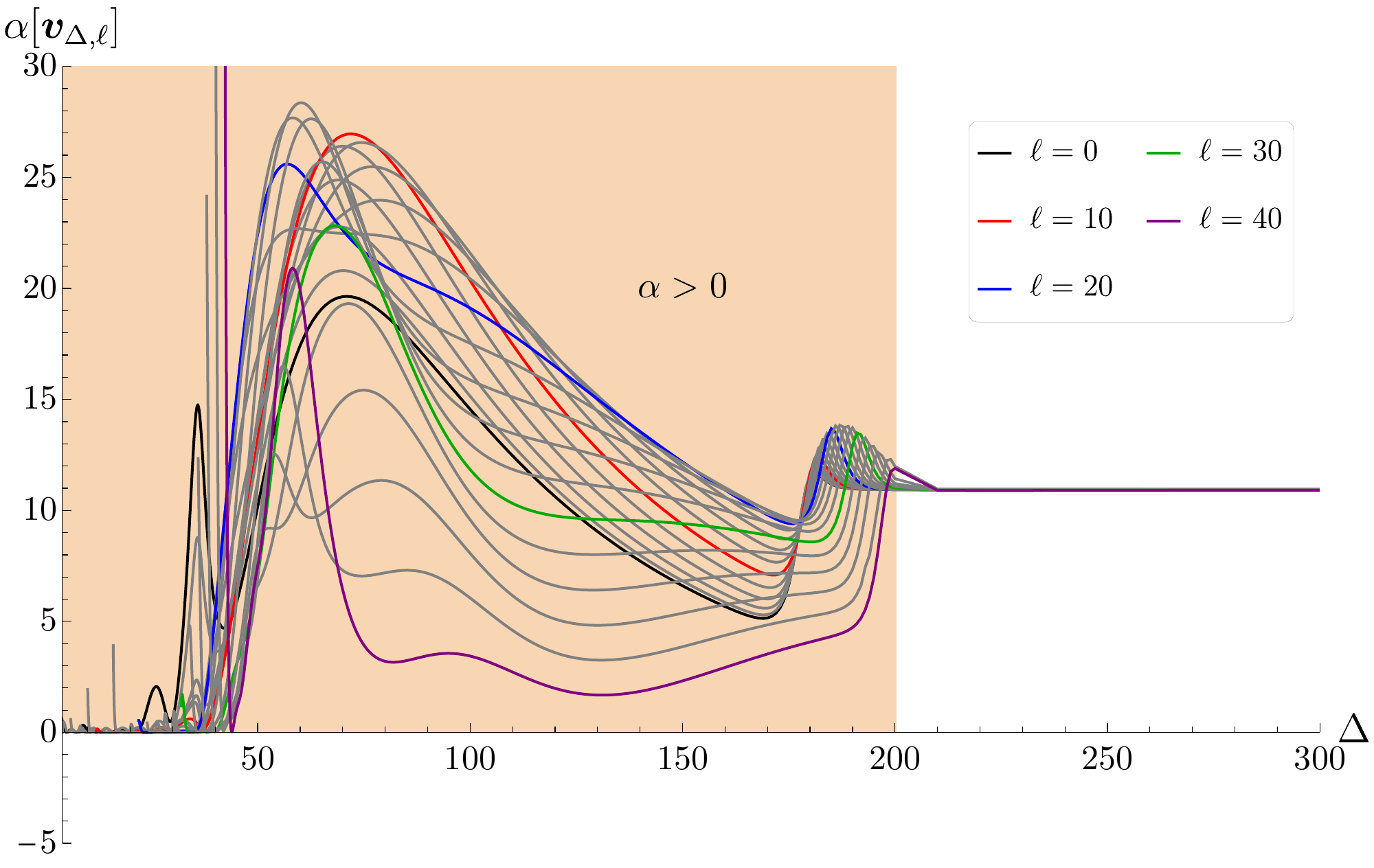}
	\caption{A functional establishing $\Delta_0 \leq 2.8688$ at the $\Z_3$ point of the $\grSU(2)$ theory. Positivity is imposed only in the yellow shaded region, but we find that the functional is positive for all $\Delta \geq \bar\Delta_\ell$ for each spin $\ell$.}
	\end{subfigure}\\[3em]
	\begin{subfigure}{\textwidth}%
	\centering
	\includegraphics[width=0.8\linewidth]{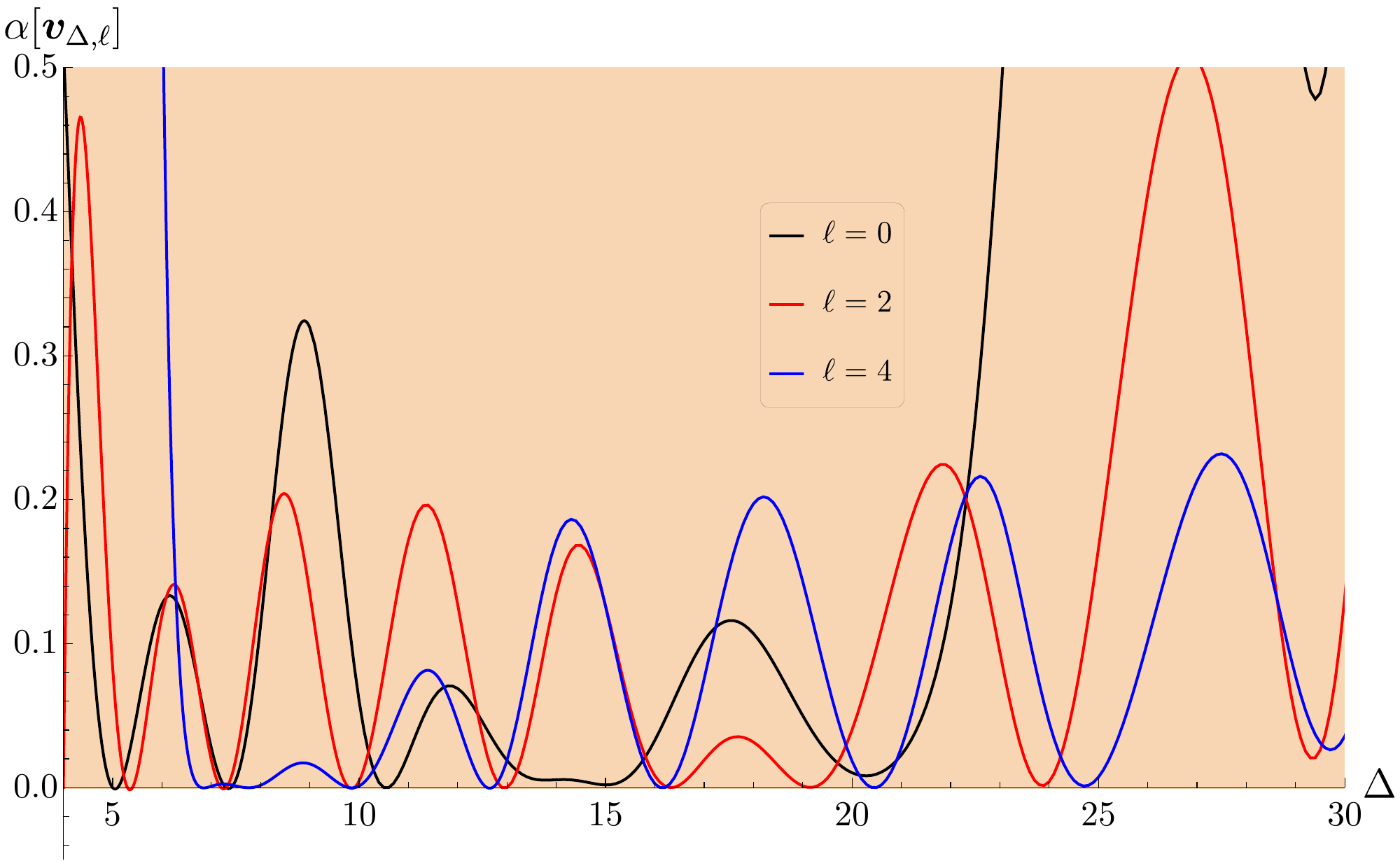}
	\caption{The functional above evaluated only for spins 0, 2, and 4, showing in more detail the low-lying extremal spectrum for these spins.}
	\label{fig:positive_functional_low}
	\end{subfigure}
	\caption{}
	\label{fig:positive_functional}
\end{figure}

In Figure \ref{fig:positive_functional}, we give an example of a positive functional calculated at $\Lambda = 39$, giving an upper bound of 2.8688 on the lowest scaling dimension for the $\grSU(2)$ theory at the $\Z_3$ symmetric point. This corresponds to the rightmost point of the $\grSU(2)$ plot in Figure \ref{fig:scaling_dimension}, and is the maximum upper bound we find throughout the conformal manifold. The bound we obtain at this point is insensitive to increases in $\Delta_\text{max}$ or the maximum spin, or decreases in $\Delta_\text{sp}$.

We used two linear program solvers, Gurobi and SDPB \cite{gurobi,Simmons-Duffin:2015qma}. Gurobi is heavily optimized for solving linear programs, and is very efficient for obtaining results at low $\Lambda$. However, it can only work at machine precision. For high enough $\Lambda$, the large dynamic range of conformal block derivatives and integrated constraints leads to numerical instability. This is most straightforwardly corrected by working at higher precision. When this is required, we use SDPB with a precision of 256 bits, and all other parameters set to their default values. The parameters used are summarized in Table \ref{tab:numerical_parameters}.

\begingroup
\renewcommand{\arraystretch}{1.3}
\begin{table}
	\centering
	\begin{tabular}{l|p{4cm}p{4cm}}
		$\Lambda$ & $\leq 29$ & $> 29$ \\
		\hline
		Solver & Gurobi & SDPB \\
		\hline
		$\Delta_\text{max}$ & 100 & 100 \\
		$\Delta_\text{sp}$ & 0.1 & 0.1 \\
		\hline
		\texttt{NumericFocus} & 3 & -- \\
		\texttt{DualReductions} & 0 & -- \\
		\hline
		\texttt{precision} & machine ($\sim 53$) & 256 \\
		\texttt{dualityGapThreshold} & -- & $10^{-30}$ \\
		\texttt{primalErrorThreshold} & -- & $10^{-30}$ \\
		\texttt{dualErrorThreshold} & -- & $10^{-30}$ \\
		\texttt{initialMatrixScalePrimal} & -- & $10^{20}$ \\
		\texttt{initialMatrixScaleDual} & -- & $10^{20}$ \\
		\texttt{feasibleCenteringParameter} & -- & $0.1$ \\		
		\texttt{infeasibleCenteringParameter} & -- & $0.3$ \\
		\texttt{stepLengthReduction} & -- & $0.7$ \\
		\texttt{minPrimalStep} & -- & $0$ \\
		\texttt{minDualStep} & -- & $0$ \\
		\texttt{maxComplementarity} & -- & $10^{100}$ \\
	\end{tabular}
	\caption{Parameters used in Gurobi and SDPB for the plots in this paper. All plots are at $\Lambda = 39$, but lower $\Lambda$ results are useful for getting preliminary estimates of bootstrap bounds.}
	\label{tab:numerical_parameters}
\end{table}
\endgroup

Empirically, we find that Gurobi is only reliable at $\Lambda \leq 29$. We use Gurobi to estimate the allowed region of scaling dimensions using lower values of $\Lambda$, and then use SDPB to obtain the exact allowed region at $\Lambda = 39$. All the bounds in this paper are obtained via binary search with a resolution of $2\times 10^{-4}$, except for plots on the bottom arc of the fundamental $\grSL(2,\Z)$ domain for which we increase the resolution to $2\times 10^{-5}$.

\bibliographystyle{ssg}
\bibliography{4dIntBoot_updated}

\begingroup\raggedright\begin{thebibliography}{10}

\bibitem{Maldacena:1997re}
J.~M. Maldacena, ``{The Large $N$ limit of superconformal field theories and
  supergravity},'' {\em Int. J. Theor. Phys.} {\bf 38} (1999) 1113--1133,
  \href{https://arxiv.org/abs/hep-th/9711200}{{\tt hep-th/9711200}}. [Adv.
  Theor. Math. Phys.2,231(1998)].

\bibitem{Witten:1998qj}
E.~Witten, ``{Anti-de Sitter space and holography},'' {\em Adv. Theor. Math.
  Phys.} {\bf 2} (1998) 253--291,
  \href{https://arxiv.org/abs/hep-th/9802150}{{\tt hep-th/9802150}}.

\bibitem{Gubser:1998bc}
S.~S. Gubser, I.~R. Klebanov, and A.~M. Polyakov, ``{Gauge theory correlators
  from noncritical string theory},'' {\em Phys. Lett.} {\bf B428} (1998)
  105--114, \href{https://arxiv.org/abs/hep-th/9802109}{{\tt hep-th/9802109}}.

\bibitem{Velizhanin:2009gv}
V.~N. Velizhanin, ``{The Non-planar contribution to the four-loop universal
  anomalous dimension in N=4 Supersymmetric Yang-Mills theory},'' {\em JETP
  Lett.} {\bf 89} (2009) 593--596, \href{https://arxiv.org/abs/0902.4646}{{\tt
  0902.4646}}.

\bibitem{Eden:2012rr}
B.~Eden, ``{Three-loop universal structure constants in N=4 susy Yang-Mills
  theory},'' \href{https://arxiv.org/abs/1207.3112}{{\tt 1207.3112}}.

\bibitem{Fleury:2019ydf}
T.~Fleury and R.~Pereira, ``{Non-planar data of $ \mathcal{N} $ = 4 SYM},''
  {\em JHEP} {\bf 03} (2020) 003, \href{https://arxiv.org/abs/1910.09428}{{\tt
  1910.09428}}.

\bibitem{Eden:2016aqo}
B.~Eden and F.~Paul, ``{Half-BPS half-BPS twist two at four loops in N=4
  SYM},'' \href{https://arxiv.org/abs/1608.04222}{{\tt 1608.04222}}.

\bibitem{Goncalves:2016vir}
V.~Gon\c{c}alves, ``{Extracting OPE coefficient of Konishi at four loops},''
  {\em JHEP} {\bf 03} (2017) 079, \href{https://arxiv.org/abs/1607.02195}{{\tt
  1607.02195}}.

\bibitem{Beisert:2010jr}
N.~Beisert {\em et.~al.}, ``{Review of AdS/CFT Integrability: An Overview},''
  {\em Lett. Math. Phys.} {\bf 99} (2012) 3--32,
  \href{https://arxiv.org/abs/1012.3982}{{\tt 1012.3982}}.

\bibitem{Gromov:2017blm}
N.~Gromov, ``{Introduction to the Spectrum of $N=4$ SYM and the Quantum
  Spectral Curve},'' \href{https://arxiv.org/abs/1708.03648}{{\tt 1708.03648}}.

\bibitem{Beem:2013sza}
C.~Beem, M.~Lemos, P.~Liendo, W.~Peelaers, L.~Rastelli, and B.~C. van Rees,
  ``{Infinite Chiral Symmetry in Four Dimensions},'' {\em Commun. Math. Phys.}
  {\bf 336} (2015), no.~3 1359--1433,
  \href{https://arxiv.org/abs/1312.5344}{{\tt 1312.5344}}.

\bibitem{Pestun:2007rz}
V.~Pestun, ``{Localization of gauge theory on a four-sphere and supersymmetric
  Wilson loops},'' {\em Commun. Math. Phys.} {\bf 313} (2012) 71--129,
  \href{https://arxiv.org/abs/0712.2824}{{\tt 0712.2824}}.

\bibitem{Hama:2012bg}
N.~Hama and K.~Hosomichi, ``{Seiberg-Witten Theories on Ellipsoids},'' {\em
  JHEP} {\bf 09} (2012) 033, \href{https://arxiv.org/abs/1206.6359}{{\tt
  1206.6359}}. [Addendum: JHEP 10, 051 (2012)].

\bibitem{Giombi:2009ds}
S.~Giombi and V.~Pestun, ``{Correlators of local operators and 1/8 BPS Wilson
  loops on S**2 from 2d YM and matrix models},'' {\em JHEP} {\bf 10} (2010)
  033, \href{https://arxiv.org/abs/0906.1572}{{\tt 0906.1572}}.

\bibitem{Giombi:2009ek}
S.~Giombi and V.~Pestun, ``{The 1/2 BPS 't Hooft loops in N=4 SYM as instantons
  in 2d Yang-Mills},'' {\em J. Phys.} {\bf A46} (2013) 095402,
  \href{https://arxiv.org/abs/0909.4272}{{\tt 0909.4272}}.

\bibitem{Pestun:2016zxk}
V.~Pestun {\em et.~al.}, ``{Localization techniques in quantum field
  theories},'' {\em J. Phys. A} {\bf 50} (2017), no.~44 440301,
  \href{https://arxiv.org/abs/1608.02952}{{\tt 1608.02952}}.

\bibitem{Beem:2013qxa}
C.~Beem, L.~Rastelli, and B.~C. van Rees, ``{The $\mathcal N=4$ Superconformal
  Bootstrap},'' {\em Phys.Rev.Lett.} {\bf 111} (2013), no.~7 071601,
  \href{https://arxiv.org/abs/1304.1803}{{\tt 1304.1803}}.

\bibitem{Beem:2016wfs}
C.~Beem, L.~Rastelli, and B.~C. van Rees, ``{More ${\mathcal N}=4$
  superconformal bootstrap},'' {\em Phys. Rev.} {\bf D96} (2017), no.~4 046014,
  \href{https://arxiv.org/abs/1612.02363}{{\tt 1612.02363}}.

\bibitem{Alday:2013opa}
L.~F. Alday and A.~Bissi, ``{The superconformal bootstrap for structure
  constants},'' {\em JHEP} {\bf 2014} (2013), no.~9
  \href{https://arxiv.org/abs/1310.3757}{{\tt 1310.3757}}.

\bibitem{Bissi:2020jve}
A.~Bissi, A.~Manenti, and A.~Vichi, ``{Bootstrapping mixed correlators in
  $\mathcal{N}=4$ Super Yang-Mills},'' {\em JHEP} {\bf 2021} (2020), no.~5
  \href{https://arxiv.org/abs/2010.15126}{{\tt 2010.15126}}.

\bibitem{Alday:2021peq}
L.~F. Alday, S.~M. Chester, and T.~Hansen, ``{Modular invariant holographic
  correlators for $\mathcal{N}=4$ SYM with general gauge group},''
  \href{https://arxiv.org/abs/2110.13106}{{\tt 2110.13106}}.

\bibitem{Baggio:2017mas}
M.~Baggio, N.~Bobev, S.~M. Chester, E.~Lauria, and S.~S. Pufu, ``{Decoding a
  Three-Dimensional Conformal Manifold},'' {\em JHEP} {\bf 02} (2018) 062,
  \href{https://arxiv.org/abs/1712.02698}{{\tt 1712.02698}}.

\bibitem{Binder:2019jwn}
D.~J. Binder, S.~M. Chester, S.~S. Pufu, and Y.~Wang, ``{$\mathcal{N}=4$
  Super-Yang-Mills Correlators at Strong Coupling from String Theory and
  Localization},'' {\em JHEP} {\bf 2019} (2019), no.~12
  \href{https://arxiv.org/abs/1902.06263}{{\tt 1902.06263}}.

\bibitem{Chester:2020dja}
S.~M. Chester and S.~S. Pufu, ``{Far beyond the planar limit in
  strongly-coupled $ \mathcal{N} $ = 4 SYM},'' {\em JHEP} {\bf 01} (2021) 103,
  \href{https://arxiv.org/abs/2003.08412}{{\tt 2003.08412}}.

\bibitem{Russo:2012ay}
J.~G. Russo and K.~Zarembo, ``{Large N Limit of N=2 SU(N) Gauge Theories from
  Localization},'' {\em JHEP} {\bf 10} (2012) 082,
  \href{https://arxiv.org/abs/1207.3806}{{\tt 1207.3806}}.

\bibitem{Russo:2013qaa}
J.~G. Russo and K.~Zarembo, ``{Evidence for Large-N Phase Transitions in N=2*
  Theory},'' {\em JHEP} {\bf 04} (2013) 065,
  \href{https://arxiv.org/abs/1302.6968}{{\tt 1302.6968}}.

\bibitem{Russo:2013kea}
J.~G. Russo and K.~Zarembo, ``{Massive N=2 Gauge Theories at Large N},'' {\em
  JHEP} {\bf 11} (2013) 130, \href{https://arxiv.org/abs/1309.1004}{{\tt
  1309.1004}}.

\bibitem{Chester:2019pvm}
S.~M. Chester, ``{Genus-2 holographic correlator on AdS$_{5}$\texttimes{}
  S$^{5}$ from localization},'' {\em JHEP} {\bf 04} (2020) 193,
  \href{https://arxiv.org/abs/1908.05247}{{\tt 1908.05247}}.

\bibitem{Chester:2019jas}
S.~M. Chester, M.~B. Green, S.~S. Pufu, Y.~Wang, and C.~Wen, ``{Modular
  invariance in superstring theory from $ \mathcal{N} $ = 4
  super-Yang-Mills},'' {\em JHEP} {\bf 11} (2020) 016,
  \href{https://arxiv.org/abs/1912.13365}{{\tt 1912.13365}}.

\bibitem{Chester:2020vyz}
S.~M. Chester, M.~B. Green, S.~S. Pufu, Y.~Wang, and C.~Wen, ``{New modular
  invariants in $ \mathcal{N} $ = 4 Super-Yang-Mills theory},'' {\em JHEP} {\bf
  04} (2021) 212, \href{https://arxiv.org/abs/2008.02713}{{\tt 2008.02713}}.

\bibitem{Lin:2015wcg}
Y.-H. Lin, S.-H. Shao, D.~Simmons-Duffin, Y.~Wang, and X.~Yin, ``{$ \mathcal{N}
  $ = 4 superconformal bootstrap of the K3 CFT},'' {\em JHEP} {\bf 05} (2017)
  126, \href{https://arxiv.org/abs/1511.04065}{{\tt 1511.04065}}.

\bibitem{Rattazzi:2008pe}
R.~Rattazzi, V.~S. Rychkov, E.~Tonni, and A.~Vichi, ``{Bounding scalar operator
  dimensions in 4D CFT},'' {\em JHEP} {\bf 0812} (2008) 031,
  \href{https://arxiv.org/abs/0807.0004}{{\tt 0807.0004}}.

\bibitem{Dolan:2001tt}
F.~Dolan and H.~Osborn, ``{Superconformal symmetry, correlation functions and
  the operator product expansion},'' {\em Nucl.Phys.} {\bf B629} (2002) 3--73,
  \href{https://arxiv.org/abs/hep-th/0112251}{{\tt hep-th/0112251}}.

\bibitem{Nekrasov:2002qd}
N.~A. Nekrasov, ``{Seiberg-Witten prepotential from instanton counting},'' {\em
  Adv. Theor. Math. Phys.} {\bf 7} (2003), no.~5 831--864,
  \href{https://arxiv.org/abs/hep-th/0206161}{{\tt hep-th/0206161}}.

\bibitem{Nekrasov:2003rj}
N.~Nekrasov and A.~Okounkov, ``{Seiberg-Witten theory and random partitions},''
  {\em Prog. Math.} {\bf 244} (2006) 525--596,
  \href{https://arxiv.org/abs/hep-th/0306238}{{\tt hep-th/0306238}}.

\bibitem{Alday:2009aq}
L.~F. Alday, D.~Gaiotto, and Y.~Tachikawa, ``{Liouville Correlation Functions
  from Four-dimensional Gauge Theories},'' {\em Lett. Math. Phys.} {\bf 91}
  (2010) 167--197, \href{https://arxiv.org/abs/0906.3219}{{\tt 0906.3219}}.

\bibitem{Alday:2010vg}
L.~F. Alday and Y.~Tachikawa, ``{Affine SL(2) conformal blocks from 4d gauge
  theories},'' {\em Lett. Math. Phys.} {\bf 94} (2010) 87--114,
  \href{https://arxiv.org/abs/1005.4469}{{\tt 1005.4469}}.

\bibitem{Alday:2016tll}
L.~F. Alday and G.~P. Korchemsky, ``{Revisiting instanton corrections to the
  Konishi multiplet},'' {\em JHEP} {\bf 12} (2016) 005,
  \href{https://arxiv.org/abs/1605.06346}{{\tt 1605.06346}}.

\bibitem{Alday:2016bkq}
L.~F. Alday and G.~P. Korchemsky, ``{On instanton effects in the operator
  product expansion},'' {\em JHEP} {\bf 05} (2017) 049,
  \href{https://arxiv.org/abs/1610.01425}{{\tt 1610.01425}}.

\bibitem{Alday:2016jeo}
L.~F. Alday and G.~P. Korchemsky, ``{Instanton corrections to twist-two
  operators},'' {\em JHEP} {\bf 06} (2017) 008,
  \href{https://arxiv.org/abs/1609.08164}{{\tt 1609.08164}}.

\bibitem{Arutyunov:2002rs}
G.~Arutyunov, S.~Penati, A.~C. Petkou, A.~Santambrogio, and E.~Sokatchev,
  ``{Nonprotected operators in N=4 SYM and multiparticle states of AdS(5)
  SUGRA},'' {\em Nucl. Phys. B} {\bf 643} (2002) 49--78,
  \href{https://arxiv.org/abs/hep-th/0206020}{{\tt hep-th/0206020}}.

\bibitem{Beisert:2003tq}
N.~Beisert, C.~Kristjansen, and M.~Staudacher, ``{The Dilatation operator of
  conformal N=4 superYang-Mills theory},'' {\em Nucl. Phys. B} {\bf 664} (2003)
  131--184, \href{https://arxiv.org/abs/hep-th/0303060}{{\tt hep-th/0303060}}.

\bibitem{Poland:2011ey}
D.~Poland, D.~Simmons-Duffin, and A.~Vichi, ``{Carving Out the Space of 4D
  CFTs},'' {\em JHEP} {\bf 1205} (2012) 110,
  \href{https://arxiv.org/abs/1109.5176}{{\tt 1109.5176}}.

\bibitem{Dolan:2000ut}
F.~A. Dolan and H.~Osborn, ``{Conformal four point functions and the operator
  product expansion},'' {\em Nucl. Phys. B} {\bf 599} (2001) 459--496,
  \href{https://arxiv.org/abs/hep-th/0011040}{{\tt hep-th/0011040}}.

\bibitem{Komargodski:2016auf}
Z.~Komargodski and D.~Simmons-Duffin, ``{The Random-Bond Ising Model in 2.01
  and 3 Dimensions},'' {\em J. Phys. A} {\bf 50} (2017), no.~15 154001,
  \href{https://arxiv.org/abs/1603.04444}{{\tt 1603.04444}}.

\bibitem{Hogervorst:2013sma}
M.~Hogervorst and S.~Rychkov, ``{Radial Coordinates for Conformal Blocks},''
  {\em Phys.Rev.} {\bf D87} (2013), no.~10 106004,
  \href{https://arxiv.org/abs/1303.1111}{{\tt 1303.1111}}.

\bibitem{Kos:2013tga}
F.~Kos, D.~Poland, and D.~Simmons-Duffin, ``{Bootstrapping the $O(N)$ vector
  models},'' {\em JHEP} {\bf 06} (2014) 091,
  \href{https://arxiv.org/abs/1307.6856}{{\tt 1307.6856}}.

\bibitem{Kos:2014bka}
F.~Kos, D.~Poland, and D.~Simmons-Duffin, ``{Bootstrapping Mixed Correlators in
  the 3D Ising Model},'' {\em JHEP} {\bf 11} (2014) 109,
  \href{https://arxiv.org/abs/1406.4858}{{\tt 1406.4858}}.

\bibitem{ElShowk:2012ht}
S.~El-Showk, M.~F. Paulos, D.~Poland, S.~Rychkov, D.~Simmons-Duffin, {\em
  et.~al.}, ``{Solving the 3D Ising Model with the Conformal Bootstrap},'' {\em
  Phys.Rev.} {\bf D86} (2012) 025022,
  \href{https://arxiv.org/abs/1203.6064}{{\tt 1203.6064}}.

\bibitem{Simmons-Duffin:2015qma}
D.~Simmons-Duffin, ``{A Semidefinite Program Solver for the Conformal
  Bootstrap},'' {\em JHEP} {\bf 06} (2015) 174,
  \href{https://arxiv.org/abs/1502.02033}{{\tt 1502.02033}}.

\bibitem{Kos:2015mba}
F.~Kos, D.~Poland, D.~Simmons-Duffin, and A.~Vichi, ``{Bootstrapping the O(N)
  Archipelago},'' {\em JHEP} {\bf 11} (2015) 106,
  \href{https://arxiv.org/abs/1504.07997}{{\tt 1504.07997}}.

\bibitem{Kos:2016ysd}
F.~Kos, D.~Poland, D.~Simmons-Duffin, and A.~Vichi, ``{Precision Islands in the
  Ising and $O(N)$ Models},'' {\em JHEP} {\bf 08} (2016) 036,
  \href{https://arxiv.org/abs/1603.04436}{{\tt 1603.04436}}.

\bibitem{Chester:2019ifh}
S.~M. Chester, W.~Landry, J.~Liu, D.~Poland, D.~Simmons-Duffin, N.~Su, and
  A.~Vichi, ``{Carving out OPE space and precise $O(2)$ model critical
  exponents},'' {\em JHEP} {\bf 06} (2020) 142,
  \href{https://arxiv.org/abs/1912.03324}{{\tt 1912.03324}}.

\bibitem{Chester:2020iyt}
S.~M. Chester, W.~Landry, J.~Liu, D.~Poland, D.~Simmons-Duffin, N.~Su, and
  A.~Vichi, ``{Bootstrapping Heisenberg Magnets and their Cubic Instability},''
  \href{https://arxiv.org/abs/2011.14647}{{\tt 2011.14647}}.

\bibitem{Losev:2003py}
A.~S. Losev, A.~Marshakov, and N.~A. Nekrasov, ``{Small instantons, little
  strings and free fermions},'' in {\em {From Fields to Strings:
  Circumnavigating Theoretical Physics: A Conference in Tribute to Ian Kogan}},
  pp.~581--621, 2003.
\newblock \href{https://arxiv.org/abs/hep-th/0302191}{{\tt hep-th/0302191}}.

\bibitem{Nakajima:2003uh}
H.~Nakajima and K.~Yoshioka, ``{Lectures on instanton counting},'' in {\em {CRM
  Workshop on Algebraic Structures and Moduli Spaces}}, 2003.
\newblock \href{https://arxiv.org/abs/math/0311058}{{\tt math/0311058}}.

\bibitem{Fucito:2015ofa}
F.~Fucito, J.~F. Morales, and R.~Poghossian, ``{Wilson loops and chiral
  correlators on squashed spheres},'' {\em JHEP} {\bf 11} (2015) 064,
  \href{https://arxiv.org/abs/1507.05426}{{\tt 1507.05426}}.

\bibitem{Collier:2017shs}
S.~Collier, P.~Kravchuk, Y.-H. Lin, and X.~Yin, ``{Bootstrapping the Spectral
  Function: On the Uniqueness of Liouville and the Universality of BTZ},'' {\em
  JHEP} {\bf 09} (2018) 150, \href{https://arxiv.org/abs/1702.00423}{{\tt
  1702.00423}}.

\bibitem{Dorigoni:2021guq}
D.~Dorigoni, M.~B. Green, and C.~Wen, ``{Exact properties of an integrated
  correlator in $ \mathcal{N} $ = 4 SU(N) SYM},'' {\em JHEP} {\bf 05} (2021)
  089, \href{https://arxiv.org/abs/2102.09537}{{\tt 2102.09537}}.

\bibitem{Kapustin:2009kz}
A.~Kapustin, B.~Willett, and I.~Yaakov, ``{Exact Results for Wilson Loops in
  Superconformal Chern-Simons Theories with Matter},'' {\em JHEP} {\bf 1003}
  (2010) 089, \href{https://arxiv.org/abs/0909.4559}{{\tt 0909.4559}}.

\bibitem{Hama:2011ea}
N.~Hama, K.~Hosomichi, and S.~Lee, ``{SUSY Gauge Theories on Squashed
  Three-Spheres},'' {\em JHEP} {\bf 1105} (2011) 014,
  \href{https://arxiv.org/abs/1102.4716}{{\tt 1102.4716}}.

\bibitem{Imamura:2012xg}
Y.~Imamura, ``{Supersymmetric theories on squashed five-sphere},'' {\em PTEP}
  {\bf 2013} (2013) 013B04, \href{https://arxiv.org/abs/1209.0561}{{\tt
  1209.0561}}.

\bibitem{Imamura:2012bm}
Y.~Imamura, ``{Perturbative partition function for squashed $S^5$},'' {\em
  PTEP} {\bf 2013} (2013), no.~7 073B01,
  \href{https://arxiv.org/abs/1210.6308}{{\tt 1210.6308}}.

\bibitem{Aharony:2008ug}
O.~Aharony, O.~Bergman, D.~L. Jafferis, and J.~Maldacena, ``{${\cal N}=6$
  superconformal Chern-Simons-matter theories, M2-branes and their gravity
  duals},'' {\em JHEP} {\bf 10} (2008) 091,
  \href{https://arxiv.org/abs/0806.1218}{{\tt 0806.1218}}.

\bibitem{Aharony:2008gk}
O.~Aharony, O.~Bergman, and D.~L. Jafferis, ``{Fractional M2-branes},'' {\em
  JHEP} {\bf 0811} (2008) 043, \href{https://arxiv.org/abs/0807.4924}{{\tt
  0807.4924}}.

\bibitem{Binder:2020ckj}
D.~J. Binder, S.~M. Chester, M.~Jerdee, and S.~S. Pufu, ``{The 3d
  $\mathcal{N}=6$ Bootstrap: From Higher Spins to Strings to Membranes},'' {\em
  JHEP} {\bf 2021} (2020), no.~5 \href{https://arxiv.org/abs/2011.05728}{{\tt
  2011.05728}}.

\bibitem{Agmon:2017xes}
N.~B. Agmon, S.~M. Chester, and S.~S. Pufu, ``{Solving M-theory with the
  Conformal Bootstrap},'' {\em JHEP} {\bf 2018} (2017), no.~6
  \href{https://arxiv.org/abs/1711.07343}{{\tt 1711.07343}}.

\bibitem{Binder:2019mpb}
D.~J. Binder, S.~M. Chester, and S.~S. Pufu, ``{AdS$_{4}$/CFT$_{3}$ from weak
  to strong string coupling},'' {\em JHEP} {\bf 01} (2020) 034,
  \href{https://arxiv.org/abs/1906.07195}{{\tt 1906.07195}}.

\bibitem{Binder:2018yvd}
D.~J. Binder, S.~M. Chester, and S.~S. Pufu, ``{Absence of $D^4 R^4$ in
  M-Theory From ABJM},'' \href{https://arxiv.org/abs/1808.10554}{{\tt
  1808.10554}}.

\bibitem{Binder:2021cif}
D.~J. Binder, S.~M. Chester, and M.~Jerdee, ``{ABJ Correlators with Weakly
  Broken Higher Spin Symmetry},'' {\em JHEP} {\bf 04} (2021) 242,
  \href{https://arxiv.org/abs/2103.01969}{{\tt 2103.01969}}.

\bibitem{Agmon:2019imm}
N.~B. Agmon, S.~M. Chester, and S.~S. Pufu, ``{The M-theory Archipelago},''
  {\em JHEP} {\bf 02} (2020) 010, \href{https://arxiv.org/abs/1907.13222}{{\tt
  1907.13222}}.

\bibitem{Chester:2020jay}
S.~M. Chester, R.~R. Kalloor, and A.~Sharon, ``{3d $ \mathcal{N} $ = 4 OPE
  coefficients from Fermi gas},'' {\em JHEP} {\bf 07} (2020) 041,
  \href{https://arxiv.org/abs/2004.13603}{{\tt 2004.13603}}.

\bibitem{Chester:2021gdw}
S.~M. Chester, R.~R. Kalloor, and A.~Sharon, ``{Squashing, Mass, and Holography
  for 3d Sphere Free Energy},'' {\em JHEP} {\bf 2021} (2021), no.~4
  \href{https://arxiv.org/abs/2102.05643}{{\tt 2102.05643}}.

\bibitem{Chester:2018lbz}
S.~M. Chester, ``{AdS$_4$/CFT$_3$ for Unprotected Operators},'' {\em JHEP} {\bf
  2018} (2018), no.~7 \href{https://arxiv.org/abs/1803.01379}{{\tt
  1803.01379}}.

\bibitem{Alday:2021ymb}
L.~F. Alday, S.~M. Chester, and H.~Raj, ``{ABJM at Strong Coupling from
  M-theory, Localization, and Lorentzian Inversion},''
  \href{https://arxiv.org/abs/2107.10274}{{\tt 2107.10274}}.

\bibitem{Chester:2018aca}
S.~M. Chester, S.~S. Pufu, and X.~Yin, ``{The M-Theory S-Matrix From ABJM:
  Beyond 11D Supergravity},'' {\em JHEP} {\bf 2018} (2018), no.~8
  \href{https://arxiv.org/abs/1804.00949}{{\tt 1804.00949}}.

\bibitem{gurobi}
{Gurobi Optimization, LLC}, ``{Gurobi Optimizer Reference Manual},'' 2021.

\end{thebibliography}\endgroup

\end{document}